\documentclass[twocolumn,traditabstract,longauth]{aa}

\usepackage{txfonts}
\usepackage{graphicx}
\usepackage{booktabs}
\usepackage{rotating}
\usepackage[breaklinks, colorlinks, citecolor=blue]{hyperref}
\usepackage{natbib}
\usepackage{url}
\usepackage{amssymb}
\bibpunct{(}{)}{;}{a}{}{,} 

\newcommand{\hi}{\ion{H}{i}}
\newcommand{\Oi}{\ion{O}{i}}
\newcommand{\Siii}{\ion{Si}{ii}}
\newcommand{\Feii}{\ion{Fe}{ii}}
\newcommand{\cmm}{${\rm cm}^{-2}$}
\newcommand{\kms}{km\,s$^{-1}$}
\newcommand{\nh}{$N_{\rm H I}$}

\newcommand{\IRAS}{{\it IRAS\/}}
\newcommand{\COBE}{{\it COBE\/}}
\newcommand{\ISO}{{\it ISO\/}}
\newcommand{\Spitzer}{{\it Spitzer\/}}
\newcommand{\Herschel}{{\it Herschel\/}}

\def\setsymbol#1#2{\expandafter\def\csname #1\endcsname{#2}}
\def\getsymbol#1{\csname #1\endcsname}

\def\Planck{{\it Planck\/}}

\def\allearlypapers{\nocite{planck2011-1.1, planck2011-1.3, planck2011-1.4, planck2011-1.5, planck2011-1.6, planck2011-1.7, planck2011-1.10, planck2011-1.10sup, planck2011-5.1a, planck2011-5.1b, planck2011-5.2a, planck2011-5.2b, planck2011-5.2c, planck2011-6.1, planck2011-6.2, planck2011-6.3a, planck2011-6.4a, planck2011-6.4b, planck2011-6.6, planck2011-7.0, planck2011-7.2, planck2011-7.3, planck2011-7.7a, planck2011-7.7b, planck2011-7.12, planck2011-7.13}}

\newbox\tablebox    \newdimen\tablewidth
\def\leaderfil{\leaders\hbox to 5pt{\hss.\hss}\hfil}
\def\tablenote#1 #2\par{\begingroup \parindent=0.8em
    \abovedisplayshortskip=0pt\belowdisplayshortskip=0pt
    \noindent
    $$\hss\vbox{\hsize\tablewidth \hangindent=\parindent \hangafter=1 \noindent
    \hbox to \parindent{\sup{\rm #1}\hss}\strut#2\strut\par}\hss$$
    \endgroup}

\def\L2{\ifmmode L_2\else $L_2$\fi}

\def\DeltaT{\ifmmode \Delta T\else $\Delta T$\fi}
\def\deltat{\ifmmode \Delta t\else $\Delta t$\fi}
\def\fknee{\ifmmode f_{\rm knee}\else $f_{\rm knee}$\fi}
\def\Fmax{\ifmmode F_{\rm max}\else $F_{\rm max}$\fi}
\def\solar{\ifmmode{\rm M}_{\mathord\odot}\else${\rm M}_{\mathord\odot}$\fi}

\def\inv{\ifmmode^{-1}\else$^{-1}$\fi}
\def\mo{\ifmmode^{-1}\else$^{-1}$\fi}
\def\sup#1{\ifmmode ^{\rm #1}\else $^{\rm #1}$\fi}
\def\expo#1{\ifmmode \times 10^{#1}\else $\times 10^{#1}$\fi}
\def\,{\thinspace}
\def\lsim{\mathrel{\raise .4ex\hbox{\rlap{$<$}\lower 1.2ex\hbox{$\sim$}}}}
\def\gsim{\mathrel{\raise .4ex\hbox{\rlap{$>$}\lower 1.2ex\hbox{$\sim$}}}}

\def\simprop{\mathrel{\raise .4ex\hbox{\rlap{$\propto$}\lower 1.2ex\hbox{$\sim$}}}}
\def\deg{\ifmmode^\circ\else$^\circ$\fi}
\def\pdeg{\ifmmode $\setbox0=\hbox{$^{\circ}$}\rlap{\hskip.11\wd0 .}$^{\circ}
          \else \setbox0=\hbox{$^{\circ}$}\rlap{\hskip.11\wd0 .}$^{\circ}$\fi}
\def\arcs{\ifmmode {^{\scriptstyle\prime\prime}}
          \else $^{\scriptstyle\prime\prime}$\fi}
\def\arcm{\ifmmode {^{\scriptstyle\prime}}
          \else $^{\scriptstyle\prime}$\fi}
\newdimen\sa  \newdimen\sb
\def\parcs{\sa=.07em \sb=.03em
     \ifmmode \hbox{\rlap{.}}^{\scriptstyle\prime\kern -\sb\prime}\hbox{\kern -\sa}
     \else \rlap{.}$^{\scriptstyle\prime\kern -\sb\prime}$\kern -\sa\fi}
\def\parcm{\sa=.08em \sb=.03em
     \ifmmode \hbox{\rlap{.}\kern\sa}^{\scriptstyle\prime}\hbox{\kern-\sb}
     \else \rlap{.}\kern\sa$^{\scriptstyle\prime}$\kern-\sb\fi}
\def\ra[#1 #2 #3.#4]{#1\sup{h}#2\sup{m}#3\sup{s}\llap.#4}
\def\dec[#1 #2 #3.#4]{#1\deg#2\arcm#3\arcs\llap.#4}
\def\deco[#1 #2 #3]{#1\deg#2\arcm#3\arcs}
\def\rra[#1 #2]{#1\sup{h}#2\sup{m}}

\def\dots{\relax\ifmmode \ldots\else $\ldots$\fi}
\def\WHzsr{\ifmmode $W\,Hz\mo\,sr\mo$\else W\,Hz\mo\,sr\mo\fi}
\def\mHz{\ifmmode $\,mHz$\else \,mHz\fi}
\def\GHz{\ifmmode $\,GHz$\else \,GHz\fi}
\def\mKs{\ifmmode $\,mK\,s$^{1/2}\else \,mK\,s$^{1/2}$\fi}
\def\muKs{\ifmmode \,\mu$K\,s$^{1/2}\else \,$\mu$K\,s$^{1/2}$\fi}
\def\muKRJs{\ifmmode \,\mu$K$_{\rm RJ}$\,s$^{1/2}\else \,$\mu$K$_{\rm RJ}$\,s$^{1/2}$\fi}
\def\muKHz{\ifmmode \,\mu$K\,Hz$^{-1/2}\else \,$\mu$K\,Hz$^{-1/2}$\fi}
\def\MJysr{\ifmmode \,$MJy\,sr\mo$\else \,MJy\,sr\mo\fi}
\def\MJysrmK{\ifmmode \,$MJy\,sr\mo$\,mK$_{\rm CMB}\mo\else \,MJy\,sr\mo\,mK$_{\rm CMB}\mo$\fi}
\def\microns{\ifmmode \,\mu$m$\else \,$\mu$m\fi}

\def\muK{\ifmmode \,\mu$K$\else \,$\mu$\hbox{K}\fi}
\def\microK{\ifmmode \,\mu$K$\else \,$\mu$\hbox{K}\fi}
\def\muW{\ifmmode \,\mu$W$\else \,$\mu$\hbox{W}\fi}
\def\kms{\ifmmode $\,km\,s$^{-1}\else \,km\,s$^{-1}$\fi}
\def\kmsMpc{\ifmmode $\,\kms\,Mpc\mo$\else \,\kms\,Mpc\mo\fi}
\setsymbol{LFI:center:frequency:70GHz:units}{70.3\,GHz}
\setsymbol{LFI:center:frequency:44GHz:units}{44.1\,GHz}
\setsymbol{LFI:center:frequency:30GHz:units}{28.5\,GHz}

\setsymbol{LFI:center:frequency:70GHz}{70.3}
\setsymbol{LFI:center:frequency:44GHz}{44.1}
\setsymbol{LFI:center:frequency:30GHz}{28.5}

\setsymbol{LFI:center:frequency:LFI18:Rad:M:units}{71.7\GHz}
\setsymbol{LFI:center:frequency:LFI19:Rad:M:units}{67.5\GHz}
\setsymbol{LFI:center:frequency:LFI20:Rad:M:units}{69.2\GHz}
\setsymbol{LFI:center:frequency:LFI21:Rad:M:units}{70.4\GHz}
\setsymbol{LFI:center:frequency:LFI22:Rad:M:units}{71.5\GHz}
\setsymbol{LFI:center:frequency:LFI23:Rad:M:units}{70.8\GHz}
\setsymbol{LFI:center:frequency:LFI24:Rad:M:units}{44.4\GHz}
\setsymbol{LFI:center:frequency:LFI25:Rad:M:units}{44.0\GHz}
\setsymbol{LFI:center:frequency:LFI26:Rad:M:units}{43.9\GHz}
\setsymbol{LFI:center:frequency:LFI27:Rad:M:units}{28.3\GHz}
\setsymbol{LFI:center:frequency:LFI28:Rad:M:units}{28.8\GHz}
\setsymbol{LFI:center:frequency:LFI18:Rad:S:units}{70.1\GHz}
\setsymbol{LFI:center:frequency:LFI19:Rad:S:units}{69.6\GHz}
\setsymbol{LFI:center:frequency:LFI20:Rad:S:units}{69.5\GHz}
\setsymbol{LFI:center:frequency:LFI21:Rad:S:units}{69.5\GHz}
\setsymbol{LFI:center:frequency:LFI22:Rad:S:units}{72.8\GHz}
\setsymbol{LFI:center:frequency:LFI23:Rad:S:units}{71.3\GHz}
\setsymbol{LFI:center:frequency:LFI24:Rad:S:units}{44.1\GHz}
\setsymbol{LFI:center:frequency:LFI25:Rad:S:units}{44.1\GHz}
\setsymbol{LFI:center:frequency:LFI26:Rad:S:units}{44.1\GHz}
\setsymbol{LFI:center:frequency:LFI27:Rad:S:units}{28.5\GHz}
\setsymbol{LFI:center:frequency:LFI28:Rad:S:units}{28.2\GHz}

\setsymbol{LFI:center:frequency:LFI18:Rad:M}{71.7}
\setsymbol{LFI:center:frequency:LFI19:Rad:M}{67.5}
\setsymbol{LFI:center:frequency:LFI20:Rad:M}{69.2}
\setsymbol{LFI:center:frequency:LFI21:Rad:M}{70.4}
\setsymbol{LFI:center:frequency:LFI22:Rad:M}{71.5}
\setsymbol{LFI:center:frequency:LFI23:Rad:M}{70.8}
\setsymbol{LFI:center:frequency:LFI24:Rad:M}{44.4}
\setsymbol{LFI:center:frequency:LFI25:Rad:M}{44.0}
\setsymbol{LFI:center:frequency:LFI26:Rad:M}{43.9}
\setsymbol{LFI:center:frequency:LFI27:Rad:M}{28.3}
\setsymbol{LFI:center:frequency:LFI28:Rad:M}{28.8}
\setsymbol{LFI:center:frequency:LFI18:Rad:S}{70.1}
\setsymbol{LFI:center:frequency:LFI19:Rad:S}{69.6}
\setsymbol{LFI:center:frequency:LFI20:Rad:S}{69.5}
\setsymbol{LFI:center:frequency:LFI21:Rad:S}{69.5}
\setsymbol{LFI:center:frequency:LFI22:Rad:S}{72.8}
\setsymbol{LFI:center:frequency:LFI23:Rad:S}{71.3}
\setsymbol{LFI:center:frequency:LFI24:Rad:S}{44.1}
\setsymbol{LFI:center:frequency:LFI25:Rad:S}{44.1}
\setsymbol{LFI:center:frequency:LFI26:Rad:S}{44.1}
\setsymbol{LFI:center:frequency:LFI27:Rad:S}{28.5}
\setsymbol{LFI:center:frequency:LFI28:Rad:S}{28.2}

\setsymbol{LFI:white:noise:sensitivity:70GHz:units}{152.6\muKs}
\setsymbol{LFI:white:noise:sensitivity:44GHz:units}{173.1\muKs}
\setsymbol{LFI:white:noise:sensitivity:30GHz:units}{146.8\muKs}

\setsymbol{LFI:white:noise:sensitivity:70GHz}{152.6}
\setsymbol{LFI:white:noise:sensitivity:44GHz}{173.1}
\setsymbol{LFI:white:noise:sensitivity:30GHz}{146.8}

\setsymbol{LFI:white:noise:sensitivity:LFI18:Rad:M:units}{512.0\muKs}
\setsymbol{LFI:white:noise:sensitivity:LFI19:Rad:M:units}{581.4\muKs}
\setsymbol{LFI:white:noise:sensitivity:LFI20:Rad:M:units}{590.8\muKs}
\setsymbol{LFI:white:noise:sensitivity:LFI21:Rad:M:units}{455.2\muKs}
\setsymbol{LFI:white:noise:sensitivity:LFI22:Rad:M:units}{492.0\muKs}
\setsymbol{LFI:white:noise:sensitivity:LFI23:Rad:M:units}{507.7\muKs}
\setsymbol{LFI:white:noise:sensitivity:LFI24:Rad:M:units}{462.2\muKs}
\setsymbol{LFI:white:noise:sensitivity:LFI25:Rad:M:units}{413.6\muKs}
\setsymbol{LFI:white:noise:sensitivity:LFI26:Rad:M:units}{478.6\muKs}
\setsymbol{LFI:white:noise:sensitivity:LFI27:Rad:M:units}{277.7\muKs}
\setsymbol{LFI:white:noise:sensitivity:LFI28:Rad:M:units}{312.3\muKs}
\setsymbol{LFI:white:noise:sensitivity:LFI18:Rad:S:units}{465.7\muKs}
\setsymbol{LFI:white:noise:sensitivity:LFI19:Rad:S:units}{555.6\muKs}
\setsymbol{LFI:white:noise:sensitivity:LFI20:Rad:S:units}{623.2\muKs}
\setsymbol{LFI:white:noise:sensitivity:LFI21:Rad:S:units}{564.1\muKs}
\setsymbol{LFI:white:noise:sensitivity:LFI22:Rad:S:units}{534.4\muKs}
\setsymbol{LFI:white:noise:sensitivity:LFI23:Rad:S:units}{542.4\muKs}
\setsymbol{LFI:white:noise:sensitivity:LFI24:Rad:S:units}{399.2\muKs}
\setsymbol{LFI:white:noise:sensitivity:LFI25:Rad:S:units}{392.6\muKs}
\setsymbol{LFI:white:noise:sensitivity:LFI26:Rad:S:units}{418.6\muKs}
\setsymbol{LFI:white:noise:sensitivity:LFI27:Rad:S:units}{302.9\muKs}
\setsymbol{LFI:white:noise:sensitivity:LFI28:Rad:S:units}{285.3\muKs}

\setsymbol{LFI:white:noise:sensitivity:LFI18:Rad:M}{512.0}
\setsymbol{LFI:white:noise:sensitivity:LFI19:Rad:M}{581.4}
\setsymbol{LFI:white:noise:sensitivity:LFI20:Rad:M}{590.8}
\setsymbol{LFI:white:noise:sensitivity:LFI21:Rad:M}{455.2}
\setsymbol{LFI:white:noise:sensitivity:LFI22:Rad:M}{492.0}
\setsymbol{LFI:white:noise:sensitivity:LFI23:Rad:M}{507.7}
\setsymbol{LFI:white:noise:sensitivity:LFI24:Rad:M}{462.2}
\setsymbol{LFI:white:noise:sensitivity:LFI25:Rad:M}{413.6}
\setsymbol{LFI:white:noise:sensitivity:LFI26:Rad:M}{478.6}
\setsymbol{LFI:white:noise:sensitivity:LFI27:Rad:M}{277.7}
\setsymbol{LFI:white:noise:sensitivity:LFI28:Rad:M}{312.3}
\setsymbol{LFI:white:noise:sensitivity:LFI18:Rad:S}{465.7}
\setsymbol{LFI:white:noise:sensitivity:LFI19:Rad:S}{555.6}
\setsymbol{LFI:white:noise:sensitivity:LFI20:Rad:S}{623.2}
\setsymbol{LFI:white:noise:sensitivity:LFI21:Rad:S}{564.1}
\setsymbol{LFI:white:noise:sensitivity:LFI22:Rad:S}{534.4}
\setsymbol{LFI:white:noise:sensitivity:LFI23:Rad:S}{542.4}
\setsymbol{LFI:white:noise:sensitivity:LFI24:Rad:S}{399.2}
\setsymbol{LFI:white:noise:sensitivity:LFI25:Rad:S}{392.6}
\setsymbol{LFI:white:noise:sensitivity:LFI26:Rad:S}{418.6}
\setsymbol{LFI:white:noise:sensitivity:LFI27:Rad:S}{302.9}
\setsymbol{LFI:white:noise:sensitivity:LFI28:Rad:S}{285.3}

\setsymbol{LFI:knee:frequency:70GHz:units}{29.5\mHz}
\setsymbol{LFI:knee:frequency:44GHz:units}{56.2\mHz}
\setsymbol{LFI:knee:frequency:30GHz:units}{113.7\mHz}

\setsymbol{LFI:knee:frequency:70GHz}{29.5}
\setsymbol{LFI:knee:frequency:44GHz}{56.2}
\setsymbol{LFI:knee:frequency:30GHz}{113.7}

\setsymbol{LFI:knee:frequency:LFI18:Rad:M:units}{16.3\mHz}
\setsymbol{LFI:knee:frequency:LFI19:Rad:M:units}{15.1\mHz}
\setsymbol{LFI:knee:frequency:LFI20:Rad:M:units}{18.7\mHz}
\setsymbol{LFI:knee:frequency:LFI21:Rad:M:units}{37.2\mHz}
\setsymbol{LFI:knee:frequency:LFI22:Rad:M:units}{12.7\mHz}
\setsymbol{LFI:knee:frequency:LFI23:Rad:M:units}{34.6\mHz}
\setsymbol{LFI:knee:frequency:LFI24:Rad:M:units}{46.2\mHz}
\setsymbol{LFI:knee:frequency:LFI25:Rad:M:units}{24.9\mHz}
\setsymbol{LFI:knee:frequency:LFI26:Rad:M:units}{67.6\mHz}
\setsymbol{LFI:knee:frequency:LFI27:Rad:M:units}{187.4\mHz}
\setsymbol{LFI:knee:frequency:LFI28:Rad:M:units}{122.2\mHz}
\setsymbol{LFI:knee:frequency:LFI18:Rad:S:units}{17.7\mHz}
\setsymbol{LFI:knee:frequency:LFI19:Rad:S:units}{22.0\mHz}
\setsymbol{LFI:knee:frequency:LFI20:Rad:S:units}{8.7\mHz}
\setsymbol{LFI:knee:frequency:LFI21:Rad:S:units}{25.9\mHz}
\setsymbol{LFI:knee:frequency:LFI22:Rad:S:units}{15.8\mHz}
\setsymbol{LFI:knee:frequency:LFI23:Rad:S:units}{129.8\mHz}
\setsymbol{LFI:knee:frequency:LFI24:Rad:S:units}{100.9\mHz}
\setsymbol{LFI:knee:frequency:LFI25:Rad:S:units}{38.9\mHz}
\setsymbol{LFI:knee:frequency:LFI26:Rad:S:units}{58.9\mHz}
\setsymbol{LFI:knee:frequency:LFI27:Rad:S:units}{104.4\mHz}
\setsymbol{LFI:knee:frequency:LFI28:Rad:S:units}{40.7\mHz}

\setsymbol{LFI:knee:frequency:LFI18:Rad:M}{16.3}
\setsymbol{LFI:knee:frequency:LFI19:Rad:M}{15.1}
\setsymbol{LFI:knee:frequency:LFI20:Rad:M}{18.7}
\setsymbol{LFI:knee:frequency:LFI21:Rad:M}{37.2}
\setsymbol{LFI:knee:frequency:LFI22:Rad:M}{12.7}
\setsymbol{LFI:knee:frequency:LFI23:Rad:M}{34.6}
\setsymbol{LFI:knee:frequency:LFI24:Rad:M}{46.2}
\setsymbol{LFI:knee:frequency:LFI25:Rad:M}{24.9}
\setsymbol{LFI:knee:frequency:LFI26:Rad:M}{67.6}
\setsymbol{LFI:knee:frequency:LFI27:Rad:M}{187.4}
\setsymbol{LFI:knee:frequency:LFI28:Rad:M}{122.2}
\setsymbol{LFI:knee:frequency:LFI18:Rad:S}{17.7}
\setsymbol{LFI:knee:frequency:LFI19:Rad:S}{22.0}
\setsymbol{LFI:knee:frequency:LFI20:Rad:S}{8.7}
\setsymbol{LFI:knee:frequency:LFI21:Rad:S}{25.9}
\setsymbol{LFI:knee:frequency:LFI22:Rad:S}{15.8}
\setsymbol{LFI:knee:frequency:LFI23:Rad:S}{129.8}
\setsymbol{LFI:knee:frequency:LFI24:Rad:S}{100.9}
\setsymbol{LFI:knee:frequency:LFI25:Rad:S}{38.9}
\setsymbol{LFI:knee:frequency:LFI26:Rad:S}{58.9}
\setsymbol{LFI:knee:frequency:LFI27:Rad:S}{104.4}
\setsymbol{LFI:knee:frequency:LFI28:Rad:S}{40.7}

\setsymbol{LFI:slope:70GHz:units}{$-1.03$\mHz}
\setsymbol{LFI:slope:44GHz:units}{$-0.89$\mHz}
\setsymbol{LFI:slope:30GHz:units}{$-0.87$\mHz}

\setsymbol{LFI:slope:70GHz}{$-1.03$}
\setsymbol{LFI:slope:44GHz}{$-0.89$}
\setsymbol{LFI:slope:30GHz}{$-0.87$}

\setsymbol{LFI:slope:LFI18:Rad:M:units}{$-1.04$\mHz}
\setsymbol{LFI:slope:LFI19:Rad:M:units}{$-1.09$\mHz}
\setsymbol{LFI:slope:LFI20:Rad:M:units}{$-0.69$\mHz}
\setsymbol{LFI:slope:LFI21:Rad:M:units}{$-1.56$\mHz}
\setsymbol{LFI:slope:LFI22:Rad:M:units}{$-1.01$\mHz}
\setsymbol{LFI:slope:LFI23:Rad:M:units}{$-0.96$\mHz}
\setsymbol{LFI:slope:LFI24:Rad:M:units}{$-0.83$\mHz}
\setsymbol{LFI:slope:LFI25:Rad:M:units}{$-0.91$\mHz}
\setsymbol{LFI:slope:LFI26:Rad:M:units}{$-0.95$\mHz}
\setsymbol{LFI:slope:LFI27:Rad:M:units}{$-0.87$\mHz}
\setsymbol{LFI:slope:LFI28:Rad:M:units}{$-0.88$\mHz}
\setsymbol{LFI:slope:LFI18:Rad:S:units}{$-1.15$\mHz}
\setsymbol{LFI:slope:LFI19:Rad:S:units}{$-1.00$\mHz}
\setsymbol{LFI:slope:LFI20:Rad:S:units}{$-0.95$\mHz}
\setsymbol{LFI:slope:LFI21:Rad:S:units}{$-0.92$\mHz}
\setsymbol{LFI:slope:LFI22:Rad:S:units}{$-1.01$\mHz}
\setsymbol{LFI:slope:LFI23:Rad:S:units}{$-0.95$\mHz}
\setsymbol{LFI:slope:LFI24:Rad:S:units}{$-0.73$\mHz}
\setsymbol{LFI:slope:LFI25:Rad:S:units}{$-1.16$\mHz}
\setsymbol{LFI:slope:LFI26:Rad:S:units}{$-0.79$\mHz}
\setsymbol{LFI:slope:LFI27:Rad:S:units}{$-0.82$\mHz}
\setsymbol{LFI:slope:LFI28:Rad:S:units}{$-0.91$\mHz}

\setsymbol{LFI:slope:LFI18:Rad:M}{$-1.04$}
\setsymbol{LFI:slope:LFI19:Rad:M}{$-1.09$}
\setsymbol{LFI:slope:LFI20:Rad:M}{$-0.69$}
\setsymbol{LFI:slope:LFI21:Rad:M}{$-1.56$}
\setsymbol{LFI:slope:LFI22:Rad:M}{$-1.01$}
\setsymbol{LFI:slope:LFI23:Rad:M}{$-0.96$}
\setsymbol{LFI:slope:LFI24:Rad:M}{$-0.83$}
\setsymbol{LFI:slope:LFI25:Rad:M}{$-0.91$}
\setsymbol{LFI:slope:LFI26:Rad:M}{$-0.95$}
\setsymbol{LFI:slope:LFI27:Rad:M}{$-0.87$}
\setsymbol{LFI:slope:LFI28:Rad:M}{$-0.88$}
\setsymbol{LFI:slope:LFI18:Rad:S}{$-1.15$}
\setsymbol{LFI:slope:LFI19:Rad:S}{$-1.00$}
\setsymbol{LFI:slope:LFI20:Rad:S}{$-0.95$}
\setsymbol{LFI:slope:LFI21:Rad:S}{$-0.92$}
\setsymbol{LFI:slope:LFI22:Rad:S}{$-1.01$}
\setsymbol{LFI:slope:LFI23:Rad:S}{$-0.95$}
\setsymbol{LFI:slope:LFI24:Rad:S}{$-0.73$}
\setsymbol{LFI:slope:LFI25:Rad:S}{$-1.16$}
\setsymbol{LFI:slope:LFI26:Rad:S}{$-0.79$}
\setsymbol{LFI:slope:LFI27:Rad:S}{$-0.82$}
\setsymbol{LFI:slope:LFI28:Rad:S}{$-0.91$}

\setsymbol{LFI:FWHM:70GHz:units}{13\parcm01}
\setsymbol{LFI:FWHM:44GHz:units}{27\parcm92}
\setsymbol{LFI:FWHM:30GHz:units}{32\parcm65}

\setsymbol{LFI:FWHM:70GHz}{13.01}
\setsymbol{LFI:FWHM:44GHz}{27.92}
\setsymbol{LFI:FWHM:30GHz}{32.65}

\setsymbol{LFI:FWHM:LFI18:units}{13\parcm39}
\setsymbol{LFI:FWHM:LFI19:units}{13\parcm01}
\setsymbol{LFI:FWHM:LFI20:units}{12\parcm75}
\setsymbol{LFI:FWHM:LFI21:units}{12\parcm74}
\setsymbol{LFI:FWHM:LFI22:units}{12\parcm87}
\setsymbol{LFI:FWHM:LFI23:units}{13\parcm27}
\setsymbol{LFI:FWHM:LFI24:units}{22\parcm98}
\setsymbol{LFI:FWHM:LFI25:units}{30\parcm46}
\setsymbol{LFI:FWHM:LFI26:units}{30\parcm31}
\setsymbol{LFI:FWHM:LFI27:units}{32\parcm65}
\setsymbol{LFI:FWHM:LFI28:units}{32\parcm66}

\setsymbol{LFI:FWHM:LFI18}{13.39}
\setsymbol{LFI:FWHM:LFI19}{13.01}
\setsymbol{LFI:FWHM:LFI20}{12.75}
\setsymbol{LFI:FWHM:LFI21}{12.74}
\setsymbol{LFI:FWHM:LFI22}{12.87}
\setsymbol{LFI:FWHM:LFI23}{13.27}
\setsymbol{LFI:FWHM:LFI24}{22.98}
\setsymbol{LFI:FWHM:LFI25}{30.46}
\setsymbol{LFI:FWHM:LFI26}{30.31}
\setsymbol{LFI:FWHM:LFI27}{32.65}
\setsymbol{LFI:FWHM:LFI28}{32.66}

\setsymbol{LFI:FWHM:uncertainty:LFI18:units}{0.170\arcm}
\setsymbol{LFI:FWHM:uncertainty:LFI19:units}{0.174\arcm}
\setsymbol{LFI:FWHM:uncertainty:LFI20:units}{0.170\arcm}
\setsymbol{LFI:FWHM:uncertainty:LFI21:units}{0.156\arcm}
\setsymbol{LFI:FWHM:uncertainty:LFI22:units}{0.164\arcm}
\setsymbol{LFI:FWHM:uncertainty:LFI23:units}{0.171\arcm}
\setsymbol{LFI:FWHM:uncertainty:LFI24:units}{0.652\arcm}
\setsymbol{LFI:FWHM:uncertainty:LFI25:units}{1.075\arcm}
\setsymbol{LFI:FWHM:uncertainty:LFI26:units}{1.131\arcm}
\setsymbol{LFI:FWHM:uncertainty:LFI27:units}{1.266\arcm}
\setsymbol{LFI:FWHM:uncertainty:LFI28:units}{1.287\arcm}

\setsymbol{LFI:FWHM:uncertainty:LFI18}{0.170}
\setsymbol{LFI:FWHM:uncertainty:LFI19}{0.174}
\setsymbol{LFI:FWHM:uncertainty:LFI20}{0.170}
\setsymbol{LFI:FWHM:uncertainty:LFI21}{0.156}
\setsymbol{LFI:FWHM:uncertainty:LFI22}{0.164}
\setsymbol{LFI:FWHM:uncertainty:LFI23}{0.171}
\setsymbol{LFI:FWHM:uncertainty:LFI24}{0.652}
\setsymbol{LFI:FWHM:uncertainty:LFI25}{1.075}
\setsymbol{LFI:FWHM:uncertainty:LFI26}{1.131}
\setsymbol{LFI:FWHM:uncertainty:LFI27}{1.266}
\setsymbol{LFI:FWHM:uncertainty:LFI28}{1.287}

\setsymbol{HFI:center:frequency:100GHz:units}{100\,GHz}
\setsymbol{HFI:center:frequency:143GHz:units}{143\,GHz}
\setsymbol{HFI:center:frequency:217GHz:units}{217\,GHz}
\setsymbol{HFI:center:frequency:353GHz:units}{353\,GHz}
\setsymbol{HFI:center:frequency:545GHz:units}{545\,GHz}
\setsymbol{HFI:center:frequency:857GHz:units}{857\,GHz}

\setsymbol{HFI:center:frequency:100GHz}{100}
\setsymbol{HFI:center:frequency:143GHz}{143}
\setsymbol{HFI:center:frequency:217GHz}{217}
\setsymbol{HFI:center:frequency:353GHz}{353}
\setsymbol{HFI:center:frequency:545GHz}{545}
\setsymbol{HFI:center:frequency:857GHz}{857}

\setsymbol{HFI:Ndetectors:100GHz}{8}
\setsymbol{HFI:Ndetectors:143GHz}{11}
\setsymbol{HFI:Ndetectors:217GHz}{12}
\setsymbol{HFI:Ndetectors:353GHz}{12}
\setsymbol{HFI:Ndetectors:545GHz}{3}
\setsymbol{HFI:Ndetectors:857GHz}{4}

\setsymbol{HFI:FWHM:Maps:100GHz:units}{9\parcm88}
\setsymbol{HFI:FWHM:Maps:143GHz:units}{7\parcm18}
\setsymbol{HFI:FWHM:Maps:217GHz:units}{4\parcm87}
\setsymbol{HFI:FWHM:Maps:353GHz:units}{4\parcm65}
\setsymbol{HFI:FWHM:Maps:545GHz:units}{4\parcm72}
\setsymbol{HFI:FWHM:Maps:857GHz:units}{4\parcm39}
\setsymbol{HFI:FWHM:Maps:100GHz}{9.88}
\setsymbol{HFI:FWHM:Maps:143GHz}{7.18}
\setsymbol{HFI:FWHM:Maps:217GHz}{4.87}
\setsymbol{HFI:FWHM:Maps:353GHz}{4.65}
\setsymbol{HFI:FWHM:Maps:545GHz}{4.72}
\setsymbol{HFI:FWHM:Maps:857GHz}{4.39}

\setsymbol{HFI:beam:ellipticity:Maps:100GHz}{1.15}
\setsymbol{HFI:beam:ellipticity:Maps:143GHz}{1.01}
\setsymbol{HFI:beam:ellipticity:Maps:217GHz}{1.06}
\setsymbol{HFI:beam:ellipticity:Maps:353GHz}{1.05}
\setsymbol{HFI:beam:ellipticity:Maps:545GHz}{1.14}
\setsymbol{HFI:beam:ellipticity:Maps:857GHz}{1.19}

\setsymbol{HFI:FWHM:Mars:100GHz:units}{9\parcm37}
\setsymbol{HFI:FWHM:Mars:143GHz:units}{7\parcm04}
\setsymbol{HFI:FWHM:Mars:217GHz:units}{4\parcm68}
\setsymbol{HFI:FWHM:Mars:353GHz:units}{4\parcm43}
\setsymbol{HFI:FWHM:Mars:545GHz:units}{3\parcm80}
\setsymbol{HFI:FWHM:Mars:857GHz:units}{3\parcm67}

\setsymbol{HFI:FWHM:Mars:100GHz}{9.37}
\setsymbol{HFI:FWHM:Mars:143GHz}{7.04}
\setsymbol{HFI:FWHM:Mars:217GHz}{4.68}
\setsymbol{HFI:FWHM:Mars:353GHz}{4.43}
\setsymbol{HFI:FWHM:Mars:545GHz}{3.80}
\setsymbol{HFI:FWHM:Mars:857GHz}{3.67}

\setsymbol{HFI:beam:ellipticity:Mars:100GHz}{1.18}
\setsymbol{HFI:beam:ellipticity:Mars:143GHz}{1.03}
\setsymbol{HFI:beam:ellipticity:Mars:217GHz}{1.14}
\setsymbol{HFI:beam:ellipticity:Mars:353GHz}{1.09}
\setsymbol{HFI:beam:ellipticity:Mars:545GHz}{1.25}
\setsymbol{HFI:beam:ellipticity:Mars:857GHz}{1.03}

\setsymbol{HFI:CMB:relative:calibration:100GHz}{$\lsim 1\%$}
\setsymbol{HFI:CMB:relative:calibration:143GHz}{$\lsim 1\%$}
\setsymbol{HFI:CMB:relative:calibration:217GHz}{$\lsim 1\%$}
\setsymbol{HFI:CMB:relative:calibration:353GHz}{$\lsim 1\%$}
\setsymbol{HFI:CMB:relative:calibration:545GHz}{}
\setsymbol{HFI:CMB:relative:calibration:857GHz}{}

\setsymbol{HFI:CMB:absolute:calibration:100GHz}{$\lsim 2\%$}
\setsymbol{HFI:CMB:absolute:calibration:143GHz}{$\lsim 2\%$}
\setsymbol{HFI:CMB:absolute:calibration:217GHz}{$\lsim 2\%$}
\setsymbol{HFI:CMB:absolute:calibration:353GHz}{$\lsim 2\%$}
\setsymbol{HFI:CMB:absolute:calibration:545GHz}{}
\setsymbol{HFI:CMB:absolute:calibration:857GHz}{}

\setsymbol{HFI:FIRAS:gain:calibration:accuracy:statistical:100GHz}{}
\setsymbol{HFI:FIRAS:gain:calibration:accuracy:statistical:143GHz}{}
\setsymbol{HFI:FIRAS:gain:calibration:accuracy:statistical:217GHz}{}
\setsymbol{HFI:FIRAS:gain:calibration:accuracy:statistical:353GHz}{2.5\%}
\setsymbol{HFI:FIRAS:gain:calibration:accuracy:statistical:545GHz}{1\%}
\setsymbol{HFI:FIRAS:gain:calibration:accuracy:statistical:857GHz}{0.5\%}

\setsymbol{HFI:FIRAS:gain:calibration:accuracy:systematic:100GHz}{}
\setsymbol{HFI:FIRAS:gain:calibration:accuracy:systematic:143GHz}{}
\setsymbol{HFI:FIRAS:gain:calibration:accuracy:systematic:217GHz}{}
\setsymbol{HFI:FIRAS:gain:calibration:accuracy:systematic:353GHz}{}
\setsymbol{HFI:FIRAS:gain:calibration:accuracy:systematic:545GHz}{7\%}
\setsymbol{HFI:FIRAS:gain:calibration:accuracy:systematic:857GHz}{7\%}

\setsymbol{HFI:FIRAS:zero:point:accuracy:100GHz:units}{0.8\MJysr}
\setsymbol{HFI:FIRAS:zero:point:accuracy:143GHz:units}{}
\setsymbol{HFI:FIRAS:zero:point:accuracy:217GHz:units}{}
\setsymbol{HFI:FIRAS:zero:point:accuracy:353GHz:units}{1.4\MJysr}
\setsymbol{HFI:FIRAS:zero:point:accuracy:545GHz:units}{2.2\MJysr}
\setsymbol{HFI:FIRAS:zero:point:accuracy:857GHz:units}{1.7\MJysr}

\setsymbol{HFI:FIRAS:zero:point:accuracy:100GHz}{0.8}
\setsymbol{HFI:FIRAS:zero:point:accuracy:143GHz}{}
\setsymbol{HFI:FIRAS:zero:point:accuracy:217GHz}{}
\setsymbol{HFI:FIRAS:zero:point:accuracy:353GHz}{1.4}
\setsymbol{HFI:FIRAS:zero:point:accuracy:545GHz}{2.2}
\setsymbol{HFI:FIRAS:zero:point:accuracy:857GHz}{1.7}

\setsymbol{HFI:unit:conversion:100GHz:units}{0.2415\MJysrmK}
\setsymbol{HFI:unit:conversion:143GHz:units}{0.3694\MJysrmK}
\setsymbol{HFI:unit:conversion:217GHz:units}{0.4811\MJysrmK}
\setsymbol{HFI:unit:conversion:353GHz:units}{0.2883\MJysrmK}
\setsymbol{HFI:unit:conversion:545GHz:units}{0.05826\MJysrmK}
\setsymbol{HFI:unit:conversion:857GHz:units}{0.002238\MJysrmK}

\setsymbol{HFI:unit:conversion:100GHz}{0.2415}
\setsymbol{HFI:unit:conversion:143GHz}{0.3694}
\setsymbol{HFI:unit:conversion:217GHz}{0.4811}
\setsymbol{HFI:unit:conversion:353GHz}{0.2883}
\setsymbol{HFI:unit:conversion:545GHz}{0.05826}
\setsymbol{HFI:unit:conversion:857GHz}{0.002238}

\setsymbol{HFI:colour:correction:alpha=-2:V1.01:100GHz}{0.9893}
\setsymbol{HFI:colour:correction:alpha=-2:V1.01:143GHz}{0.9759}
\setsymbol{HFI:colour:correction:alpha=-2:V1.01:217GHz}{1.0007}
\setsymbol{HFI:colour:correction:alpha=-2:V1.01:353GHz}{1.0028}
\setsymbol{HFI:colour:correction:alpha=-2:V1.01:545GHz}{1.0019}
\setsymbol{HFI:colour:correction:alpha=-2:V1.01:857GHz}{0.9889}

\setsymbol{HFI:colour:correction:alpha=0:V1.01:100GHz}{1.0008}
\setsymbol{HFI:colour:correction:alpha=0:V1.01:143GHz}{1.0148}
\setsymbol{HFI:colour:correction:alpha=0:V1.01:217GHz}{0.9909}
\setsymbol{HFI:colour:correction:alpha=0:V1.01:353GHz}{0.9888}
\setsymbol{HFI:colour:correction:alpha=0:V1.01:545GHz}{0.9878}
\setsymbol{HFI:colour:correction:alpha=0:V1.01:857GHz}{1.0014}

\begin{document}

\title{Planck Early Results. XXIV. Dust in the diffuse interstellar medium and the Galactic halo}

\author{\small
Planck Collaboration:
A.~Abergel\inst{46}
\and
P.~A.~R.~Ade\inst{71}
\and
N.~Aghanim\inst{46}
\and
M.~Arnaud\inst{57}
\and
M.~Ashdown\inst{55, 4}
\and
J.~Aumont\inst{46}
\and
C.~Baccigalupi\inst{69}
\and
A.~Balbi\inst{28}
\and
A.~J.~Banday\inst{75, 7, 62}
\and
R.~B.~Barreiro\inst{52}
% ...
}
\institute{\small
Aalto University Mets\"{a}hovi Radio Observatory, Mets\"{a}hovintie 114, FIN-02540 Kylm\"{a}l\"{a}, Finland\\
\and
Agenzia Spaziale Italiana Science Data Center, c/o ESRIN, via Galileo Galilei, Frascati, Italy\\
\and
Astroparticule et Cosmologie, CNRS (UMR7164), Universit\'{e} Denis Diderot Paris 7, B\^{a}timent Condorcet, 10 rue A. Domon et L\'{e}onie Duquet, Paris, France\\
\and
Astrophysics Group, Cavendish Laboratory, University of Cambridge, J J Thomson Avenue, Cambridge CB3 0HE, U.K.\\
\and
Atacama Large Millimeter/submillimeter Array, ALMA Santiago Central Offices, Alonso de Cordova 3107, Vitacura, Casilla 763 0355, Santiago, Chile\\
\and
CITA, University of Toronto, 60 St. George St., Toronto, ON M5S 3H8, Canada\\
\and
CNRS, IRAP, 9 Av. colonel Roche, BP 44346, F-31028 Toulouse cedex 4, France\\
\and
California Institute of Technology, Pasadena, California, U.S.A.\\
\and
DAMTP, University of Cambridge, Centre for Mathematical Sciences, Wilberforce Road, Cambridge CB3 0WA, U.K.\\
\and
DSM/Irfu/SPP, CEA-Saclay, F-91191 Gif-sur-Yvette Cedex, France\\
\and
DTU Space, National Space Institute, Juliane Mariesvej 30, Copenhagen, Denmark\\
\and
D\'{e}partement de physique, de g\'{e}nie physique et d'optique, Universit\'{e} Laval, Qu\'{e}bec, Canada\\
\and
Departamento de F\'{\i}sica, Universidad de Oviedo, Avda. Calvo Sotelo s/n, Oviedo, Spain\\
\and
Department of Astronomy and Astrophysics, University of Toronto, 50 Saint George Street, Toronto, Ontario, Canada\\
\and
Department of Physics \& Astronomy, University of British Columbia, 6224 Agricultural Road, Vancouver, British Columbia, Canada\\
\and
Department of Physics, Gustaf H\"{a}llstr\"{o}min katu 2a, University of Helsinki, Helsinki, Finland\\
\and
Department of Physics, Princeton University, Princeton, New Jersey, U.S.A.\\
\and
Department of Physics, Purdue University, 525 Northwestern Avenue, West Lafayette, Indiana, U.S.A.\\
\and
Department of Physics, University of California, Berkeley, California, U.S.A.\\
\and
Department of Physics, University of California, One Shields Avenue, Davis, California, U.S.A.\\
\and
Department of Physics, University of California, Santa Barbara, California, U.S.A.\\
\and
Department of Physics, University of Illinois at Urbana-Champaign, 1110 West Green Street, Urbana, Illinois, U.S.A.\\
\and
Dipartimento di Fisica G. Galilei, Universit\`{a} degli Studi di Padova, via Marzolo 8, 35131 Padova, Italy\\
\and
Dipartimento di Fisica, Universit\`{a} La Sapienza, P. le A. Moro 2, Roma, Italy\\
\and
Dipartimento di Fisica, Universit\`{a} degli Studi di Milano, Via Celoria, 16, Milano, Italy\\
\and
Dipartimento di Fisica, Universit\`{a} degli Studi di Trieste, via A. Valerio 2, Trieste, Italy\\
\and
Dipartimento di Fisica, Universit\`{a} di Ferrara, Via Saragat 1, 44122 Ferrara, Italy\\
\and
Dipartimento di Fisica, Universit\`{a} di Roma Tor Vergata, Via della Ricerca Scientifica, 1, Roma, Italy\\
\and
Discovery Center, Niels Bohr Institute, Blegdamsvej 17, Copenhagen, Denmark\\
\and
Dpto. Astrof\'{i}sica, Universidad de La Laguna (ULL), E-38206 La Laguna, Tenerife, Spain\\
\and
European Southern Observatory, ESO Vitacura, Alonso de Cordova 3107, Vitacura, Casilla 19001, Santiago, Chile\\
\and
European Space Agency, ESAC, Planck Science Office, Camino bajo del Castillo, s/n, Urbanizaci\'{o}n Villafranca del Castillo, Villanueva de la Ca\~{n}ada, Madrid, Spain\\
\and
European Space Agency, ESTEC, Keplerlaan 1, 2201 AZ Noordwijk, The Netherlands\\
\and
Helsinki Institute of Physics, Gustaf H\"{a}llstr\"{o}min katu 2, University of Helsinki, Helsinki, Finland\\
\and
INAF - Osservatorio Astrofisico di Catania, Via S. Sofia 78, Catania, Italy\\
\and
INAF - Osservatorio Astronomico di Padova, Vicolo dell'Osservatorio 5, Padova, Italy\\
\and
INAF - Osservatorio Astronomico di Roma, via di Frascati 33, Monte Porzio Catone, Italy\\
\and
INAF - Osservatorio Astronomico di Trieste, Via G.B. Tiepolo 11, Trieste, Italy\\
\and
INAF/IASF Bologna, Via Gobetti 101, Bologna, Italy\\
\and
INAF/IASF Milano, Via E. Bassini 15, Milano, Italy\\
\and
INRIA, Laboratoire de Recherche en Informatique, Universit\'{e} Paris-Sud 11, B\^{a}timent 490, 91405 Orsay Cedex, France\\
\and
IPAG: Institut de Plan\'{e}tologie et d'Astrophysique de Grenoble, Universit\'{e} Joseph Fourier, Grenoble 1 / CNRS-INSU, UMR 5274, Grenoble, F-38041, France\\
\and
Imperial College London, Astrophysics group, Blackett Laboratory, Prince Consort Road, London, SW7 2AZ, U.K.\\
\and
Infrared Processing and Analysis Center, California Institute of Technology, Pasadena, CA 91125, U.S.A.\\
\and
Institut N\'{e}el, CNRS, Universit\'{e} Joseph Fourier Grenoble I, 25 rue des Martyrs, Grenoble, France\\
\and
Institut d'Astrophysique Spatiale, CNRS (UMR8617) Universit\'{e} Paris-Sud 11, B\^{a}timent 121, Orsay, France\\
\and
Institut d'Astrophysique de Paris, CNRS UMR7095, Universit\'{e} Pierre \& Marie Curie, 98 bis boulevard Arago, Paris, France\\
\and
Institute of Astronomy and Astrophysics, Academia Sinica, Taipei, Taiwan\\
\and
Institute of Astronomy, University of Cambridge, Madingley Road, Cambridge CB3 0HA, U.K.\\
\and
Institute of Theoretical Astrophysics, University of Oslo, Blindern, Oslo, Norway\\
\and
Instituto de Astrof\'{\i}sica de Canarias, C/V\'{\i}a L\'{a}ctea s/n, La Laguna, Tenerife, Spain\\
\and
Instituto de F\'{\i}sica de Cantabria (CSIC-Universidad de Cantabria), Avda. de los Castros s/n, Santander, Spain\\
\and
Jet Propulsion Laboratory, California Institute of Technology, 4800 Oak Grove Drive, Pasadena, California, U.S.A.\\
\and
Jodrell Bank Centre for Astrophysics, Alan Turing Building, School of Physics and Astronomy, The University of Manchester, Oxford Road, Manchester, M13 9PL, U.K.\\
\and
Kavli Institute for Cosmology Cambridge, Madingley Road, Cambridge, CB3 0HA, U.K.\\
\and
LERMA, CNRS, Observatoire de Paris, 61 Avenue de l'Observatoire, Paris, France\\
\and
Laboratoire AIM, IRFU/Service d'Astrophysique - CEA/DSM - CNRS - Universit\'{e} Paris Diderot, B\^{a}t. 709, CEA-Saclay, F-91191 Gif-sur-Yvette Cedex, France\\
\and
Laboratoire Traitement et Communication de l'Information, CNRS (UMR 5141) and T\'{e}l\'{e}com ParisTech, 46 rue Barrault F-75634 Paris Cedex 13, France\\
\and
Laboratoire de Physique Subatomique et de Cosmologie, CNRS/IN2P3, Universit\'{e} Joseph Fourier Grenoble I, Institut National Polytechnique de Grenoble, 53 rue des Martyrs, 38026 Grenoble cedex, France\\
\and
Laboratoire de l'Acc\'{e}l\'{e}rateur Lin\'{e}aire, Universit\'{e} Paris-Sud 11, CNRS/IN2P3, Orsay, France\\
\and
Lawrence Berkeley National Laboratory, Berkeley, California, U.S.A.\\
\and
Max-Planck-Institut f\"{u}r Astrophysik, Karl-Schwarzschild-Str. 1, 85741 Garching, Germany\\
\and
MilliLab, VTT Technical Research Centre of Finland, Tietotie 3, Espoo, Finland\\
\and
NRAO, P.O. Box 2, Rt 28/92, Green Bank, WV 24944-0002, U.S.A.\\
\and
National University of Ireland, Department of Experimental Physics, Maynooth, Co. Kildare, Ireland\\
\and
Niels Bohr Institute, Blegdamsvej 17, Copenhagen, Denmark\\
\and
Observational Cosmology, Mail Stop 367-17, California Institute of Technology, Pasadena, CA, 91125, U.S.A.\\
\and
Optical Science Laboratory, University College London, Gower Street, London, U.K.\\
\and
SISSA, Astrophysics Sector, via Bonomea 265, 34136, Trieste, Italy\\
\and
SUPA, Institute for Astronomy, University of Edinburgh, Royal Observatory, Blackford Hill, Edinburgh EH9 3HJ, U.K.\\
\and
School of Physics and Astronomy, Cardiff University, Queens Buildings, The Parade, Cardiff, CF24 3AA, U.K.\\
\and
Space Sciences Laboratory, University of California, Berkeley, California, U.S.A.\\
\and
Spitzer Science Center, 1200 E. California Blvd., Pasadena, California, U.S.A.\\
\and
Stanford University, Dept of Physics, Varian Physics Bldg, 382 Via Pueblo Mall, Stanford, California, U.S.A.\\
\and
Universit\'{e} de Toulouse, UPS-OMP, IRAP, F-31028 Toulouse cedex 4, France\\
% ...
}

\abstract { 
This paper presents the first results of comparison of \Planck\ dust maps
at 353, 545 and 857\,GHz, along with \IRAS\ data at 3000 (100\,$\mu$m) and
5000\,GHz (60\,$\mu$m), with Green Bank Telescope 21-cm observations of
\hi\ in 14 fields covering more than 800\,deg$^2$ at high Galactic
latitude. The main goal of this study is to estimate the far-infrared to
sub-millimetre emissivity of dust in the diffuse local interstellar
medium (ISM) and in the intermediate-velocity (IVC) and high-velocity clouds (HVC) of the Galactic halo.
Galactic dust emission for fields with average \hi\ column density lower than
$2 \times 10^{20}$\,cm$^{-2}$ is well correlated with 21-cm emission
because in such diffuse areas the hydrogen is predominantly in the
neutral atomic phase. The residual emission in these fields, once the
\hi-correlated emission is removed, is consistent with the expected
statistical properties of the cosmic infrared background fluctuations.
The brighter fields in our sample, with an average \hi\ column density greater than $2 \times 10^{20}$\,cm$^{-2}$, 
show significant excess dust emission compared to the \hi\ column density.  
Regions of excess lie in organized structures that suggest the presence of hydrogen in molecular form,
though they are not always correlated with CO emission.
In the higher \hi\ column density fields the excess emission at 857 GHz is about 40\% of that coming from the \hi, 
but over all the high latitude fields surveyed the molecular mass faction is about 10\%. 
Dust emission from IVCs is detected with high significance by this correlation analysis.  
Its spectral properties are consistent with, compared to the local ISM values, significantly hotter dust ($T\sim 20$\,K), lower sub-millimeter dust opacity normalized per H-atom, and a relative abundance of very small grains to large grains about four times higher. These results are compatible with expectations for clouds that are part of the Galactic fountain in which there is dust shattering and fragmentation. Correlated dust emission in HVCs is not detected; the average of the 99.9\% confidence upper limits to the emissivity is 0.15 times the local ISM value at 857 and 3000\,GHz, in accordance with gas phase evidence for lower metallicity and depletion in these clouds.
Unexpected anti-correlated variations of the dust temperature and emission cross-section per H atom are identified in the local ISM and IVCs, a trend that continues into molecular environments. This suggests that dust growth through aggregation, seen in molecular clouds, is active much earlier in the cloud condensation and star formation processes. 
}


\titlerunning{\textit{Planck} Early Results. XXIV. Dust in the diffuse interstellar medium and the Galactic halo}

\authorrunning{Planck collaboration}

\maketitle

\allearlypapers

\section{Introduction}

\Planck\footnote{\Planck\ (http://www.esa.int/\Planck ) is a project of the European Space Agency (ESA) with instruments provided by two scientific consortia funded by ESA member states (in particular the lead countries France and Italy), with contributions from NASA (USA) and telescope reflectors provided by a collaboration between ESA and a scientific consortium led and funded by Denmark.}\ \citep{tauber2010a, planck2011-1.1} is the third-generation
space mission to measure the anisotropy of the cosmic microwave
background (CMB).  It observes the sky in nine frequency bands covering
30--857\,GHz with high sensitivity and angular resolution from
31\arcm\ to 5\arcm.  The Low Frequency Instrument (LFI;
\citealt{Mandolesi2010, Bersanelli2010, planck2011-1.4}) covers the 30,
44, and 70\,GHz bands with amplifiers cooled to 20\,\hbox{K}.  The High
Frequency Instrument (HFI; \citealt{Lamarre2010, planck2011-1.5}) covers
the 100, 143, 217, 353, 545, and 857\,GHz bands with bolometers cooled
to 0.1\,\hbox{K}.  \Planck's sensitivity, angular resolution, and
frequency coverage make it a powerful instrument for Galactic and
extragalactic astrophysics as well as cosmology.  
This paper presents the first results of the analysis of
\Planck\ observations of the diffuse interstellar medium (ISM) at high
Galactic latitude.

From the pioneering work of Spitzer and Field
\citep{spitzer1956,field1965,field1969}, observations of the diffuse ISM
including intermediate and high-velocity clouds (IVCs and HVCs) have
been the basis of our understanding of the dynamical
interplay between ISM phases and the disk-halo connection in relation to
star formation.  Space-based observations have given us spectacular
perspectives on the diffuse Galactic infrared emission, which highlight
the role of dust not only as a tracer of the diffuse ISM but also as an
agent in its evolution.

The Infrared Astronomical Satellite (\IRAS) 
revealed the intricate morphology of infrared cirrus
\citep{low1984} and prompted a wide range of observations.  The cirrus
is inferred to be inhomogeneous turbulent dusty clouds with dense
CO-emitting gas intermixed with cold (CNM) and warm (WNM) neutral atomic
gas and also diffuse H$_2$.  From imaging by the \Spitzer\ Space Telescope, 
and very recently by the \Herschel\ Space Observatory, 
their structure is now known to extend to much smaller
angular scales than observable at the \IRAS\ resolution
\cite[]{ingalls2004,martin2010,miville-deschenes2010}.  Observations
from \IRAS, the Infrared Space Observatory (\ISO), and \Spitzer\ have also been used to characterize
changes in the spectral energy distribution (SED) from mid- to far-IR
wavelengths, which have been interpreted as evidence for variations in
the abundance of small stochastically-heated dust particles.  The
correlation with \hi\ spectroscopic data suggests that interstellar
turbulence may play a role in changing the dust size distribution
\cite[]{miville-deschenes2002}.

\begin{table*}
\begin{center}
\begin{tabular}{lccccccccc}\toprule
& & & & \multicolumn{2}{c}{LVC} & \multicolumn{2}{c}{IVC} & \multicolumn{2}{c}{HVC} \\ \cmidrule(lr){5-6} \cmidrule(lr){7-8} \cmidrule(lr){9-10} 
Field & l & b & Area & $\langle N_{\rm HI} \rangle $ & $v$ & $\langle N_{\rm HI} \rangle $ & $v$ & $\langle N_{\rm HI} \rangle $ & $v$ \\ 
& (deg) & (deg) & (deg$^2$) & (10$^{19}$ cm$^{-2}$) & (\kms) & (10$^{19}$ cm$^{-2}$) & (\kms) & (10$^{19}$ cm$^{-2}$) & (\kms)\\  \midrule
AG & $164.8$ & $65.5$ & $26.4$ & $5.3 \pm 0.3$ & $-18.7\pm9.7$ & $9.45 \pm 0.19$ & $-51.3\pm11.4$ & $3.8 \pm 0.3$ & $-107.2\pm17.8$\\
BOOTES & $58.0$ & $68.6$ & $49.1$ & $6.93 \pm 0.12$ & $-5.2\pm7.8$ & $3.88 \pm 0.13$ & $-34.6\pm11.9$ & $0.16 \pm 0.11$ & $-87.5\pm10.3$\\
DRACO & $92.3$ & $38.5$ & $26.4$ & $6.79 \pm 0.12$ & $-1.3\pm4.6$ & $11.5 \pm 0.2$ & $-24.1\pm8.9$ & $3.4 \pm 0.3$ & $-141.8\pm27.0$\\
G86 & $88.0$ & $59.1$ & $26.4$ & $8.81 \pm 0.13$ & $-0.1\pm6.5$ & $10.25 \pm 0.17$ & $-35.3\pm9.6$ & $0.36 \pm 0.15$ & $-93.7\pm17.8$\\
MC & $56.6$ & $-81.5$ & $30.7$ & $8.4 \pm 0.2$ & $-0.7\pm4.8$ & $5.22 \pm 0.13$ & $-18.3\pm7.1$ & $6.2 \pm 0.4$ & $-106.5\pm26.3$\\
N1 & $85.3$ & $44.3$ & $26.4$ & $6.4 \pm 0.2$ & $4.3\pm10.1$ & $2.84 \pm 0.19$ & $-23.1\pm7.2$ & $3.0 \pm 0.3$ & $-112.8\pm16.9$\\
NEP & $96.4$ & $30.0$ & $146.5$ & $26.16 \pm 0.19$ & $-2.2\pm11.3$ & $14.41 \pm 0.18$ & $-42.0\pm14.1$ & $1.10 \pm 0.16$ & $-108.4\pm11.6$\\
POL & $124.9$ & $27.5$ & $60.6$ & $62.6 \pm 0.5$ & $-7.3\pm15.0$ & $5.0 \pm 0.3$ & $-69.2\pm12.6$ & --- & ---\\
POLNOR & $125.0$ & $37.4$ & $60.6$ & $39.4 \pm 0.4$ & $-2.4\pm17.3$ & $3.4 \pm 0.2$ & $-63.2\pm9.7$ & --- & ---\\
SP & $132.3$ & $47.5$ & $26.4$ & $6.1 \pm 0.2$ & $-2.7\pm10.8$ & $3.8 \pm 0.2$ & $-51.8\pm11.3$ & $1.8 \pm 0.2$ & $-138.5\pm11.9$\\
SPC & $135.6$ & $29.3$ & $102.1$ & $30.3 \pm 0.4$ & $-8.3\pm15.2$ & $2.7 \pm 0.2$ & $-62.0\pm8.2$ & $0.9 \pm 0.2$ & $-186.4\pm14.8$\\
SPIDER & $134.9$ & $40.0$ & $103.8$ & $18.5 \pm 0.2$ & $6.0\pm7.1$ & $8.4 \pm 0.3$ & $-39.3\pm18.7$ & $0.27 \pm 0.18$ & $-121.7\pm20.3$\\
UMA & $144.2$ & $38.5$ & $80.7$ & $27.9 \pm 0.3$ & $3.5\pm7.9$ & $9.1 \pm 0.3$ & $-49.4\pm12.0$ & $0.9 \pm 0.3$ & $-153.3\pm12.1$\\
UMAEAST & $155.7$ & $37.0$ & $61.3$ & $31.7 \pm 0.3$ & $0.3\pm6.4$ & $9.5 \pm 0.3$ & $-48.5\pm13.2$ & $2.9 \pm 0.4$ & $-171.8\pm12.3$\\
\bottomrule
\end{tabular}
\end{center}
\caption{\label{table:mainHI} The \hi\ fields. Columns 2-4 give the central Galactic coordinates and size of each field. Columns 5-10 give the average \hi\ column density and its uncertainty (see Section~\ref{sec:noise_hi}), and the average LSR velocity and HWHM for each \hi\ component.}
\end{table*}

Since the breakthrough discoveries made with the Cosmic Background
Explorer (\COBE), the study of dust and the diffuse ISM structure has
also become an integral part of the analysis of the Cosmic Microwave
Background (CMB) and the Cosmic Infrared extragalactic Background
(CIB). Our ability to model the spatial and spectral distribution of the
infrared cirrus emission could limit our ability to achieve the
cosmological goals of \Planck, as well as of present balloon-borne and
ground-based CMB experiments.

Accordingly, the \Planck\ survey was designed to provide an
unprecedented view of the structure of the diffuse ISM and its dust
content.  \Planck\ extends to sub-millimeter (submm) wavelengths the
detailed mapping of the infrared cirrus by the \IRAS\ survey.  Its
sensitivity to faint Galactic cirrus emission is limited only by the
astrophysical noise associated with the anisotropy of the CIB.  The
\Planck\ survey is a major step forward from \IRAS\ for two main reasons.
First, by extending the spectral coverage to submm wavelengths, \Planck\
allows us to probe the full SED of thermal emission from the large dust
grains that are the bulk of the dust mass.  Second, the dust
temperatures obtained via submm SEDs also help us to disentangle the
effects of dust column density, dust heating and dust emission cross-section 
on the brightness of the dust emission.

The scientific motivation of this paper is to trace the structure of
the diffuse ISM, including its elusive diffuse H$_2$ component, 
H$^+$ components, and the evolution of interstellar dust grains within
the local ISM and the Galactic halo.  We analyze the \Planck\ data in
selected fields which cover the full range of hydrogen column
densities from high Galactic latitude cirrus, observed away from dark
molecular clouds such as, e.g., Taurus \citep{planck2011-7.13}.
For all of our fields, we have deep 21-cm spectroscopic observations obtained with
the Green Bank Telescope (GBT).  Our data analysis makes use of, and explores, the
dust/gas correlation by spatially correlating \Planck\ and \IRAS\ data
with \hi\ observations.  More specifically our study extends previous
work on the diffuse ISM SED carried out with 7$^\circ$ resolution FIRAS (Far Infrared Absolute Spectrophotometer) data
\citep{boulanger1996} or with 5\arcmin\ resolution 100\,$\mu$m \IRAS\ data
\citep{reach1998a}.

The paper is organized as follows. In Sect.~\ref{sec:21cm} we describe the
21-cm data and the construction of the column density map for each
\hi\ component.  In Sect.~\ref{sec:planckiras} we describe the \Planck\ and
\IRAS\ data.  Section \ref{sec:correlation} describes the main analysis of
the paper: the determination of the \hi\ emissivities from 353 to
5000\,GHz (60 to 850\,$\mu$m).  Results are presented in
Sect.~\ref{sec:results} followed by a discussion of some implications in
Sect.~\ref{sec:discussion}.  Conclusions wrap up the paper in
Sect.~\ref{sec:conclusion}.

\begin{figure}
\begin{center}
\includegraphics[width=0.8\linewidth, draft=false, angle=0]{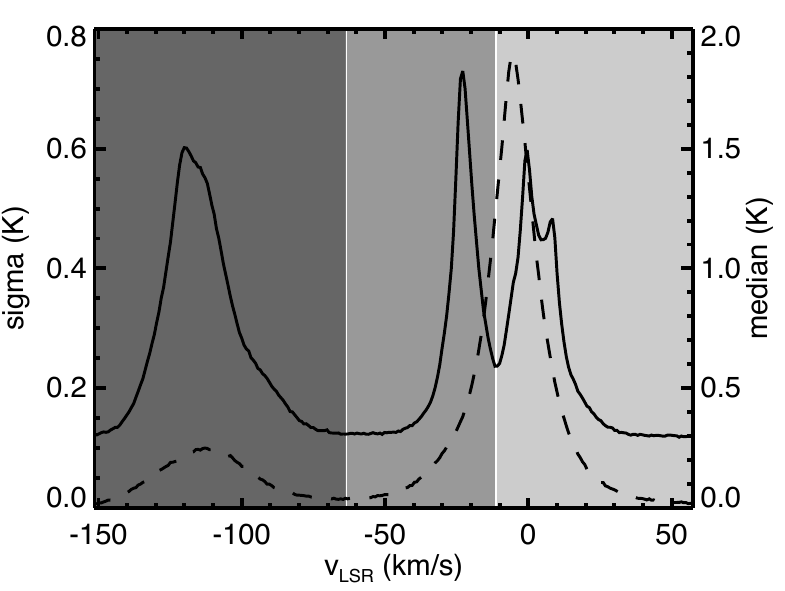}
\caption{\label{fig:N1_HI} The median 21-cm spectrum (dashed) and the standard-deviation
  spectrum (solid) of the N1 field; the shaded backgrounds 
  show the LSR velocity ranges used to estimate LVC, IVC and HVC components 
  (from light to dark grey).}
\end{center}
\end{figure}

\section{21-cm data}

\label{sec:21cm}

\subsection{The Green Bank Telescope cirrus survey}

{The 21-cm \hi\ spectra exploited here were obtained with the 100-meter
Green Bank Telescope (GBT) over the period 2005 to 2010, as part of a
high-latitude survey of 14 fields (for details, see Martin et al., in prep).
The total area mapped is about 825\,deg$^2$.
The adopted names, central coordinates, and
sizes of the 14 GBT fields are given in Table~\ref{table:mainHI}.  
The brighter fields of the sample, the ones
that cover the range of column densities that spans the \hi-H$_2$ transition, are located in the 
North Celestial Loop region, covering the Polaris flare (POL and POLNOR), the Ursa Major cirrus (UMA and UMAEAST),
the bridge between the two (SPIDER) and the interior of the loop (SPC). The largest field in our sample (NEP) is
centered on the north ecliptic pole, a region of high coverage for \Planck. Two fields were targeted
for their specific IVC (DRACO and G86). The faintest fields of the sample were selected either because
the HVC has a major contribution to the total hydrogen column density (AG, MC, SP) or because they are known 
CIB targets (N1 and BOOTES).
}

The spectra were taken with on-the-fly mapping.  The primary beam of the
GBT at 21-cm has a full-width half-maximum (FWHM) of 9.1\arcm, and so
the integration time (4\,s) and telescope scan rate were chosen to sample
every 3.5\arcm, more finely than the Nyquist interval, 3.86\arcm.  The
beam is only slightly broadened to 9.4\arcm\ in the in-scan direction.
Scans were made moving the telescope in one direction (Galactic
longitude or Right Ascension), with steps of 3.5\arcm\ in the orthogonal
coordinate direction before the subsequent reverse scan.

Data were recorded with the GBT spectrometer by in-band frequency
switching, yielding spectra with a local standard of rest (LSR) 
velocity coverage $-450 \leq v_{\rm
  LSR} \leq +355$\,\kms\ at a resolution of 0.80\,\kms.  Spectra were
calibrated, corrected for stray radiation, and placed on a brightness
temperature ($T_{\rm b}$) scale as described in 
\citet{blagrave2010,boothroyd2011}.  
A third-order polynomial was fit to the emission-free
regions of the spectra to remove any residual instrumental baseline.
The spectra were gridded on the equiareal SFL 
\citep[Sanson-Flamsteed --][]{calabretta2002}
projection to produce a data cube.  Some regions were mapped two or three times.

With the broad spectral coverage, all \hi\ components from local gas to
HVCs are accessible.  The total column density \nh\ ranges from $0.6
\times 10^{20}$\,\cmm\ in the SP field to $10 \times 10^{20}$\,\cmm\ in
POL and NEP; the average column density per field ranges from $1.1
\times 10^{20}$\,\cmm\ in BOOTES to $6.3 \times 10^{20}$\,\cmm\ in POL.

\subsection{The \hi\ components}

\label{sec:HIcomponents}

{
We use channel maps from the 21-cm GBT spectra to produce maps of the
\hi\ column density in different velocity ranges. 
For convenience we have separated the emission into three components for
each field: Low-velocity cloud (LVC), IVC, and HVC.
The selection of the velocity range for each component is based on
inspection of both the median 21-cm spectrum and the spectrum made from 
the standard deviation of each channel map. An example of these two
spectra for the N1 field is shown in 
Fig.~\ref{fig:N1_HI}. The dashed and solid lines show the median and
standard deviation of the brightness temperature as a function of
velocity.  
The standard-deviation spectrum, more sensitive to the 
structure within channel maps, 
is used here to establish the velocity range of the three components in cases 
where velocity components are blended in the median spectrum.
The three shaded backgrounds in Fig.~\ref{fig:N1_HI} show
the velocity ranges used to calculate the \hi\ column density of the
three components in this field.
}

The brightness temperature of each velocity channel was converted to
column density, assuming an opacity correction for \hi\ gas with a
constant spin temperature $T_{\rm s}$, to provide an estimate of the total
\hi\ column density of each component:
\begin{equation}
N_{\rm HI}(x,y) = A \times T_{\rm s} \sum_v - \ln \left(1-
\frac{T_{\rm b}(x,y,v)}{T_{\rm s}} \right)\Delta v,
\end{equation}
where $A=1.823\times 10^{18}$\,\cmm\,(K \kms)$^{-1}$. 
In the
optically-thin case (assuming $T_{\rm s} \gg T_{\rm b}$) this reduces to
$N_{\rm HI}(x,y) = A \sum_v T_{\rm b}(x,y,v) \, \Delta v$.

\begin{figure*}
\centering
\setlength{\unitlength}{\textwidth}
\begin{picture}(1.0,1.31)
\put(0.05,0.88){\includegraphics[scale=0.97]{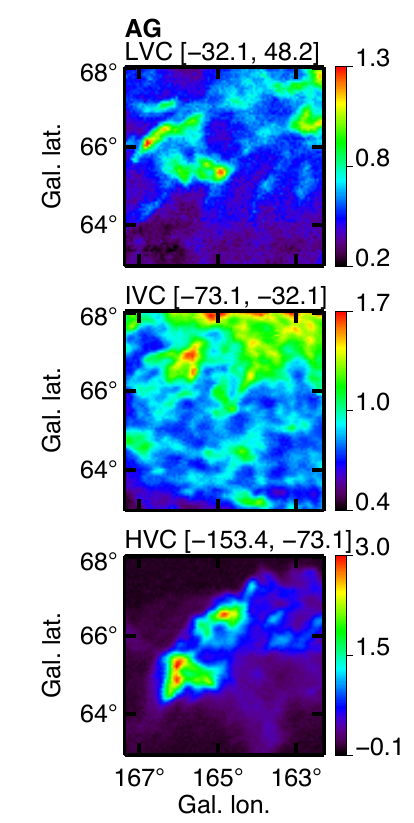}}
\put(0.275,0.88){\includegraphics[scale=0.97]{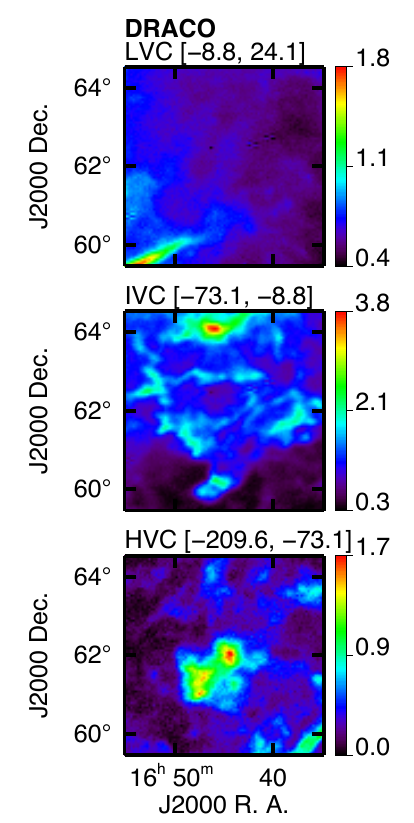}}
\put(0.5,0.88){\includegraphics[scale=0.97]{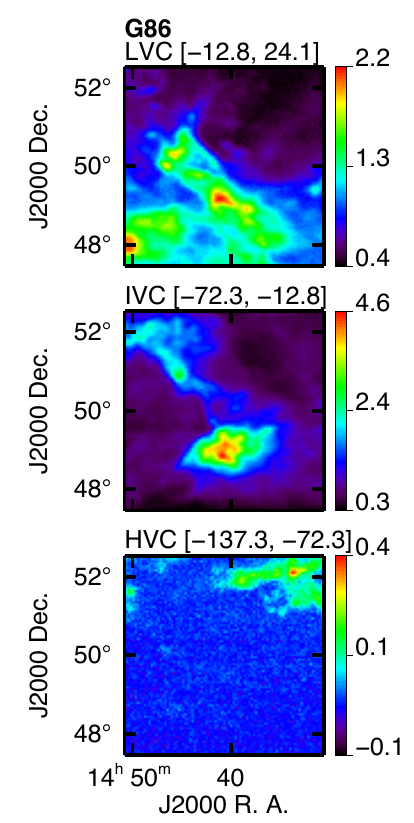}}
\put(0.725,0.88){\includegraphics[scale=0.97]{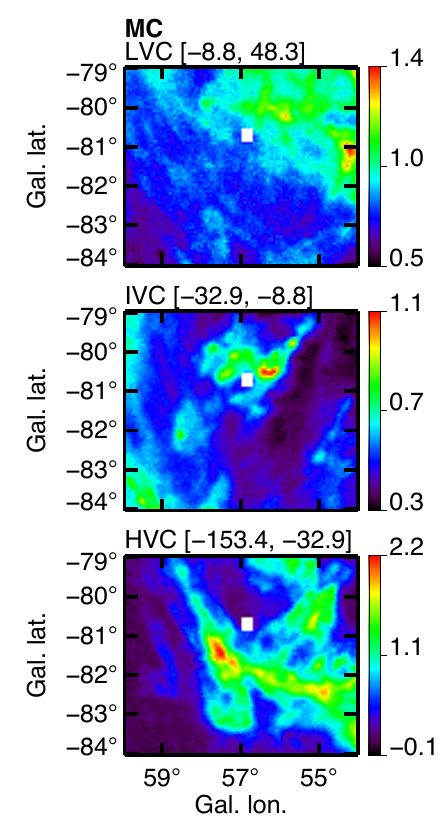}}
\put(0.05,0.0){\includegraphics[scale=0.97]{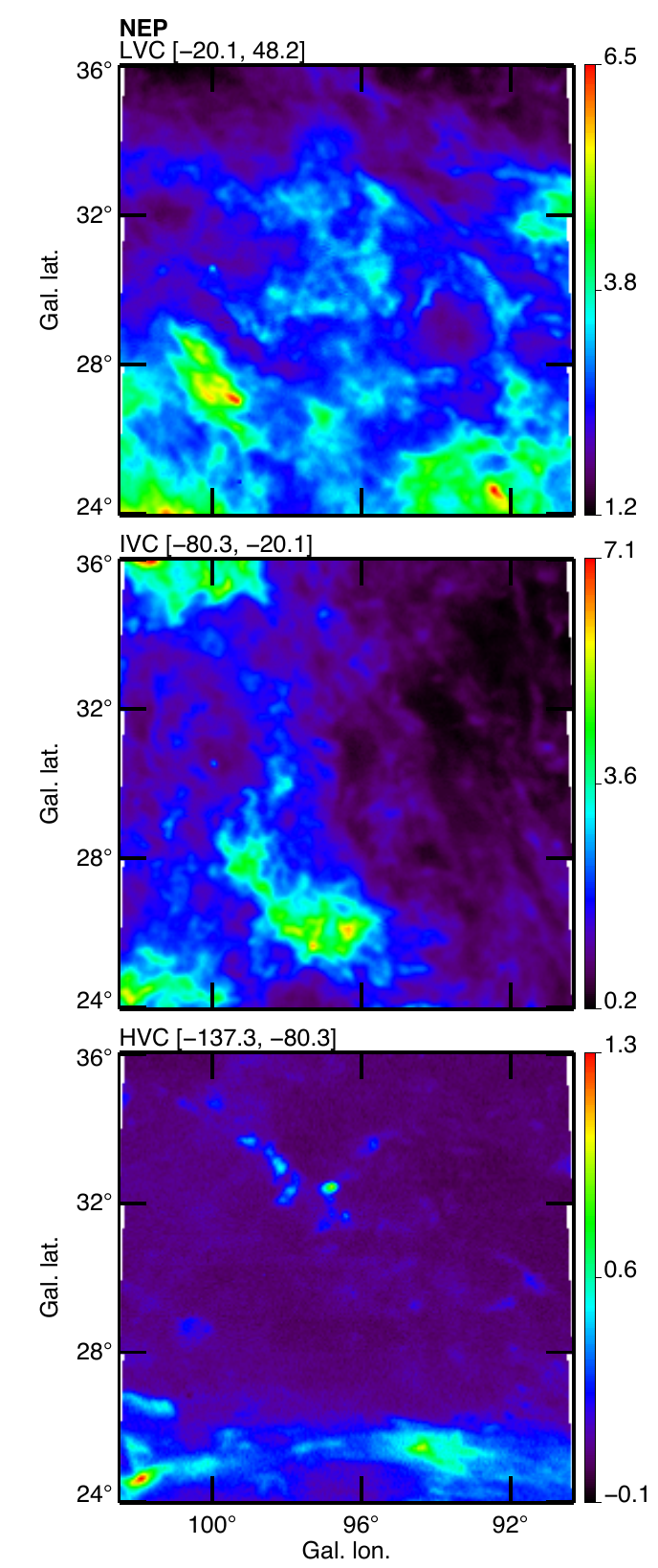}}
\put(0.5,0.43){\includegraphics[scale=0.97]{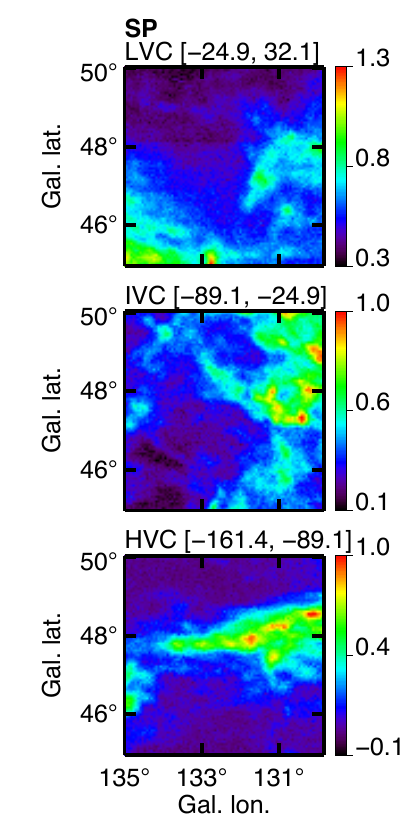}}
\put(0.725,0.43){\includegraphics[scale=0.97]{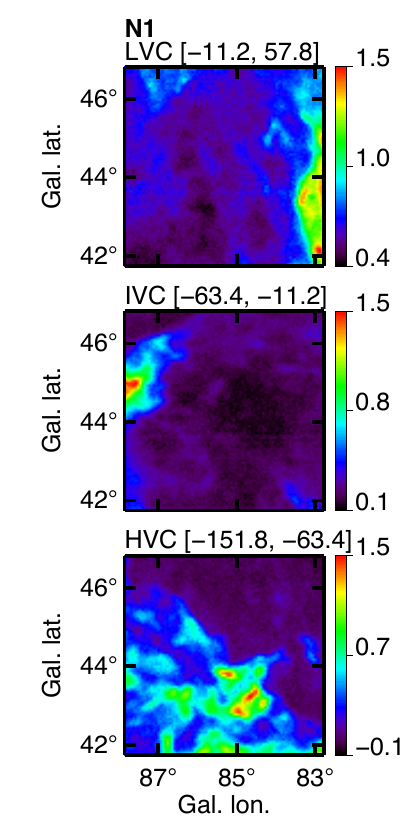}}
\put(0.53,0.){\includegraphics[scale=0.97]{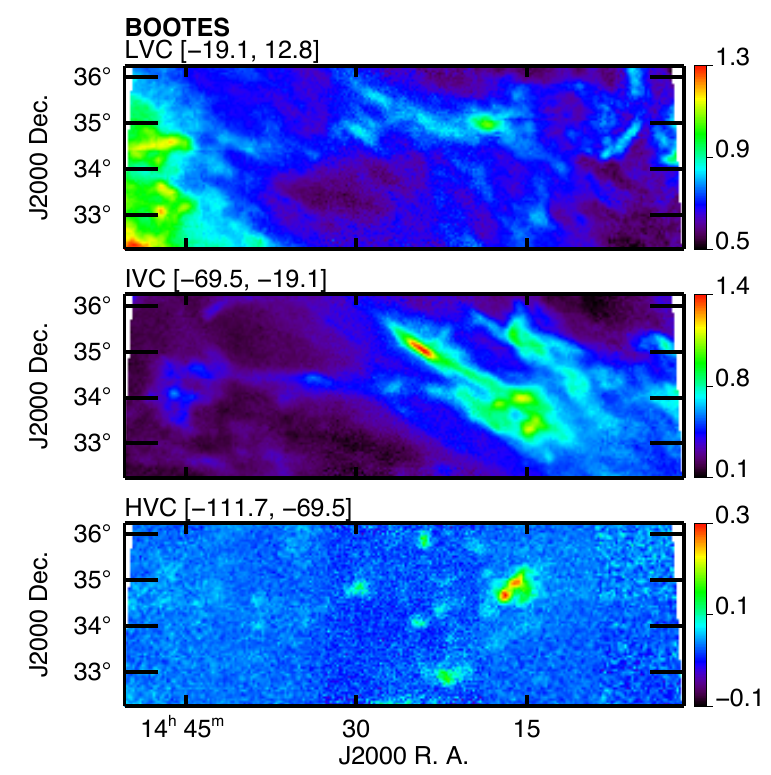}}
\end{picture}
\caption{\label{fig:maps_nhi_1} \hi\ column density maps in units of
$10^{20}$\,cm$^{-2}$
for the AG, DRACO, G86, MC, NEP, SP, N1 and BOOTES fields. 
To show the full detail, the range is different for each \hi\ component for a given field.
The LSR velocity range used to compute each \hi\ component map
is given in brackets (unit is \kms).}
\end{figure*}

\begin{figure*}
\centering
\setlength{\unitlength}{\textwidth}
\begin{picture}(1,1.28)
\put(0.08,0.76){\includegraphics[scale=0.97]{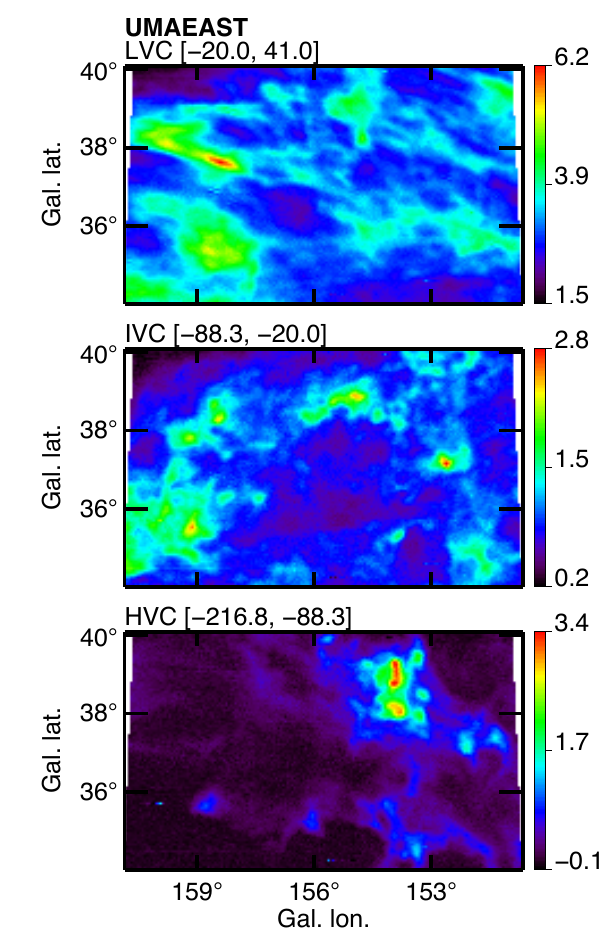}}
\put(0.42,0.76){\includegraphics[scale=0.97]{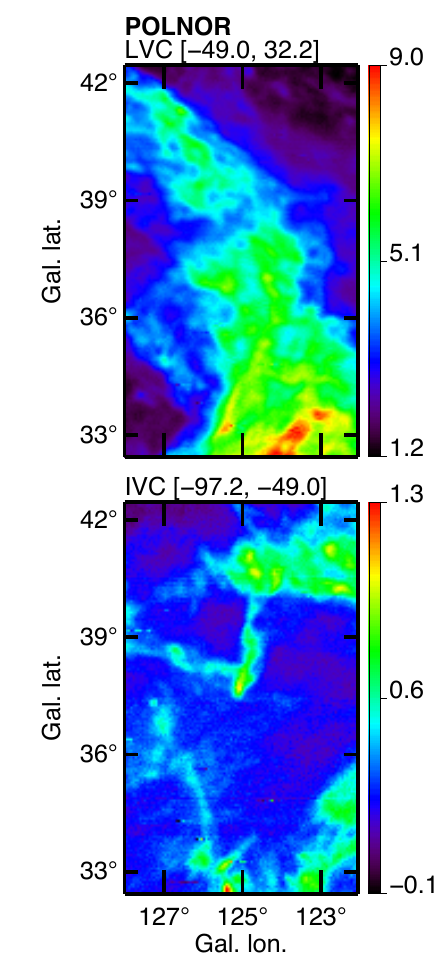}}
\put(0.67,0.76){\includegraphics[scale=0.97]{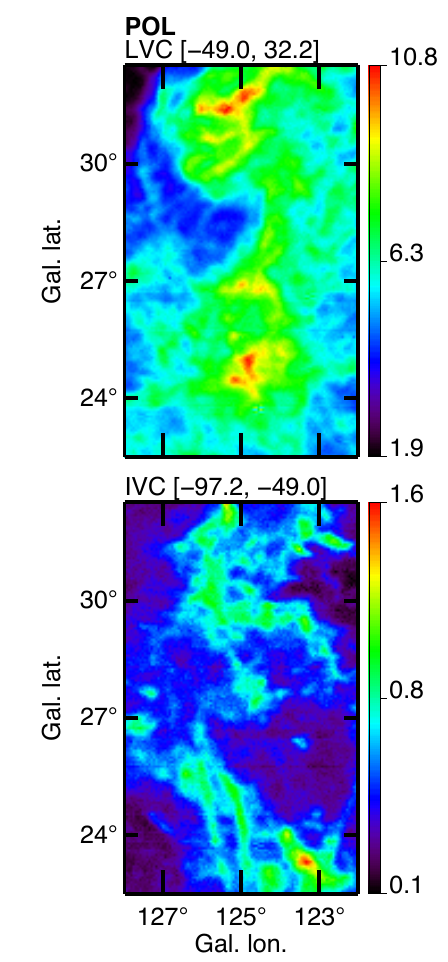}}
\put(0.00,0.0){\includegraphics[scale=0.97]{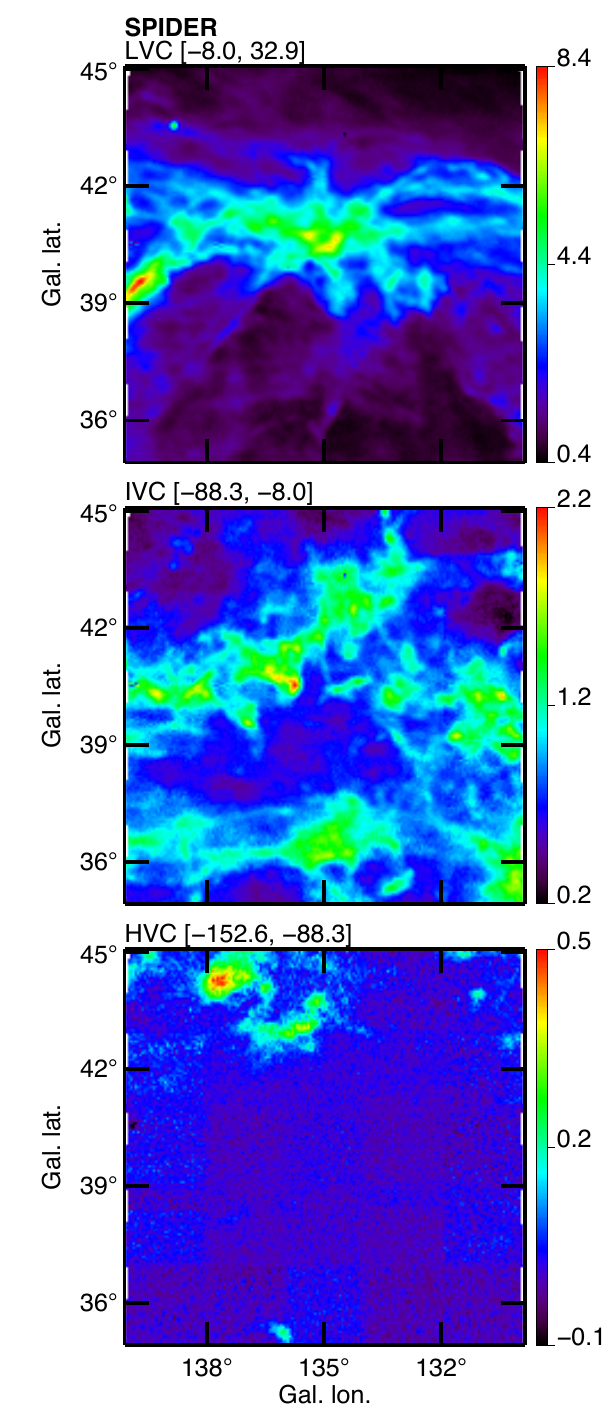}}
\put(0.33,0.0){\includegraphics[scale=0.97]{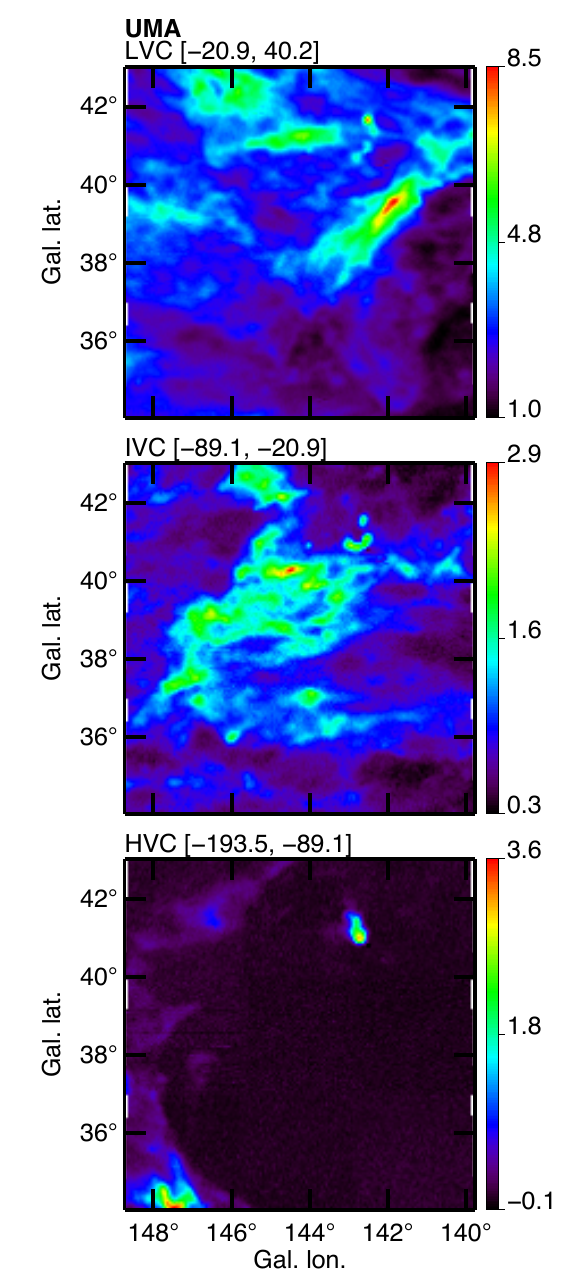}}
\put(0.64,0.0){\includegraphics[scale=0.97]{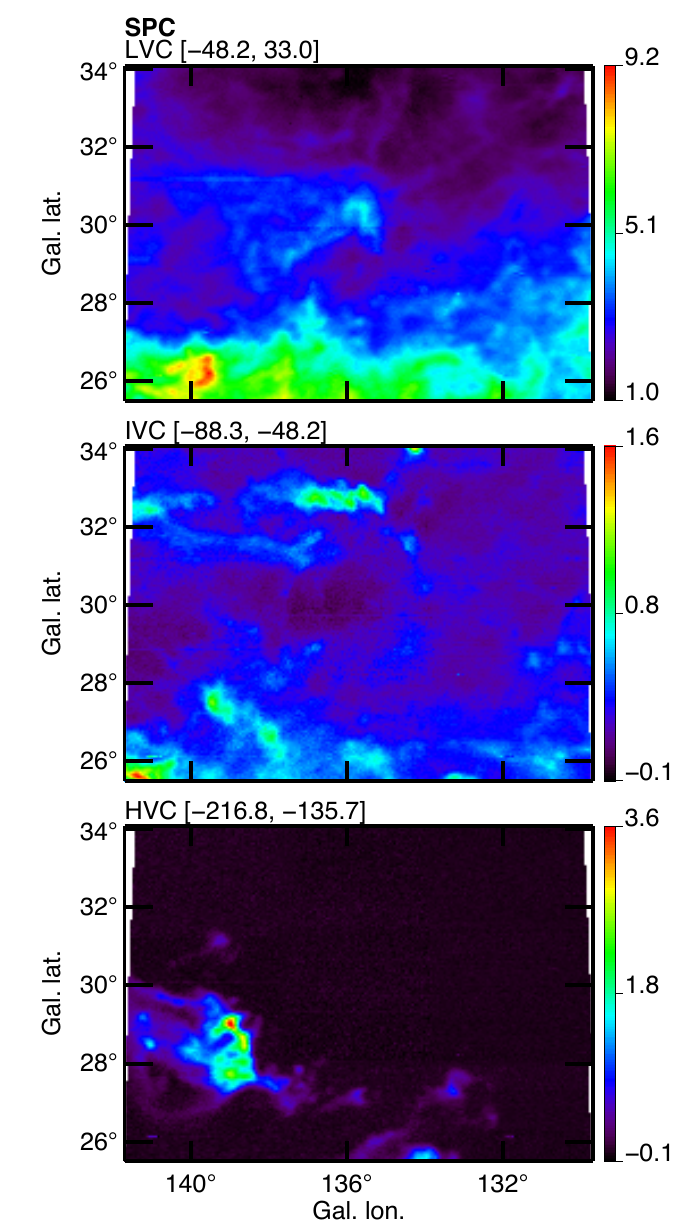}}
\end{picture}
\caption{\label{fig:maps_nhi_2}  Like Fig.~\ref{fig:maps_nhi_1} for fields
UMAEAST, POLNOR, POL, SPIDER, UMA and SPC. 
The M81-M82 complex can be seen in the UMA field 
near $l=142.5^\circ$, $b=41.0^\circ$ in the three \hi\ components. It was masked
in the analysis (see Fig.~\ref{fig:mask_1}).}
\end{figure*}

We used $T_{\rm s} = 80$\,K which is compatible with 
the collisional temperature found from the
H$_2$ observations for column densities near $10^{20}$\,\cmm\ 
\citep{gillmon2006,wakker2006}.  
It is also similar to the average \hi\ spin temperature (column density weighted) found by \cite{heiles2003}.
$T_{\rm s}$ will be higher
for the WNM, but for these high latitude diffuse lines of sight $T_{\rm b} \ll
80$\,K for the broad WNM lines, and so adopting the wrong $T_{\rm s}$ is of no
consequence.
For most fields the correction is less than 3\% compared to the
optically-thin assumption.  Indeed very few of our 21-cm spectra reach
brightness temperatures above 40\,K -- only 3\% of the spectra in POL
field, the brightest one in the sample. For these ``extreme'' cases the
opacity correction to the column density reaches 35\%. 
Figures~\ref{fig:maps_nhi_1} and \ref{fig:maps_nhi_2} show 
the \hi\ column density maps of all fields, in units of $10^{20}$\,cm$^{-2}$.

\subsection{Uncertainties in \nh}

The main analysis presented here relies on a correlation analysis
between far-infrared/submm brightness and \nh\ of the components deduced
from 21-cm emission. In order to estimate properly the uncertainties of
the deduced correlation coefficients, we need to evaluate the
uncertainty of the values of \nh\ for the \hi\ components.  The method
described in Appendix~\ref{sec:noise_hi} takes advantage of the fact
that the GBT observations were obtained in two polarisations. The
difference between these spectra gives a direct estimate of the
uncertainty in each channel, which can be integrated over the appropriate
velocity ranges, providing a column density uncertainty for each \hi\ component.
The average \hi\ column densities and uncertainties, expressed in
units of 10$^{20}$\,cm$^{-2}$, are given in Table~\ref{table:mainHI} for
the three \hi\ components of each field. {This table gives also 
the average velocity of each \hi\ component and an 
estimate of the half-width at half-maximum of the 21-cm feature.}

\begin{figure*}
\begin{center}
\includegraphics[scale=1, draft=false, angle=0]{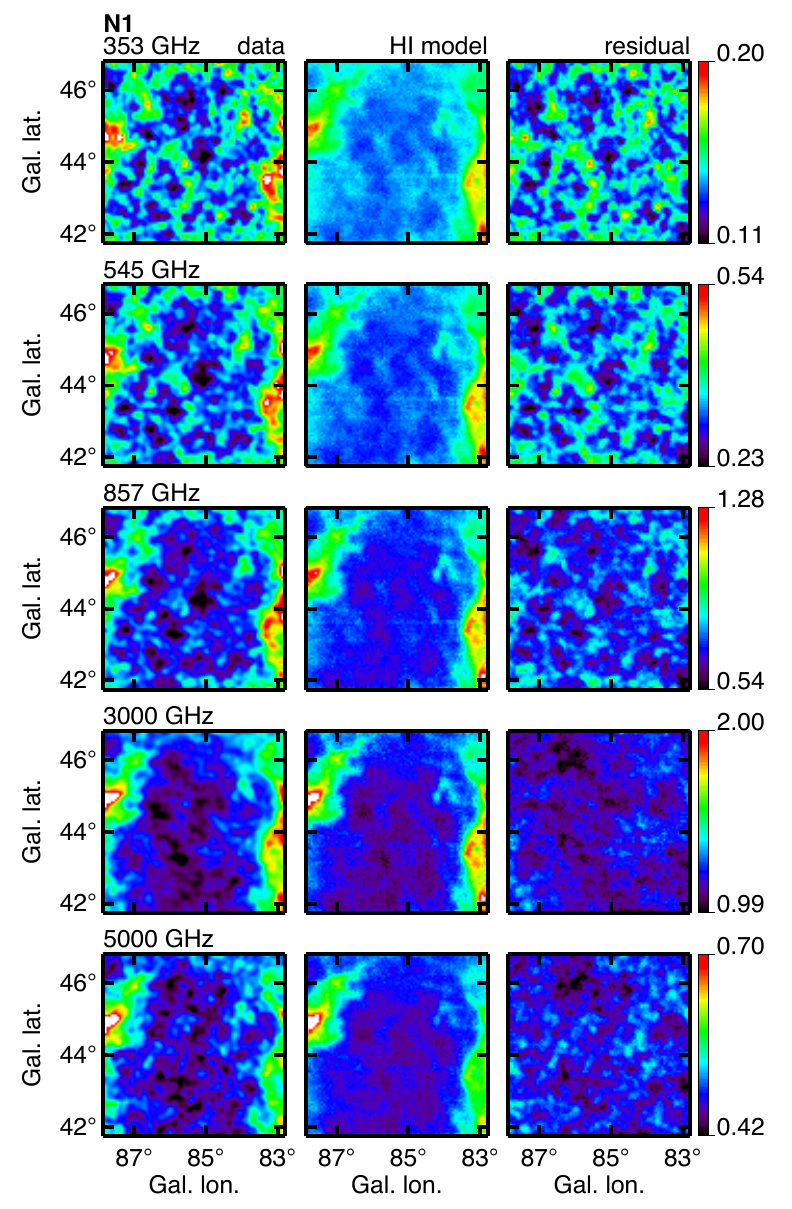}
\includegraphics[scale=1, draft=false, angle=0]{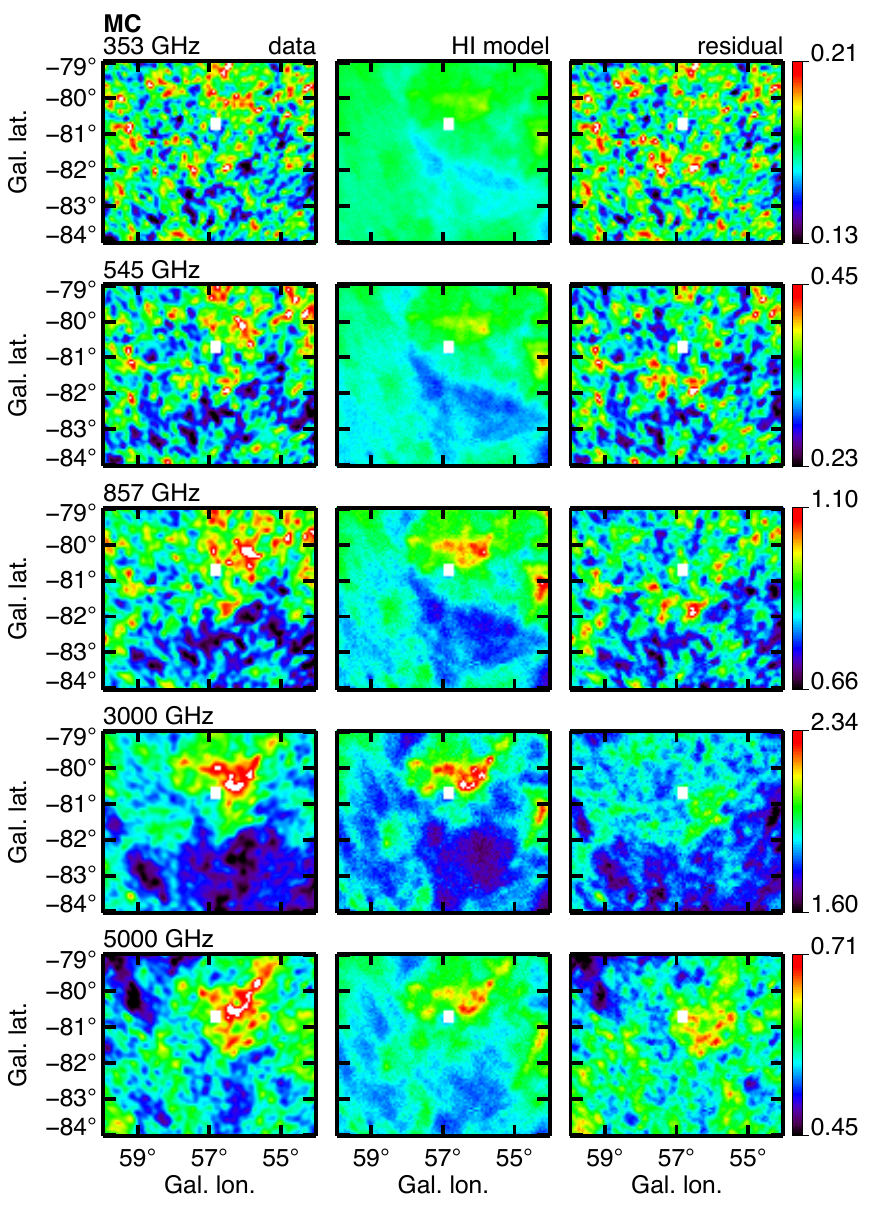}
\includegraphics[scale=1, draft=false, angle=0]{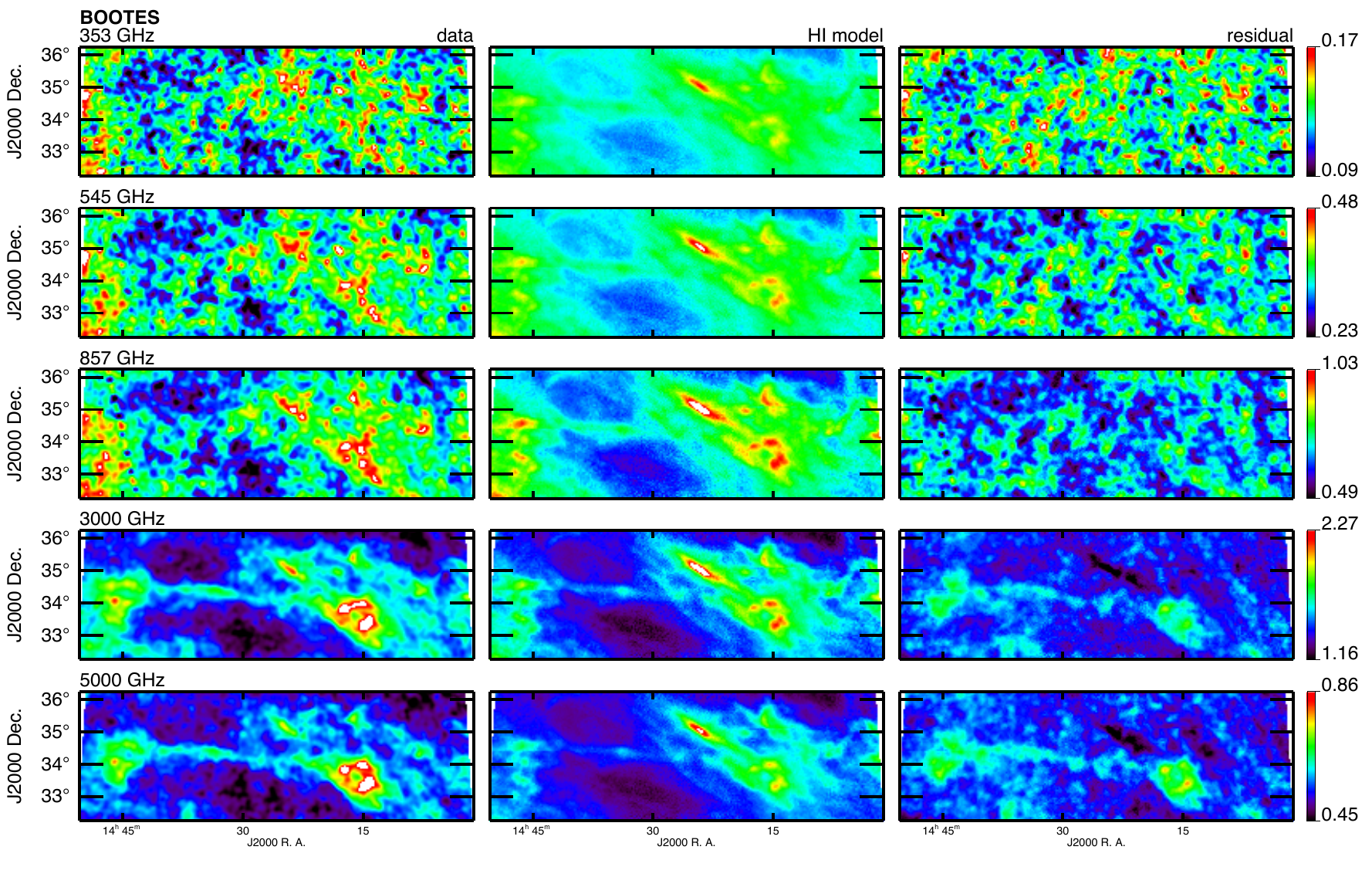}
\caption{\label{fig:N1_dust} Dust/gas correlation in N1 (top-left), 
  MC (top-right) and BOOTES (bottom):
  \Planck\ and \IRAS\ raw maps (left column), 
  the model of the dust emission based on the \hi\ observations 
  (middle: {\it \hi\ model}, $\sum_{i=1}^3 \epsilon_\nu^i N_{\rm HI}^{i}(x,y)$, Eq.~\ref{eq:model})
  and the residual emission
  (right, see Eq.~\ref{eq:residu}). Units are MJy\,sr$^{-1}$. Data are described in Sect.~\ref{sec:planckiras} and model and 
residual in Sect.~\ref{sec:correlation}.}
\end{center}
\end{figure*}

\section{\Planck\ and \IRAS}

\label{sec:planckiras}

\subsection{Map construction}

Our analysis uses infrared to submm data at 3000 and 5000\,GHz (100 and
60\,$\mu$m, respectively) from \IRAS\ (IRIS,
\citealp{miville-deschenes2005a}) and at 353, 545, and 857\,GHz (850,
550, and 350\,$\mu$m, respectively) from \Planck\ (DR2 release;
\citealp{planck2011-1.7}), beginning with maps in Healpix form
\cite[]{gorski2005} with $N_{\rm side}=2048$ (pixel size of 1.7\arcmin).  
We concentrated here on the three highest frequencies of \Planck\ to
avoid the significant contamination from residual CMB fluctuations and
interstellar emission other than thermal dust (CO, synchrotron,
free-free and spinning dust).

\begin{figure*}
\begin{center}
\includegraphics[scale=1, draft=false, angle=0]{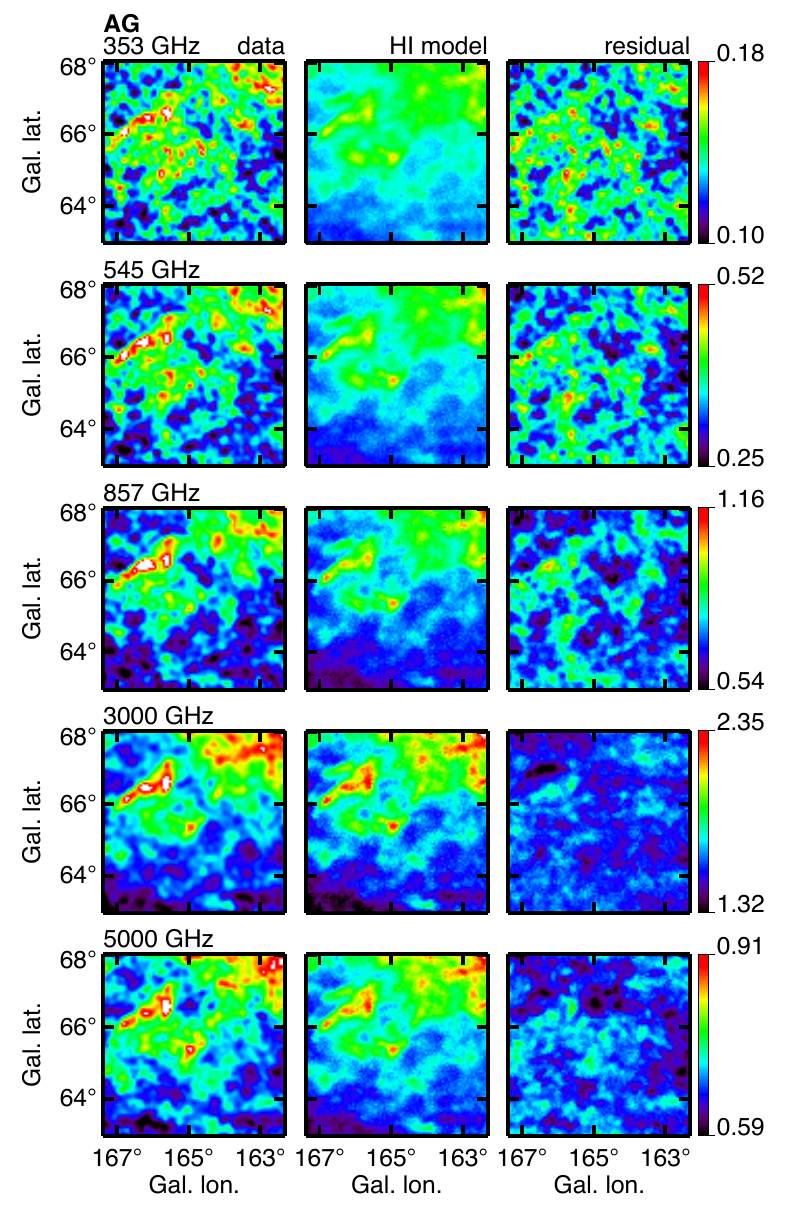}
\includegraphics[scale=1, draft=false, angle=0]{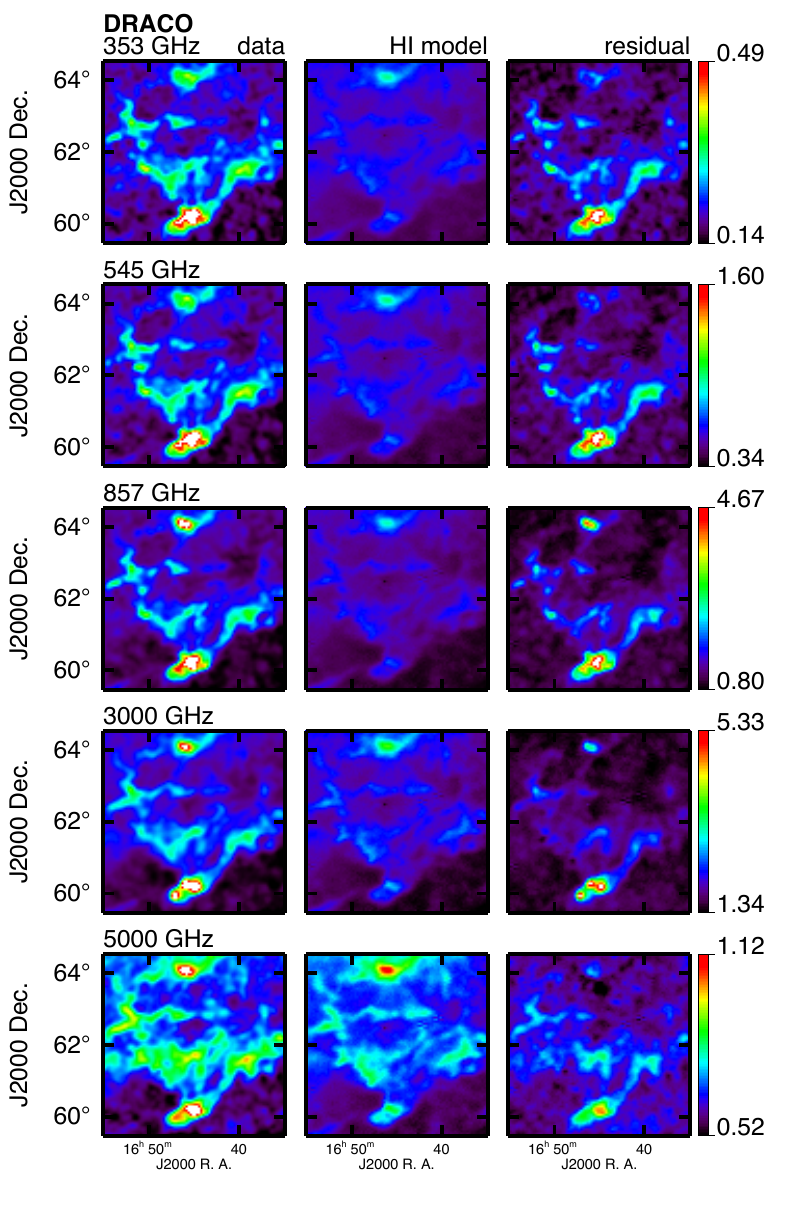}
\includegraphics[scale=1, draft=false, angle=0]{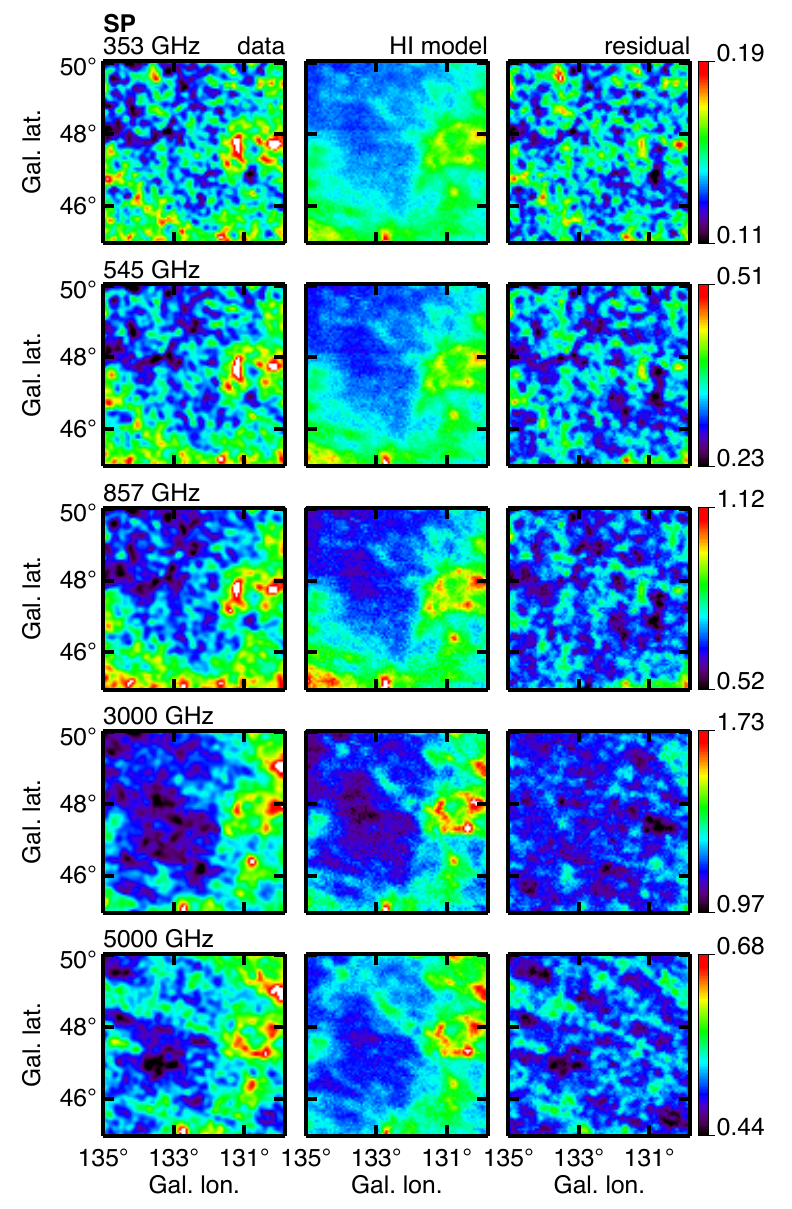}
\includegraphics[scale=1, draft=false, angle=0]{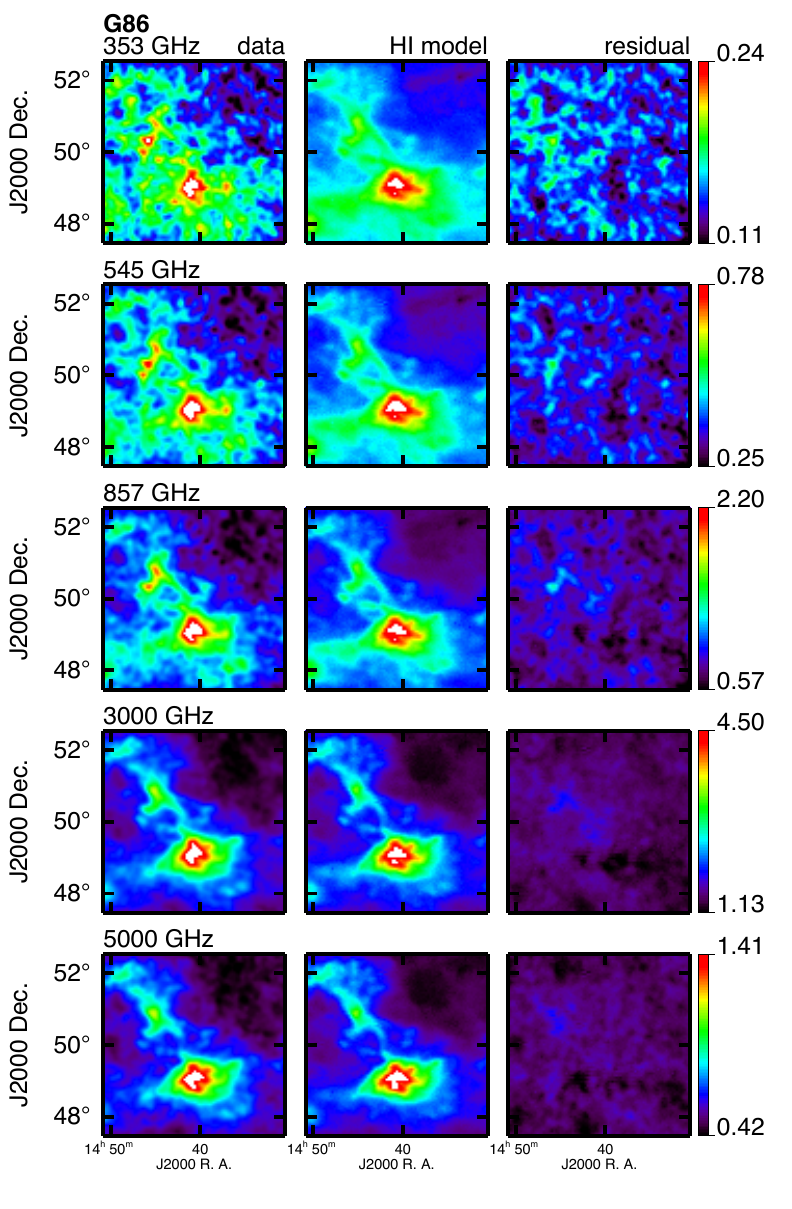}
\caption{\label{fig:AG_dust} Like Fig.~\ref{fig:N1_dust}, for AG (top-left), DRACO (top-right), SP (bottom-left) and G86 (bottom-right).}
\end{center}
\end{figure*}

\begin{figure*}
\centering
\includegraphics[scale=1, draft=false, angle=0]{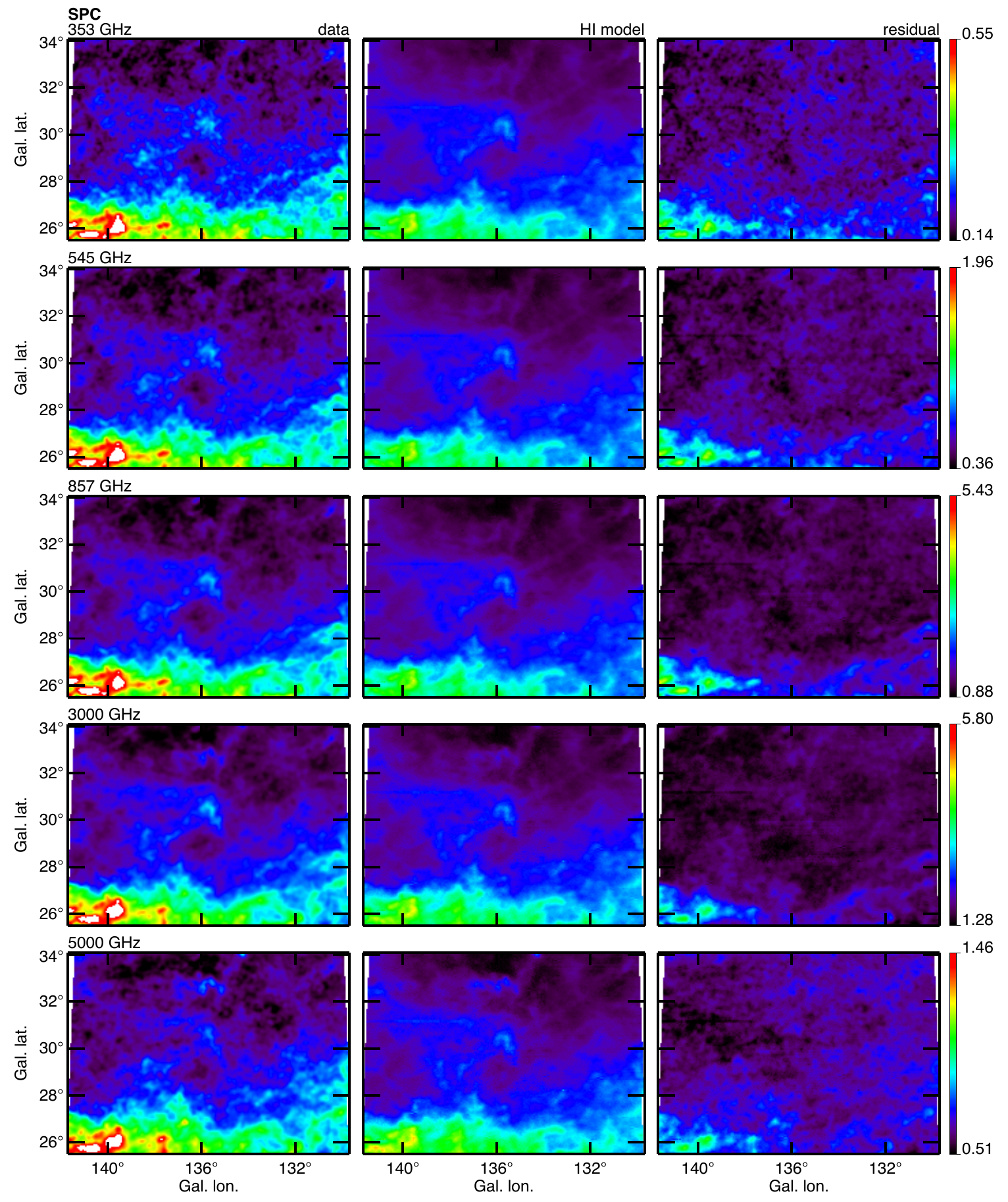}
\caption{\label{fig:SPC_dust} Like Fig.~\ref{fig:N1_dust}, for field SPC. }
\end{figure*}

To obtain infrared-submm maps corresponding to each GBT field we first
projected each Healpix map, using the nearest neighbour method, onto SFL
grids with a pixel size of 1.7\arcmin.  Each grid was centred on a given GBT
field with a size 10\% larger in each direction in order to avoid edge
effects in subsequent convolution steps.  Each SFL map was then converted to
MJy\,sr$^{-1}$ and point sources were removed and replaced by
interpolation of the surrounding map\footnote{For \Planck\ channels we
  removed only point sources identified in the ERCSC \cite[]{planck2011-1.10}. 
  For \IRAS\ maps we used the source removal method described in
  \cite{miville-deschenes2005a}.}.  The map was then convolved to bring
it to the GBT 9.4\arcm\ resolution and finally projected, using bi-linear
interpolation, on the actual GBT grid (3.5\arcmin\,pixel$^{-1}$).  
The \Planck\ and \IRAS\ maps for our fields are shown in 
Figs~\ref{fig:N1_dust} to \ref{fig:NEP_dust}. 
As will be discussed in Sect.~\ref{sec:correlation}, these
figures also show models of this emission based on \hi\ observations
in a masked subset of the map, and the residual map on subtracting this model from the entire field. 
The residuals are largest for those areas in the map not used to constrain the model.

\subsection{Dust brightness uncertainty}

To estimate the noise level of the \Planck\ and \IRAS\ maps we used the
method described in Sect.~5.1 of \cite{miville-deschenes2005a}.  For both
data sets we used the difference of maps of the same region of the sky
obtained with different sub-samples of the data.  These difference maps,
properly weighted by their coverage maps, provide an estimate of the
statistical properties of the noise.  For \Planck\ the noise was
estimated using the difference of the first and second half ring maps \cite[]{planck2011-1.7}.
In the case of \IRAS, each ISSA plate is the combination of up to three
maps built from independent observations over the life of the
satellite. We built difference maps from these three sets of maps.  The
procedure used to estimate the \Planck\ and \IRAS\ noise levels at the
GBT resolution is detailed in Appendix~\ref{sec:noise_hfi_iris}.  The
noise levels for each field and each frequency are given in
Table~\ref{table:noise_hfi_iris}.

\section{Dust--HI correlation}

\label{sec:correlation}

\subsection{Model}

Many studies, mostly using the \IRAS\ and \COBE\ data compared with
various 21-cm surveys
\cite[]{boulanger1988,joncas1992,jones1995,boulanger1996,arendt1998,reach1998a,lockman2005,miville-deschenes2005},
have revealed the strong correlation between far-infrared/submm dust
emission and 21-cm integrated emission $W_{\rm HI}$\footnote{equivalent to
  optically-thin \hi\ column density.} at high Galactic latitudes.  In
particular \cite{boulanger1996} studied this relation over the whole
high Galactic latitude sky.  They reported a tight dust--\hi\ correlation
for $W_{\rm HI}<250$\,K\,km\,s$^{-1}$, corresponding to $N_{\rm HI}<4.6\times
10^{20}$\,cm$^{-2}$. For higher column densities the dust emission
systematically exceeds that expected by extrapolating the correlation.
Examining specific high Galactic latitude regions, \cite{arendt1998} and
\cite{reach1998a} found infrared excesses with respect to \nh, with a
threshold varying from 1.5 to $5.0 \times 10^{20}$\,cm$^{-2}$.

Part of this excess is due to the effect of 21-cm
self-absorption that produces a systematic underestimate of the column
density when deduced with the optically thin assumption.  
Even though this effect is only at a level of a few percent in our case because of the low column densities, 
applying an opacity correction (see Sect.~\ref{sec:HIcomponents}) helps to limit this systematic effect.

Most of the infrared/submm excess is usually attributed to dust
associated with hydrogen in molecular form.  This hypothesis is in
accordance with UV absorption measurements that show a sudden increase
of the H$_2$ absorption at $N_{\rm H} = (3-5) \times 10^{20}$\,cm$^{-2}$
\cite[]{savage1977,gillmon2006}, roughly the threshold for departure
from the linear correlation between dust emission and \nh.  It is also
observed that the pixels showing evidence of excess are spatially
correlated and correspond to, or at least are in the vicinity of, known
molecular clouds traced by CO emission.
See also the discussion in Sect.~\ref{sec:transition}.

\begin{figure*}
\centering
\includegraphics[scale=1, draft=false, angle=0]{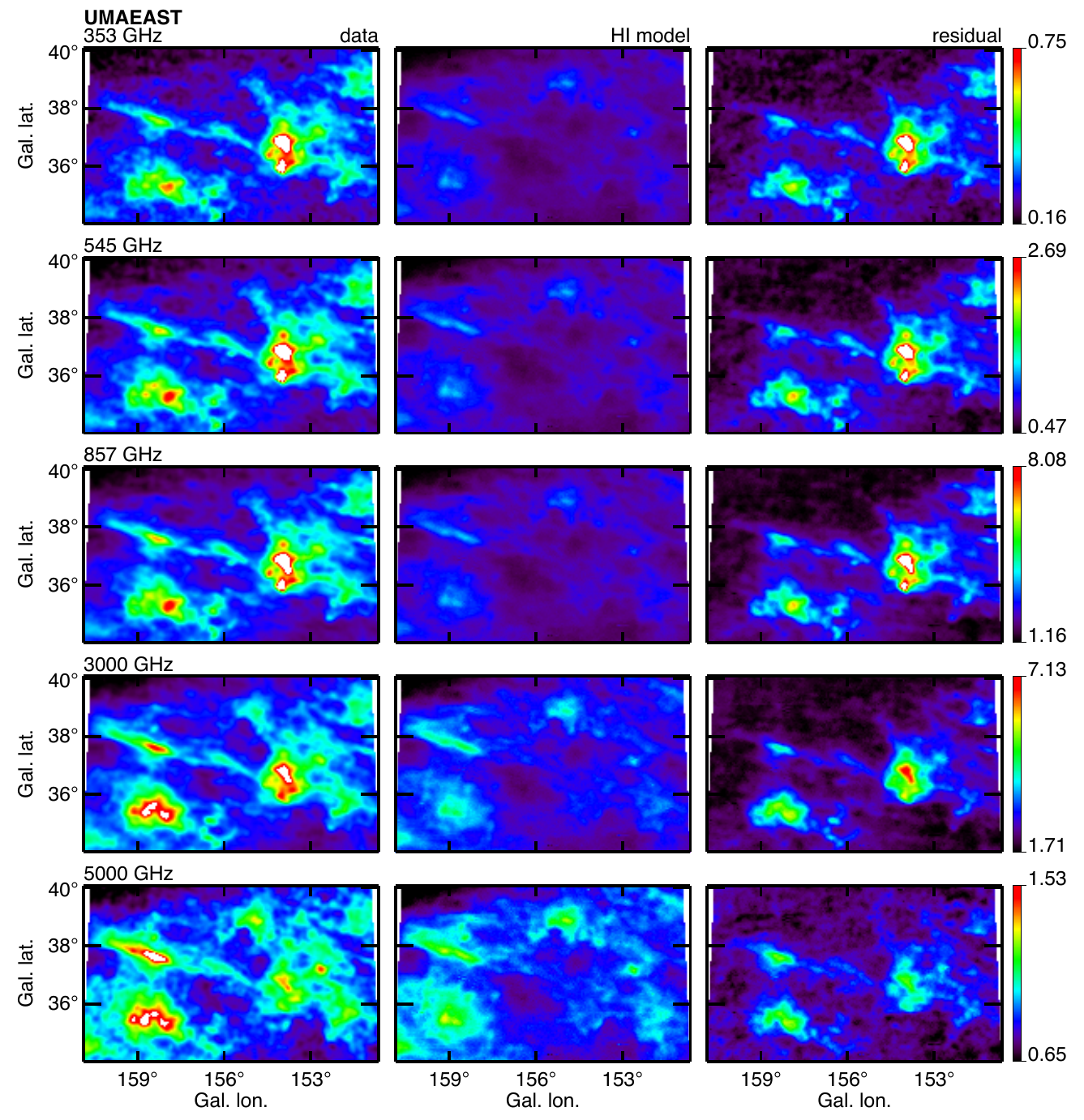}
\caption{\label{fig:UMAEAST_dust} Like Fig.~\ref{fig:N1_dust}, for field UMAEAST.}
\end{figure*}

\begin{figure*}
\centering
\includegraphics[scale=1, draft=false, angle=0]{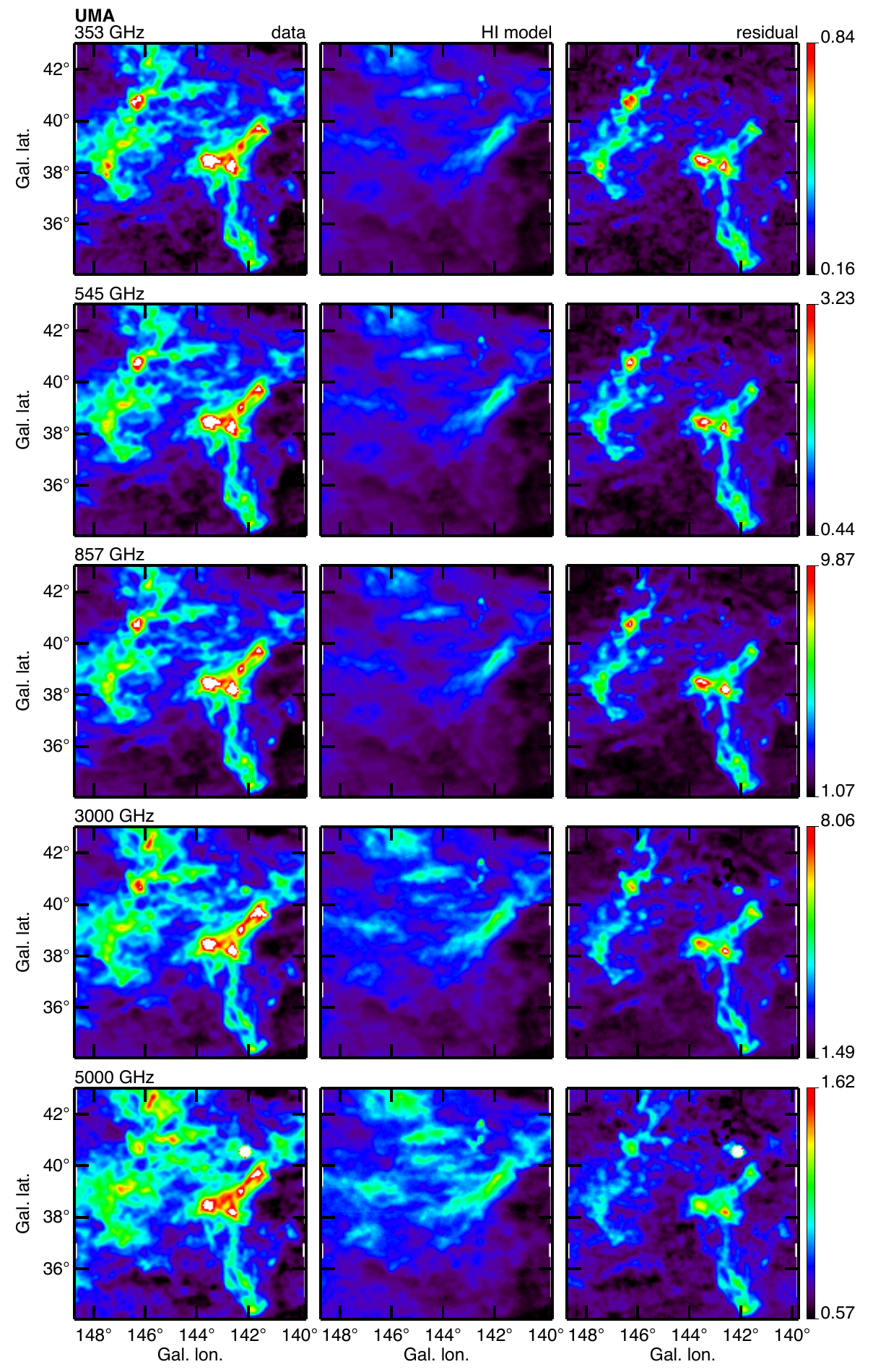}
\caption{\label{fig:UMA_dust} Like Fig.~\ref{fig:N1_dust}, for field UMA.}
\end{figure*}

\begin{figure*}
\centering
\includegraphics[scale=1, draft=false, angle=0]{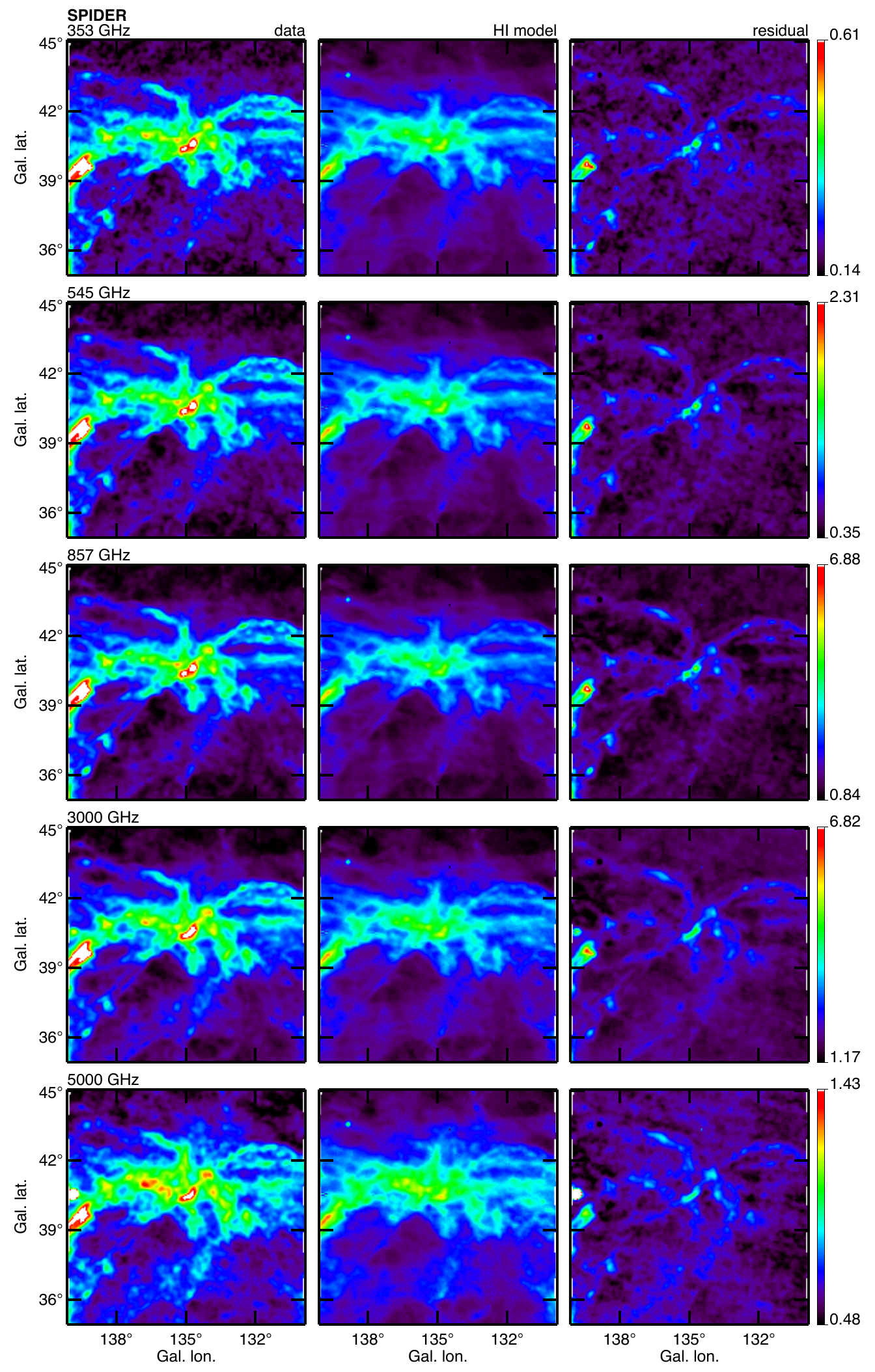}
\caption{\label{fig:SPIDER_dust} Like Fig.~\ref{fig:N1_dust}, for field SPIDER.}
\end{figure*}

\begin{figure*}
\centering
\includegraphics[scale=1, draft=false, angle=0]{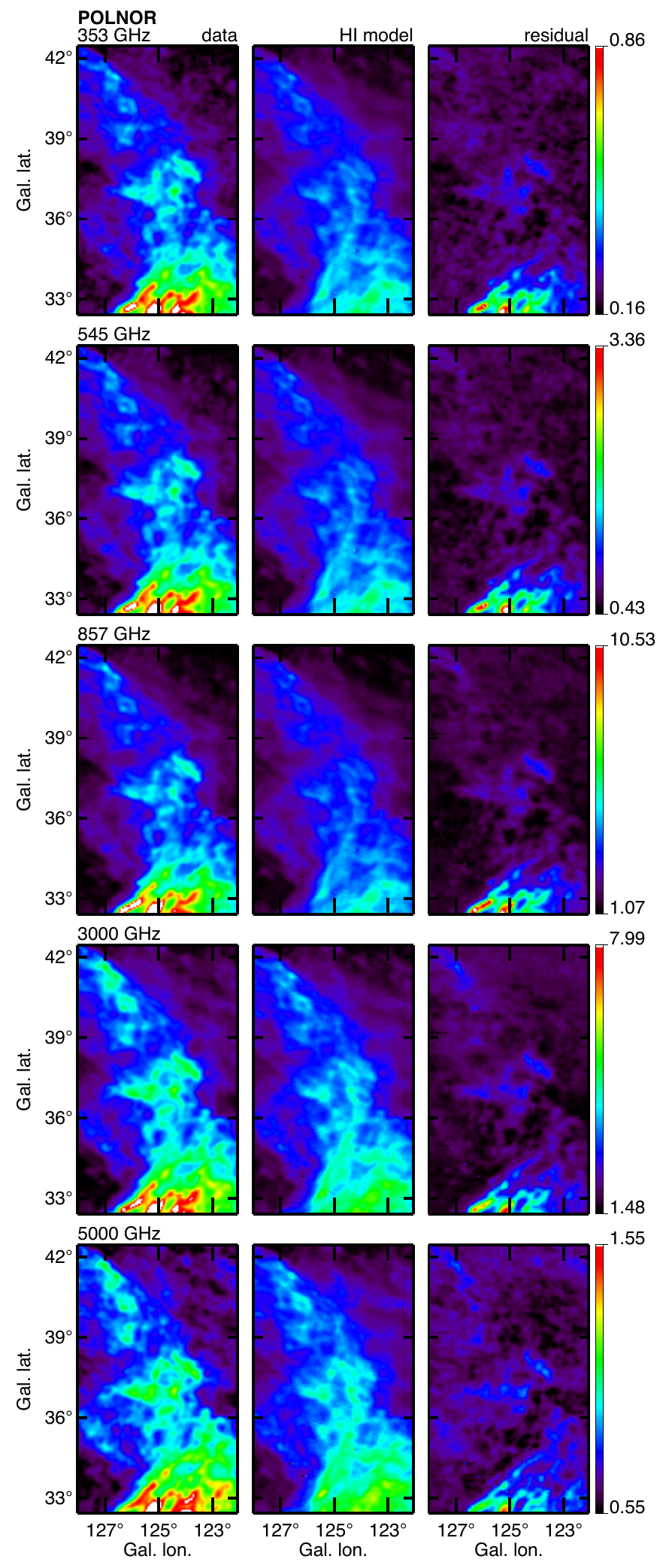}
\includegraphics[scale=1, draft=false, angle=0]{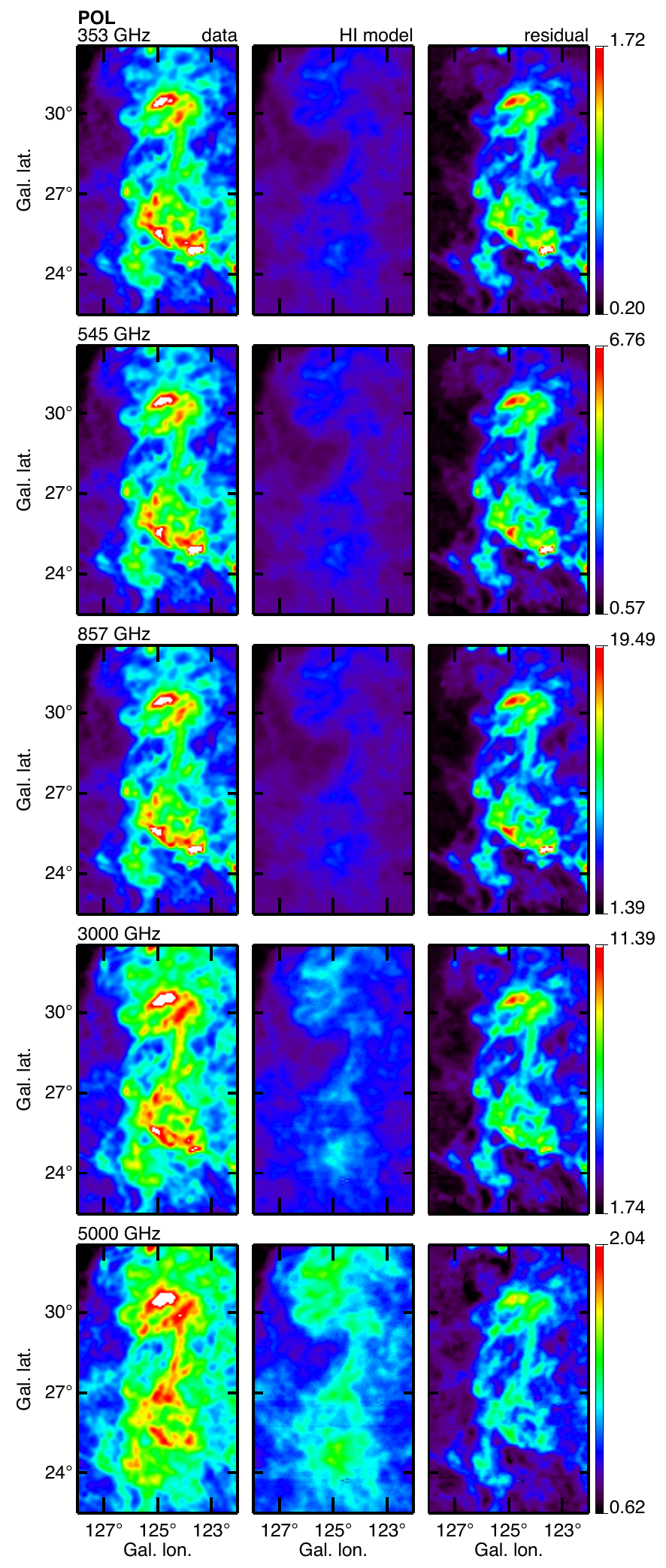}
\caption{\label{fig:POL_dust} Like Fig.~\ref{fig:N1_dust}, for fields POLNOR (left) and POL (right).}
\end{figure*}

\begin{figure*}
\centering
\includegraphics[scale=1, draft=false, angle=0]{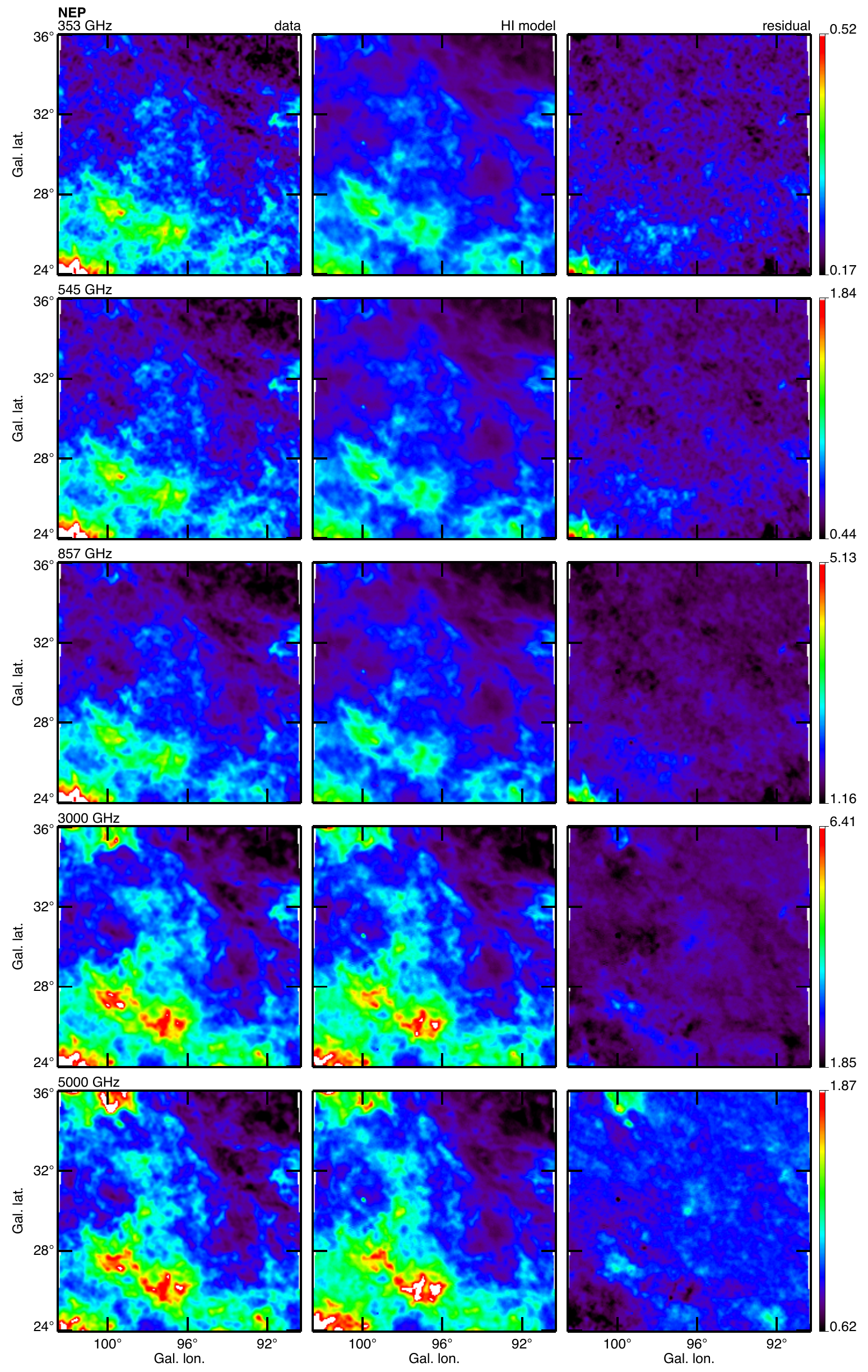}
\caption{\label{fig:NEP_dust} Like Fig.~\ref{fig:N1_dust}, for field NEP, the largest.}
\end{figure*}

A third source of this excess emission could be dust associated with the
Warm Ionized Medium (WIM) but detection of this component is difficult 
\citep{arendt1998,lagache2000a} because 
there is no direct tracer of the ionized gas column density;
H$\alpha$ depends on the square of the electron density and
part of the structure seen in H$\alpha$ might be back-scattering of diffuse Galactic emission
on dust and not photons produced within cirrus clouds \citep{witt2010}.

Finally, in the most diffuse regions of the high-latitude sky, the
fluctuations of the CIB are a significant fraction of the brightness
fluctuations in the infrared/submm.  With a power spectrum flatter
($k^{-1}$) than that of the interstellar dust emission ($k^{-3}$) 
\citep{miville-deschenes2002b,miville-deschenes2007a,lagache2007,planck2011-6.6},
the CIB anisotropies contribute mostly at small angular scales,
producing statistically homogeneous brightness fluctuations over any
observed field, like an instrumental noise. 
Furthermore, because the CIB is unrelated to interstellar
emission, the CIB fluctuations cannot be responsible for the excess of
infrared emission seen at moderate to high \nh\ column density.

In the analysis presented here we go a few steps further than the
previous studies by: 1) allowing for different dust emissivities for the
local ISM (i.e., LVC), IVC, and HVC components; 2) applying an opacity correction to
the 21-cm brightness temperatures in order to compute a more reliable
\nh; and 3) taking into account explicitly the CIB fluctuations which
turn out to dominate the uncertainties in the derived emissivities.

In some fields (like G86 with strong IVC emission) the distinctive
morphology of the IVC column density map can be seen clearly in the
line-of-sight integrated dust emission map \citep[see Figs.~\ref{fig:maps_nhi_1} and \ref{fig:AG_dust} and][]{martin1994}, but even
faint signals can be brought out by formal correlation analysis.
We use the following model:
\begin{equation}
\label{eq:model}
I_\nu(x,y) = \sum_{i=1}^3 \epsilon_\nu^i N_{\rm HI}^{i}(x,y) + R_\nu(x,y) +
Z_\nu,
\end{equation}
where $I_\nu(x,y)$ is the dust map at frequency $\nu$ (\IRAS\ or
\Planck), $\epsilon_\nu^i$ the emissivity of \hi\ component $i$ (LVC, IVC and HVC), 
and $Z_\nu$ is the zero level of the map.  $R_\nu$ represents not only the
contribution from noise in the data but also any emission in the
\IRAS\ and \Planck\ bands that is not correlated with \nh\, including
the CIB anisotropies and potential dust emission coming from molecular
or ionized gas.  In this model we assume that the three HI components
$N_{\rm HI}^{i}(x,y)$ have a constant emissivity $\epsilon_\nu^i$ over the
field.  Any spatial variations of the emissivity would also contribute
to fluctuations in $R_\nu(x,y)$.

\subsection{Estimating the dust emissivities}

To estimate the parameters $\epsilon_\nu^i$ and constant $Z_\nu$ we used
the IDL function {\em regress} which, in the case of a general linear
least-squares fit, solves the following equation \cite[]{press1995}:
\begin{equation}
\label{eq:inverse1}
a = (A^TA)^{-1} \times (A^T b),
\end{equation}
where $a$ is the vector of the parameters $\epsilon_\nu^i$ and $b$ is a
vector of the $N$ \IRAS\ or \Planck\ data points from the map, divided
by their respective error:
\begin{equation}
b_i = \frac{I_\nu(i)}{\sigma_\nu}.
\end{equation}
$A$ is an $N \times M$ matrix that includes the \nh\ values of the $M$
\hi\ components,
\begin{equation}
\label{eq:inverse3}
A_{ij} = \frac{N_{\rm HI}^j(i)}{\sigma_\nu}.
\end{equation}
{\em Regress} uses a Gaussian elimination method for the inversion.

\begin{figure}[ht]
\centering
\includegraphics[width=0.95\linewidth, draft=false, angle=0]{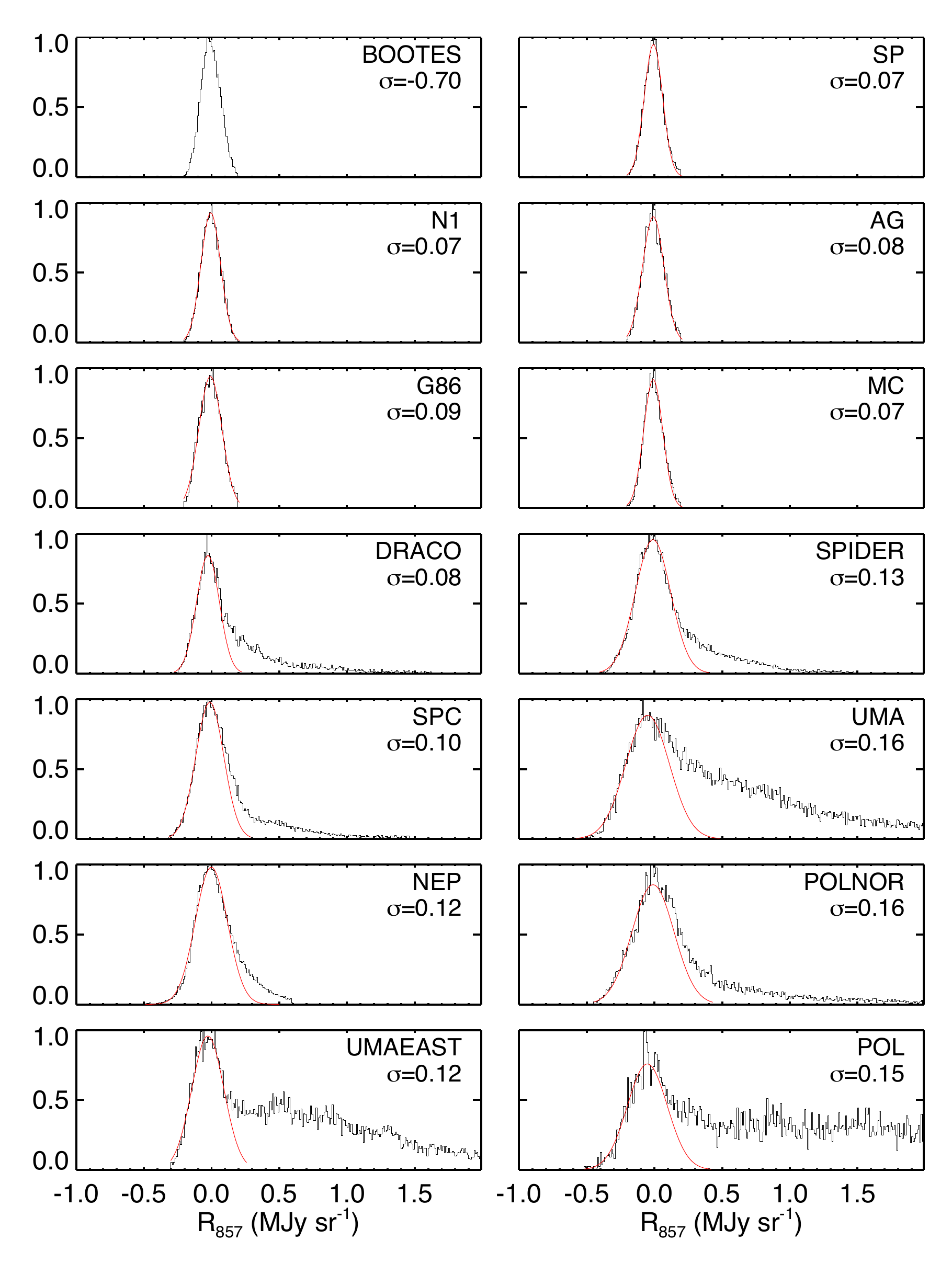}
\caption{\label{fig:pdf_mask_residu} Normalised PDF of the residual
  $R_{857}$ of the dust--gas correlation at 857\,GHz for each field after
  convergence of the masking procedure.  The red curve is the result of
  a Gaussian fit to the lower, rising part of the PDF. }
\end{figure}

For the model described in Eq.~\ref{eq:model}, the least-squares fit
method provides a maximum-likelihood estimation of the parameters
$\epsilon_\nu^i$ provided that the residual term $R_\nu(x,y)$ is
uncorrelated with $N_{\rm HI}^i$ and its fluctuations are normally
distributed (i.e., white noise).  In addition, in order that the
parameter estimates not be biased, the uncertainties on $N_{\rm HI}^i$ have
to be comparatively small; we will show (see Sect.~\ref{sec:monte_carlo})
that this last condition is satisfied for our data.  On the other hand,
we will also show that the residual term $R_\nu(x,y)$ is clearly not
compatible with white noise.  Even for the most diffuse fields in our
sample, where the CIB fluctuations dominate the residual emission and
the Probability Density Function (PDF) of $R_\nu(x,y)$ is normally
distributed, the condition that there be no (not even chance)
correlation with \nh\ is not satisfied, because the power spectrum of
the CIB is not white, but rather like $k^{-1}$ \citep{planck2011-6.6}.
For brighter fields, where spatial variation of the dust emission with
respect to the \hi\ templates is expected (due to the presence of
molecules, a poor \hi\ opacity correction, or spatial variation of dust properties), the residual is not
even normally distributed.

To limit the influence of these effects, and in order to focus on
estimating the dust emissivity of the \hi\ components, we relied on a
masking procedure to flag and remove obvious outliers with respect to
the correlation, and on Monte-Carlo simulations to estimate the
$\epsilon_\nu^i$ uncertainties and bias (see Sect.~\ref{sec:monte_carlo}).

\subsection{Masking}

In order to limit the effect of lines of sight with significant
``excess'' dust emission that is not associated with \hi\ gas, most
previous authors used only data points with \nh\ lower than a given
threshold to stay in a regime of linear correlation.  This thresholding
was motivated by the fact that above some \nh\ the extinction and
self-shielding of H$_2$ are strong enough to limit photo-dissociation,
whereas below the threshold the hydrogen is mostly atomic.  The
threshold depends sensitively on local gas density and temperature but
for typical interstellar conditions for CNM gas ($n=100$\,cm$^{-3}$,
$T=80$\,K, $G=1$, where $G$ is the scaling factor of the InterStellar Radiation Field (ISRF) as defined
by \cite{mathis1983}), it is about $N_{\rm HI}=2.5 \times 10^{20}$\,cm$^{-2}$
\cite[]{reach1998a}.
Others have used a quadratic function for $\epsilon_\nu$ \cite[]{fixsen1998}
based on the idea that the H$_2$ column density depends (at least
dimensionally) on $N_{\rm HI}^2$ \cite[]{reach1994}.
Both methods introduce a bias in the parameter estimation that is
difficult to quantify.

Instead of applying an arbitrary cutoff in $N_{\rm HI}$, \cite{arendt1998}
used an iterative method to exclude data points above a cut along lines
perpendicular to the fit, in order to arrive at a stable solution for
$\epsilon_\nu$.  We used a similar approach by iteratively masking out
data points that would produce a positively-skewed residual.  That way
we expect to keep pixels in the maps that correspond to lines of sight
where the dust emission is dominated by the atomic \hi\ components.

The PDF of residual map, $R$, defined as
\begin{equation}
\label{eq:residu}
R_\nu(x,y) \equiv I_\nu(x,y) - \sum_{i=1}^3 \epsilon_\nu^i
N_{\rm HI}^{i}(x,y) - Z_\nu,
\end{equation}
is used to estimate the mask.  We used the \Planck\ 857-GHz channel
because it has the best signal-to-noise ratio and is less sensitive than
the \IRAS\ channels to dust temperature-induced emissivity variations.
For the most diffuse fields in our sample (AG, MC, N1, BOOTES, G86, SP)
the PDF of $R_{857}$ is very close to a Gaussian, which suggests that the
model described by Eq.~\ref{eq:model} is the right one in such low
column density regions.

Accordingly, for the first iteration of the masking process for each field, 
we performed the multi-variate linear
regression based on Equations~\ref{eq:inverse1} to \ref{eq:inverse3}
using only the faintest 10\% pixels in the 857~GHz map. 
For further iterations this threshold was relaxed, bringing in more pixels in the map compatible with the iterated model.

As discussed above, the presence of dust emission associated with
molecular gas can positively skew the PDF, and empirically the PDF of
$R_{857}$ is indeed positively skewed for the eight remaining fields (Fig.~\ref{fig:pdf_mask_residu}). 
To determine the set of pixels to be retained, we used a Gaussian fit to
the lower, rising part of the PDF, up to the PDF maximum, and estimated
the $\sigma$ (see the red curves in Fig.~\ref{fig:pdf_mask_residu}).
With the above motivation, we assume that the lower part of the PDF is
representative of pixels where the fit works well (i.e., for these
pixels the residual is normally distributed).  Using the $\sigma$ fit
only to this part of the PDF, we apply a threshold in $R_{857}$ by
masking out all pixels with $R_{857} > 3\times \sigma$ away from the
mean.  We iteratively recompute the parameters and mask to converge on a
stable solution. The PDFs of $R_{857}$ obtained at the end of the process are shown in
Fig.~\ref{fig:pdf_mask_residu}.

\begin{table*}
\begin{center}
\begin{tabular}{lcccccc} \toprule
Field & HI & $\epsilon_{353}$ & $\epsilon_{545}$ & $\epsilon_{857}$ & $\epsilon_{3000}$ & $\epsilon_{5000}$ \\ \midrule
AG & LVC & $0.034 \pm 0.007$ & $0.14 \pm 0.02$ & $0.39 \pm 0.05$ & $0.68 \pm 0.04$ & $0.181 \pm 0.018$\\
 & IVC & $0.020 \pm 0.005$ & $0.075 \pm 0.016$ & $0.22 \pm 0.03$ & $0.58 \pm 0.03$ & $0.161 \pm 0.013$\\
 & HVC & $0.004 \pm 0.003$ & $0.011 \pm 0.008$ & $0.018 \pm 0.015$ & $0.037 \pm 0.015$ & $0.009 \pm 0.006$\\
BOOTES & LVC & $0.045 \pm 0.010$ & $0.17 \pm 0.03$ & $0.54 \pm 0.06$ & $0.95 \pm 0.08$ & $0.23 \pm 0.03$\\
 & IVC & $0.029 \pm 0.007$ & $0.12 \pm 0.02$ & $0.37 \pm 0.04$ & $0.87 \pm 0.05$ & $0.275 \pm 0.020$\\
 & HVC & $0.02 \pm 0.02$ & $0.09 \pm 0.07$ & $0.20 \pm 0.13$ & $-0.30 \pm 0.18$ & $-0.13 \pm 0.07$\\
DRACO & LVC & $0.043 \pm 0.011$ & $0.18 \pm 0.04$ & $0.49 \pm 0.07$ & $0.57 \pm 0.07$ & $0.08 \pm 0.02$\\
 & IVC & $0.042 \pm 0.004$ & $0.168 \pm 0.013$ & $0.48 \pm 0.03$ & $0.70 \pm 0.03$ & $0.167 \pm 0.009$\\
 & HVC & $0.007 \pm 0.006$ & $0.032 \pm 0.018$ & $0.07 \pm 0.04$ & $0.10 \pm 0.04$ & $0.025 \pm 0.013$\\
G86 & LVC & $0.033 \pm 0.004$ & $0.146 \pm 0.012$ & $0.45 \pm 0.02$ & $0.71 \pm 0.03$ & $0.165 \pm 0.008$\\
 & IVC & $0.0151 \pm 0.0019$ & $0.070 \pm 0.006$ & $0.238 \pm 0.013$ & $0.643 \pm 0.015$ & $0.206 \pm 0.004$\\
 & HVC & $-0.05 \pm 0.02$ & $-0.19 \pm 0.07$ & $-0.36 \pm 0.16$ & $-0.24 \pm 0.19$ & $-0.09 \pm 0.05$\\
MC & LVC & $0.031 \pm 0.010$ & $0.15 \pm 0.03$ & $0.45 \pm 0.06$ & $0.78 \pm 0.07$ & $0.18 \pm 0.03$\\
 & IVC & $0.008 \pm 0.009$ & $0.02 \pm 0.03$ & $0.16 \pm 0.05$ & $0.70 \pm 0.06$ & $0.17 \pm 0.03$\\
 & HVC & $-0.006 \pm 0.002$ & $-0.020 \pm 0.007$ & $-0.047 \pm 0.015$ & $-0.030 \pm 0.017$ & $0.019 \pm 0.007$\\
N1 & LVC & $0.056 \pm 0.007$ & $0.215 \pm 0.020$ & $0.58 \pm 0.04$ & $0.86 \pm 0.03$ & $0.166 \pm 0.011$\\
 & IVC & $0.039 \pm 0.007$ & $0.15 \pm 0.02$ & $0.41 \pm 0.04$ & $0.72 \pm 0.03$ & $0.213 \pm 0.012$\\
 & HVC & $0.005 \pm 0.004$ & $0.015 \pm 0.014$ & $0.04 \pm 0.03$ & $-0.01 \pm 0.02$ & $-0.001 \pm 0.007$\\
NEP & LVC & $0.0420 \pm 0.0014$ & $0.163 \pm 0.005$ & $0.470 \pm 0.011$ & $0.664 \pm 0.013$ & $0.141 \pm 0.005$\\
 & IVC & $0.0197 \pm 0.0012$ & $0.080 \pm 0.004$ & $0.236 \pm 0.010$ & $0.666 \pm 0.013$ & $0.229 \pm 0.004$\\
 & HVC & $-0.021 \pm 0.009$ & $-0.09 \pm 0.03$ & $-0.22 \pm 0.07$ & $-0.43 \pm 0.10$ & $-0.02 \pm 0.03$\\
POL & LVC & $0.0519 \pm 0.0019$ & $0.203 \pm 0.007$ & $0.57 \pm 0.02$ & $0.455 \pm 0.018$ & $0.102 \pm 0.004$\\
 & IVC & $0.056 \pm 0.012$ & $0.24 \pm 0.05$ & $0.61 \pm 0.13$ & $0.72 \pm 0.11$ & $0.18 \pm 0.03$\\
 & HVC & --- & --- & --- & --- & ---\\
POLNOR & LVC & $0.0476 \pm 0.0012$ & $0.200 \pm 0.004$ & $0.612 \pm 0.012$ & $0.538 \pm 0.012$ & $0.088 \pm 0.003$\\
 & IVC & $0.023 \pm 0.007$ & $0.08 \pm 0.02$ & $0.20 \pm 0.07$ & $0.51 \pm 0.07$ & $0.156 \pm 0.016$\\
 & HVC & --- & --- & --- & --- & ---\\
SP & LVC & $0.063 \pm 0.008$ & $0.25 \pm 0.03$ & $0.72 \pm 0.05$ & $0.59 \pm 0.04$ & $0.094 \pm 0.016$\\
 & IVC & $0.029 \pm 0.008$ & $0.11 \pm 0.02$ & $0.29 \pm 0.05$ & $0.77 \pm 0.04$ & $0.229 \pm 0.016$\\
 & HVC & $-0.003 \pm 0.007$ & $-0.02 \pm 0.02$ & $-0.07 \pm 0.04$ & $-0.12 \pm 0.04$ & $-0.042 \pm 0.013$\\
SPC & LVC & $0.0365 \pm 0.0017$ & $0.140 \pm 0.005$ & $0.401 \pm 0.011$ & $0.411 \pm 0.011$ & $0.086 \pm 0.004$\\
 & IVC & $0.015 \pm 0.008$ & $0.06 \pm 0.02$ & $0.18 \pm 0.05$ & $0.54 \pm 0.05$ & $0.197 \pm 0.020$\\
 & HVC & $-0.004 \pm 0.005$ & $-0.005 \pm 0.017$ & $0.00 \pm 0.04$ & $-0.05 \pm 0.04$ & $-0.005 \pm 0.013$\\
SPIDER & LVC & $0.0474 \pm 0.0014$ & $0.200 \pm 0.004$ & $0.602 \pm 0.012$ & $0.570 \pm 0.013$ & $0.093 \pm 0.004$\\
 & IVC & $0.030 \pm 0.003$ & $0.107 \pm 0.012$ & $0.30 \pm 0.03$ & $0.59 \pm 0.03$ & $0.162 \pm 0.009$\\
 & HVC & $-0.056 \pm 0.016$ & $-0.17 \pm 0.05$ & $-0.52 \pm 0.14$ & $-0.80 \pm 0.16$ & $-0.10 \pm 0.04$\\
UMA & LVC & $0.049 \pm 0.002$ & $0.211 \pm 0.007$ & $0.62 \pm 0.02$ & $0.563 \pm 0.017$ & $0.098 \pm 0.005$\\
 & IVC & $0.030 \pm 0.005$ & $0.12 \pm 0.02$ & $0.37 \pm 0.06$ & $0.58 \pm 0.05$ & $0.151 \pm 0.012$\\
 & HVC & $-0.013 \pm 0.005$ & $-0.064 \pm 0.019$ & $-0.18 \pm 0.05$ & $-0.12 \pm 0.04$ & $-0.006 \pm 0.011$\\
UMAEAST & LVC & $0.031 \pm 0.003$ & $0.147 \pm 0.008$ & $0.48 \pm 0.02$ & $0.566 \pm 0.016$ & $0.106 \pm 0.005$\\
 & IVC & $0.059 \pm 0.004$ & $0.219 \pm 0.014$ & $0.59 \pm 0.04$ & $0.65 \pm 0.03$ & $0.191 \pm 0.009$\\
 & HVC & $0.007 \pm 0.004$ & $0.024 \pm 0.013$ & $0.06 \pm 0.03$ & $0.04 \pm 0.03$ & $0.008 \pm 0.008$\\ \bottomrule
\end{tabular}
\end{center}
\caption{\label{table:emissivities} Emissivities of each \hi\
component at 353, 545, 857, 3000, and 5000\,GHz. Units are
MJy\,sr$^{-1}$/$10^{20}$\,cm$^{-2}$. The uncertainties were obtained
using Monte-Carlo simulations (see Section~\ref{sec:monte_carlo}).}
\end{table*}

\begin{figure*}
\centering
\setlength{\unitlength}{\textwidth}
\begin{picture}(1,1.25)
\put(0.0,0.88){\includegraphics[scale=0.98]{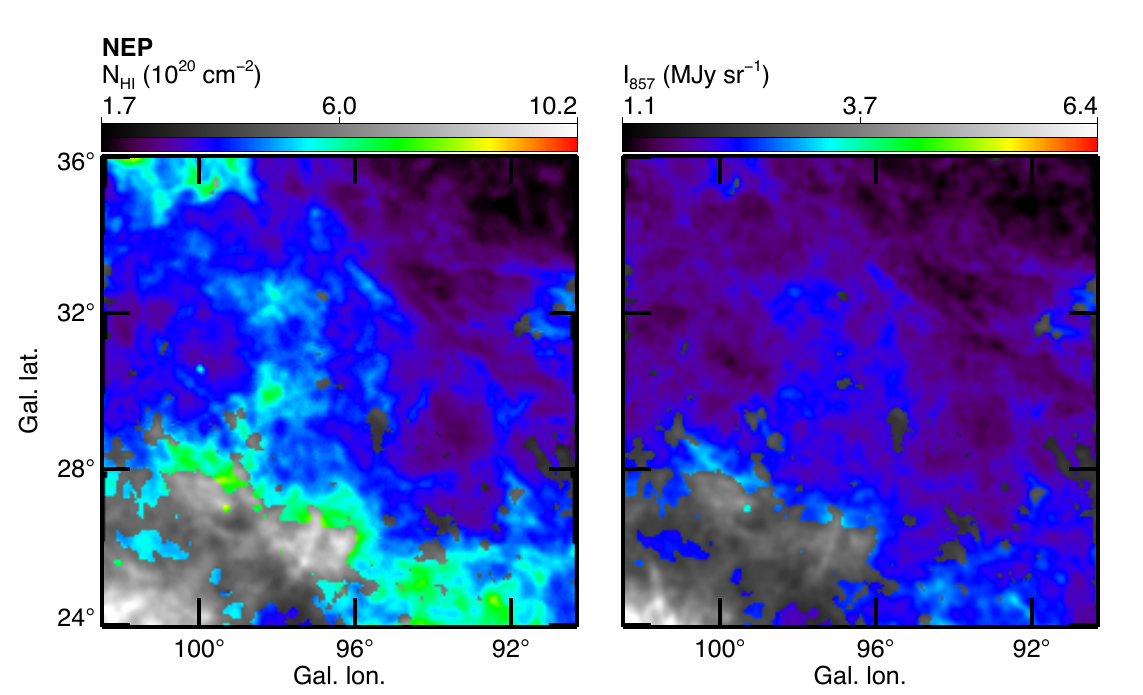}}
\put(0.65,0.92){\includegraphics[scale=0.98]{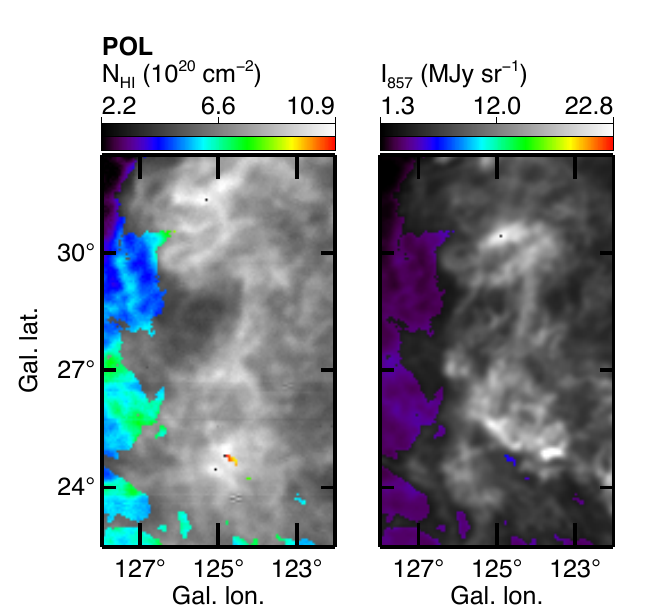}}
\put(0.0,0.58){\includegraphics[scale=0.98]{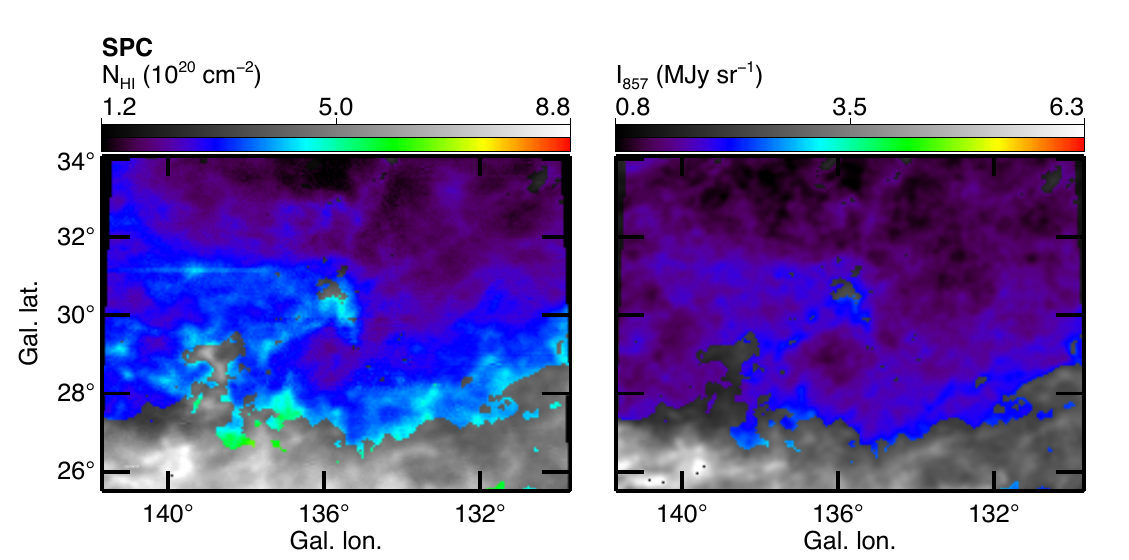}}
\put(0.65,0.58){\includegraphics[scale=0.98]{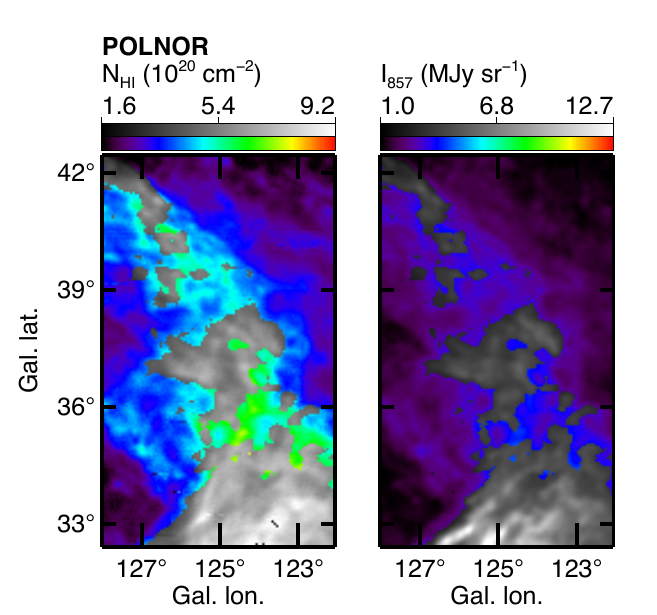}}
\put(0.0,0.25){\includegraphics[scale=0.98]{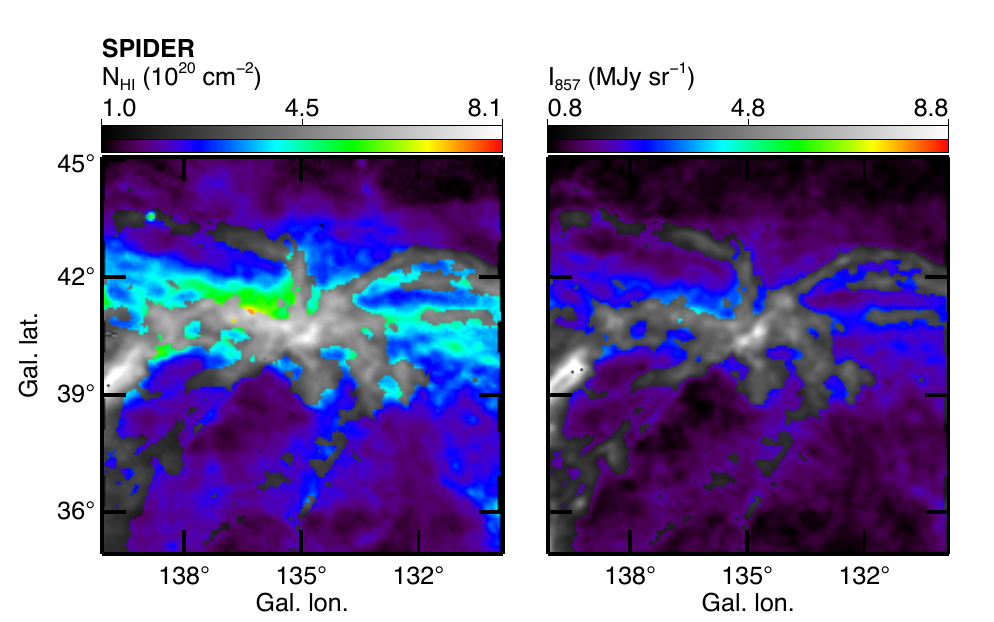}}
\put(0.53,0.25){\includegraphics[scale=0.98]{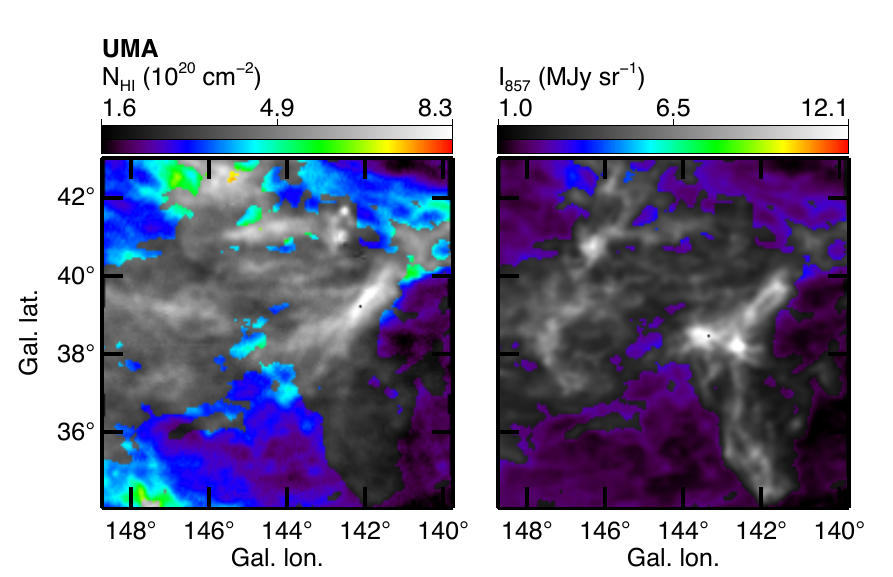}}
\put(0.0,0.0){\includegraphics[scale=0.98]{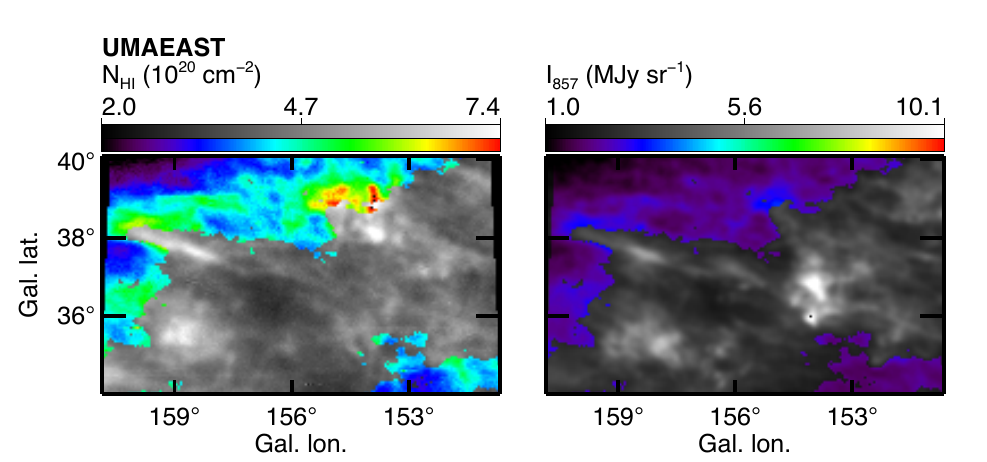}}
\put(0.65,0.){\includegraphics[scale=0.98]{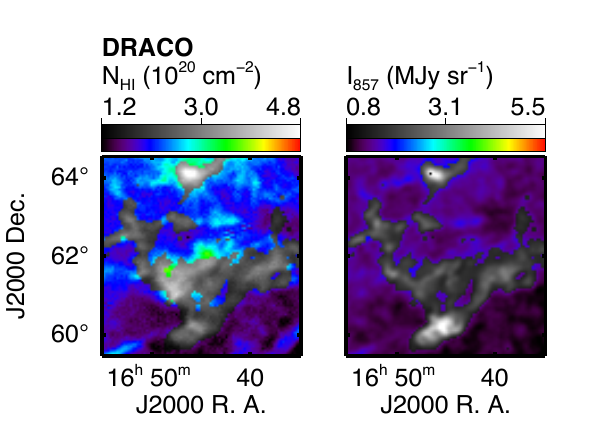}}
\end{picture}
\caption{\label{fig:mask_1}  Mask for the eight fields where more than 1\% of the pixels were excluded
(NEP=17\%, POL=83\%, SPC=26\%, POLNOR=30\%, SPIDER=24\%,  UMA=56\%, UMAEAST=66\%, DRACO=29\%). 
For each field the left image is the total \hi\ integrated emission and the
right image is the \Planck\ 857\,GHz map. The regions in greyscale were excluded from the correlation analysis.
Note the relationship of the masks to the regions of high residual in Figs.~\ref{fig:AG_dust} to \ref{fig:NEP_dust}.}
\end{figure*}

For the six faintest fields, the masking excluded less than 1\% of the points.
For these fields the mask has no significant effect on the estimated parameters.
For the eight other fields, the masking method excluded from 17 to 83\% of the pixels. 
The masks for these eight fields are shown in Fig.~\ref{fig:mask_1}.
In these cases the masking has a significant effect on the result, but we have checked that the estimated parameters are
similar to the ones obtained with a $N_{\rm HI} \leq 4 \times 10^{20}$\,cm$^{-2}$ threshold. 
In fact the masking method used here allows us to keep pixels that would have been excluded 
by a simple $N_{\rm HI}$ thresholding even though they
do not depart significantly from the linear correlation.

\begin{figure*}
\begin{center}
\includegraphics[width=0.49\linewidth, draft=false, angle=0]{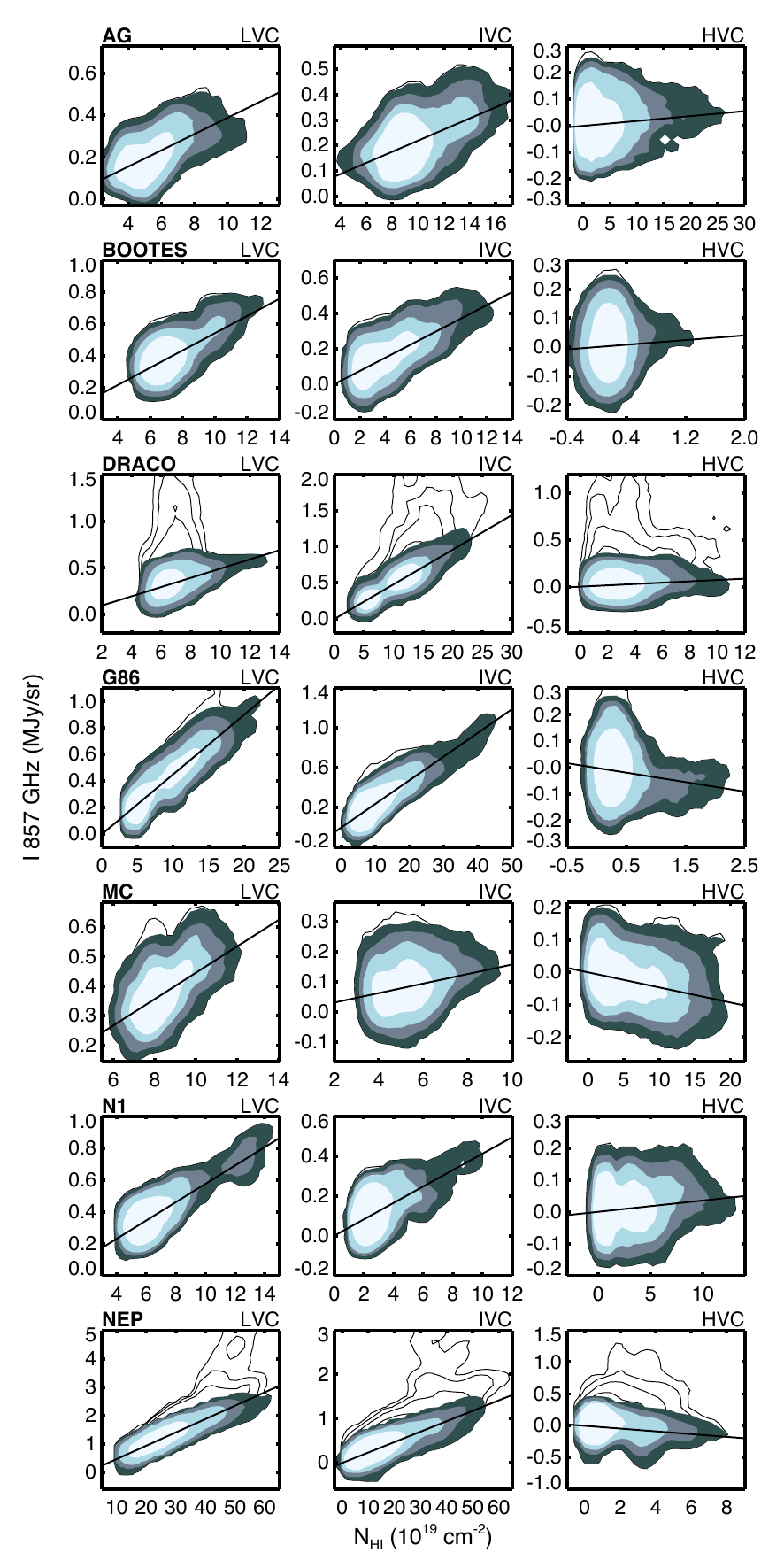}
\includegraphics[width=0.49\linewidth, draft=false, angle=0]{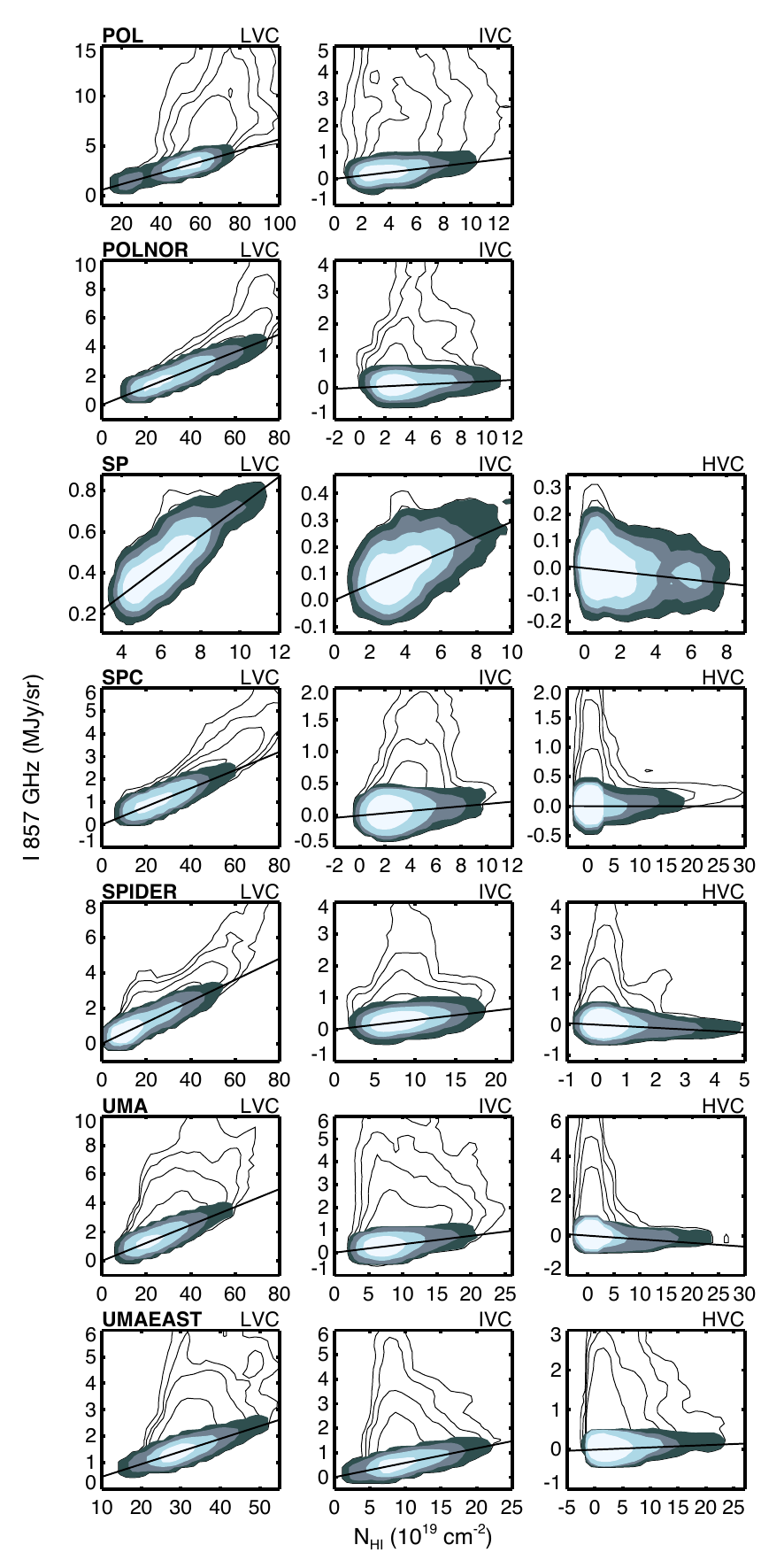}
\caption{\label{fig:tt_plot_1} $I_{857}$ vs.\ $N_{\rm HI}$ scatter plots,
visualising the dust--gas correlation for each of the three \hi\
components, across a row. The fields POL and POLNOR do not have an HVC component. 
For a given \hi\ component, the remaining 857\,GHz
emission, once the contribution of the other two \hi\ components has
been removed, is plotted as a function of \nh\ of that component (i.e., 
$I_\nu - \sum_{j \ne i} \epsilon_j
N_{\rm HI}^j -Z_\nu$ as a function of $N_{\rm HI}^i$).  The contours show quartiles of the density of 
points in each scatter plot. The filled contours show the data points used in
the correlation analysis (i.e., data points not masked out) while the open contours show all data points.}
\end{center}
\end{figure*}

Table~\ref{table:emissivities} provides
the $\epsilon_{\nu}^i$ values for each field/component/frequency.  In
order to visualize the results, Fig.~\ref{fig:tt_plot_1} gives scatter plots together with the line of slope
$\epsilon_{857}^i$ for each component, field by field.  Specifically,
for each HI component $i$ we plotted $I_\nu - \sum_{j \ne i} \epsilon_j
N_{\rm HI}^j -Z_\nu$ as a function of $N_{\rm HI}^i$.

\subsection{Statistics of the residual}

Figs~\ref{fig:N1_dust} to \ref{fig:NEP_dust} show the \IRAS\ and
\Planck\ maps, together with the \hi\ correlated emission and the
residual maps $R_\nu$ for all our fields.
For the six faintest fields 
(N1, SP, BOOTES, AG, MC, and G86 -- see Figs.~\ref{fig:N1_dust}-\ref{fig:AG_dust}),
the structure in the residuals is, even by visual inspection, clearly spatially correlated
between frequencies, especially in the \Planck\ bands. 
It is dominated by small scale structures 
with equally negative and positive brightness fluctuations.
The structure of the residual for brighter fields 
is also clearly correlated between frequencies but in these cases the residual is mostly positive
(i.e., they are excesses with respect to the \hi). These residuals also show larger coherent structures
than in the fainter fields.

\begin{figure}
\begin{center}
\includegraphics[width=\linewidth, draft=false, angle=0]{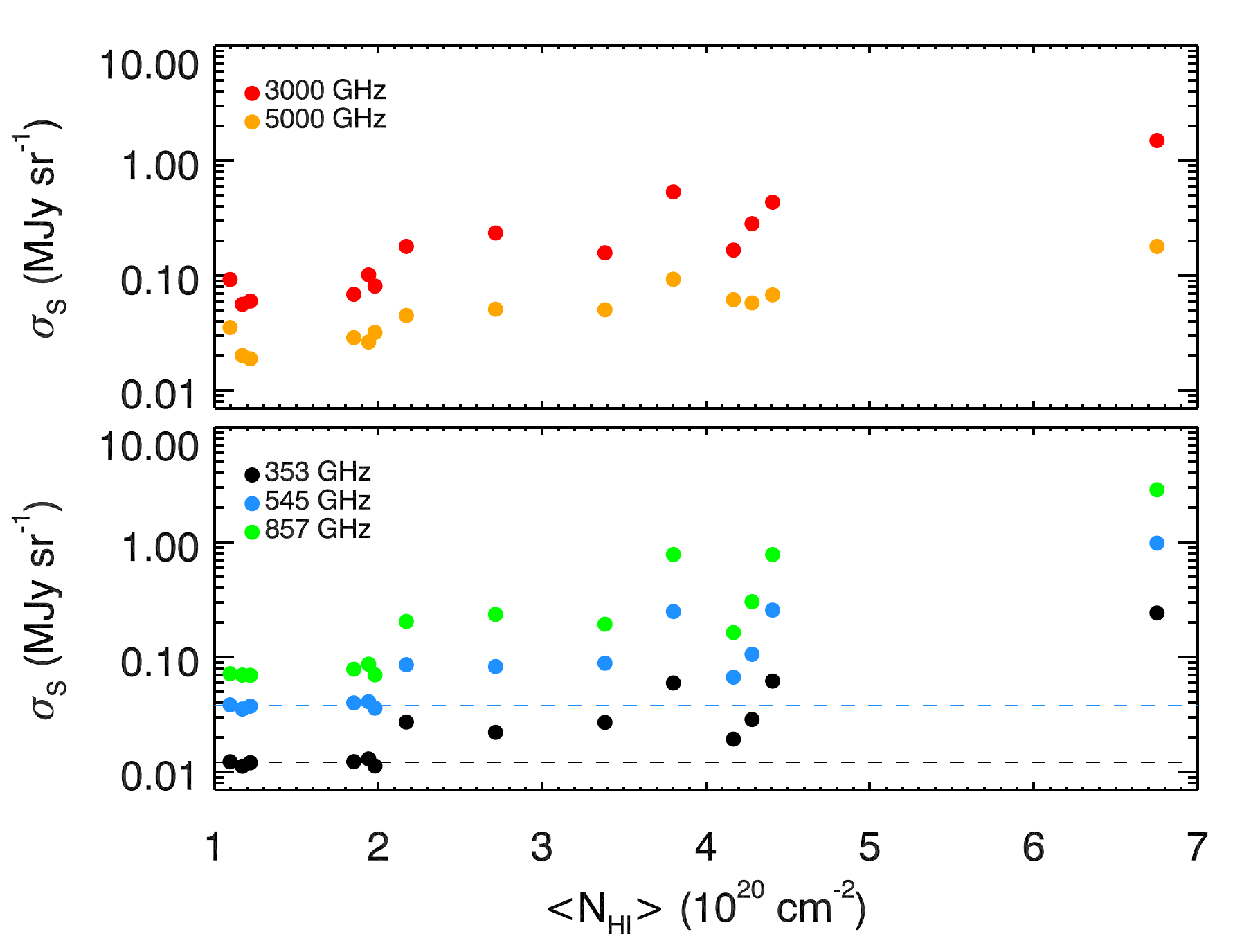}
\caption{\label{fig:pdf_residu} Value of $\sigma_S$ obtained from the
  standard deviation $\sigma_R$ of the residual maps $R_\nu(x,y)$ from
  which contributions from the \IRAS\ or \Planck\ and the GBT noise were
  removed quadratically (see Eq.~\ref{eq:sigma}), plotted as a function
  of the average HI column density of each field (sum of LVC, IVC and HVC).  
The dashed line is
  the average of $\sigma_{S}$ for fields with $N_{\rm HI} < 2 \times
  10^{20}$\,cm$^{-2}$ (see Table~\ref{table:sigma}).  }
\end{center}
\end{figure}

The rms (about the mean) of the residual $R$ can be approximated as:
\begin{equation}
\label{eq:sigma}
\sigma_{R} = \sqrt{ \sigma_{S}^2 + (\sigma_{\rm noise}^{\rm dust})^2 +
\sum_{i=1}^3 (\epsilon_\nu^i \delta N_{\rm HI}^{i})^2 },
\end{equation}
where noise in the \IRAS\ or \Planck\ data and that induced by
uncertainties in \nh\ are explicitly accounted for and $\sigma_{S}$
includes all other contributions not in the model, including CIB
anisotropies and dust emission associated with molecular gas.  After
quadratic subtraction 
Fig.~\ref{fig:pdf_residu} shows the value of $\sigma_{S}$ at each
frequency, as a function of the average \nh\ density for each field.
For fields with a median column density lower than $2\times
10^{20}$\,cm$^{-2}$, the PDFs of the residual emission all have skewness
and kurtosis values compatible with a Gaussian distribution. Furthermore
the width of these PDFs shows very small scatter from field to field
(see Fig.~\ref{fig:pdf_residu}). This is another indication that, for
such diffuse fields, the model is a good description of the data; i.e.,
the Galactic dust emission is largely dominated by the \hi\ components,
with very limited spatial variations of the emissivity across a given
field.

The dashed line gives the average level of $\sigma_{S}$ for those fields
with $N_{\rm HI}< 2 \times 10^{20}$\,cm$^{-2}$.  These quantities, together
with the average \Planck\ or \IRAS\ and GBT noise contributions to
$\sigma_R$, are summarized in Table~\ref{table:sigma}.  It is clear from
Table~\ref{table:sigma} and Fig.~\ref{fig:pdf_residu} that even for
faint fields $\sigma_{S}$ is the main contributor to the rms of the
residual, and therefore the main contributor to the dispersion in the
dust--gas correlation diagrams in these fields.  This is in accordance
with the findings of \cite{planck2011-6.6} who concluded that the CIB
fluctuations dominate $\sigma_{S}$ at 353\,GHz and higher frequencies.

At higher column densities, the PDF of the residual shows positive
skewness and an rms that increases with \nh, significantly exceeding the
level of CIB anisotropies.  These aspects of the PDF indicate one or
more extra components that contribute to the dust emission and are not
taken into account in our model.  
Contributions to the residual that grow with \nh\ are compatible with the presence of a molecular gas component.
This could also come from spatial variations of the dust emissivity for given
\hi\ components or from inadequately-separated \hi\ components.

\begin{table}
\begin{center}
\begin{tabular}{cccc}\toprule
$\nu$ & $\sigma_{\rm S}$ & $\sigma_{\rm noise}^{\rm dust}$ & $\sigma_{\rm noise}^{\rm HI}$\\ 
(GHz) & (MJy\,sr$^{-1}$) & (MJy\,sr$^{-1}$) & (MJy\,sr$^{-1}$)\\ \midrule
353 & $0.0120 \pm 0.0006$ & $0.0060 \pm 0.0012$ & $0.0014 \pm 0.0006$ \\
545 & $0.038 \pm 0.002$ & $0.0096 \pm 0.0018$ & $0.006 \pm 0.003$ \\
857 & $0.074 \pm 0.006$ & $0.0097 \pm 0.0019$ & $0.016 \pm 0.007$ \\
3000 & $0.077 \pm 0.017$ & $0.028 \pm 0.004$ & $0.022 \pm 0.005$ \\
5000 & $0.027 \pm 0.006$ & $0.0146 \pm 0.0016$ & $0.0053 \pm 0.0010$ \\ \bottomrule
\end{tabular}
\end{center}
\caption{\label{table:sigma} Average of the standard deviations of the
  sky residual $\sigma_{\rm S}$ for the six fields with $\langle N_{\rm HI}
  \rangle$ lower than $2\times 10^{20}$\,cm$^{-2}$ (AG, BOOTES, G86, MC,
  N1, and SP).  $\sigma_{\rm noise}^{\rm dust}$ is the average level of noise
  rms in the \IRAS\ and \Planck\ maps at the GBT resolution (i.e., noise level in maps
  convolved at 9.4\arcm) computed for
  the 14 fields. $\sigma_{\rm noise}^{\rm HI}$ gives the average GBT noise 
  converted to MJy\,sr$^{-1}$ (see Eq.~\ref{eq:sigma}), for
  the 14 fields.  All uncertainties are the 1$\sigma$ value of each sample.}
\end{table}

\subsection{Monte-Carlo method to determine emissivity uncertainties}

\label{sec:monte_carlo}

The statistical uncertainties estimated for the emissivities by the
least-squares fit method are accurate only for the case of white
Gaussian noise in $I_\nu$ and (sufficiently) low noise in the $N_{\rm HI}$
components.  The noise in $I_\nu$ includes \IRAS\ or
\Planck\ instrumental noise, CIB anisotropies, and various interstellar
contaminants, and for the most diffuse fields in our sample, its PDF is close to Gaussian.
However, its power spectrum is certainly not white.  First, at the
angular scales of our observations (from 9\arcm\ to a few degrees), the
power spectrum of the CIB anisotropies is $P(k) \propto k^{-1}$
\citep{planck2011-6.6}.  Second, the spatial variation of the coverage
and the convolution to the GBT resolution both introduce spatial
structure in the noise that modifies its power spectrum.
The addition of all those noise sources produces a net noise term $n_\nu$ on $I_\nu$ that is not white.
Because of the random chance correlation of $n_\nu$  with the \nh\ components, the uncertainties on the
parameters estimated from the least-squares are systematically underestimated (the least-squares fit is
not optimal).
In addition, the \nh\ maps are not noise free; significant noise on the
independent variable in a least-squares fit produces a systematic bias
in the solution.
Furthermore, imperfect opacity correction of the \hi\ spectra, the
presence of molecular gas, and spatial variations of the dust properties
will also produce non-random fluctuations in the residual map.  For all
of these reasons, an analysis of Monte-Carlo simulations is required for
proper estimation of the uncertainties and biases in the
$\epsilon_\nu^i$.

To generate simulations for each field, we adopted the \nh\ maps
obtained from the 21-cm observations as templates of the dust emission.
We built dust maps $I_\nu^\prime$ for each frequency $\nu$ by adding up
these \nh\ maps multiplied by their respective estimated emissivities
$\epsilon_\nu^i$ (just as in computing the residual),
to which we added realizations of the \IRAS\ or \Planck\ noise $n_\nu$
appropriate to the field\footnote{We assumed white noise maps for both
  \IRAS\ and \Planck, each divided by the square root of their coverage
  map.}  at a level compatible with Table~\ref{table:sigma} once convolved at GBT
resolution, and of the sky residual $a_\nu$ with a $k^{-1}$ power
spectrum at a level to reproduce $\sigma_S$ within the mask\footnote{For
  the faintest fields this will produce a map with the statistical
  properties of the CIB anisotropies. For brighter fields, where the
  residual also includes interstellar contributions, we assumed it also
  follows a $k^{-1}$ power spectrum, but with a greater normalisation
  (i.e., $\sigma_S$) to take into account these effects.}:
\begin{equation}
I_\nu^\prime = \sum_{i=1}^3 \epsilon_\nu^i N_{\rm HI}^{i} + a_\nu + n_\nu.
\end{equation}
Before fitting these simulated data we added white noise to each
\nh\ component at the level estimated for the GBT data
(Table~\ref{table:mainHI}).
We then carried out the least-squares fit, using the mask already
estimated for each field.
From a thousand such simulations for each field, we obtained the
statistics of the recovered $\epsilon^{i\prime}_\nu$ so that we could
determine if our original fit (fed into the simulation) was biased and
could compare the dispersion of the parameters to the statistical
uncertainties returned by the least-squares fitting procedure.

Table~\ref{table:monte_carlo} summarizes the results of the Monte-Carlo
simulations.  Due to random correlation between $n_\nu$ and
the \nh\ components, we find the Monte-Carlo-derived uncertainty in
$\epsilon_\nu^i$ is several times higher than the analytically-derived
uncertainty.  The results are reported in the table in terms of
$\phi_\nu$, the ratio of $\sigma(\epsilon^{i\prime}_\nu)$, the standard
deviation of the emissivities recovered in the simulations, to
$\delta\epsilon_\nu^i$, the standard deviation expected from the linear
fit performed by {\em Regress}.  Therefore, in what follows we use the
values of $\sigma(\epsilon^{i\prime}_\nu)$ as the uncertainties in
$\epsilon_\nu^i$ (see Table~\ref{table:emissivities}).  

Finally, these simulations make it possible for us to estimate any bias
in $\epsilon_\nu^i$ which could arise from the noise in \nh.
Table~\ref{table:monte_carlo} also provides the bias in \%: $b_\nu= 100
(\langle \epsilon^{\prime}_\nu \rangle - \epsilon_\nu) / \epsilon_\nu$.
Except for some (undetected) HVCs, the bias is only at the few percent level; in
what follows we made no correction for it.

\begin{sidewaystable*}
\centering
\vspace*{+18cm}
\begin{tabular}{llccccccccccccccc} \toprule
Field & HI & $\delta_{353}$ & $\delta_{545}$ & $\delta_{857}$ & $\delta_{3000}$ & $\delta_{5000}$ 
& $\phi_{353}$ & $\phi_{545}$ & $\phi_{857}$ & $\phi_{3000}$ & $\phi_{5000}$ 
& $b_{353}$ & $b_{545}$ & $b_{857}$ & $b_{3000}$ & $b_{5000}$  \\ \midrule
AG & LVC & $0.98$ & $0.98$ & $0.99$ & $0.98$ & $1.03$ & $6.35$ & $6.75$ & $6.57$ & $5.91$ & $6.41$ & $-3.05$ & $-3.99$ & $-4.12$ & $-4.49$ & $-4.46$\\
 & IVC & $0.99$ & $0.99$ & $1.01$ & $0.99$ & $1.04$ & $7.02$ & $6.88$ & $7.69$ & $6.61$ & $7.06$ & $1.67$ & $2.12$ & $1.75$ & $0.74$ & $0.37$\\
 & HVC & $1.00$ & $1.00$ & $1.02$ & $1.00$ & $1.05$ & $7.66$ & $7.78$ & $7.98$ & $7.60$ & $7.78$ & $-0.05$ & $5.81$ & $7.94$ & $7.16$ & $3.47$\\
BOOTES & LVC & $0.99$ & $0.99$ & $1.01$ & $0.96$ & $0.93$ & $10.39$ & $11.63$ & $12.41$ & $11.27$ & $10.87$ & $-1.50$ & $-1.72$ & $-1.65$ & $-1.69$ & $-1.40$\\
 & IVC & $1.00$ & $1.00$ & $1.01$ & $0.97$ & $0.93$ & $12.37$ & $13.14$ & $13.64$ & $13.24$ & $11.86$ & $-0.41$ & $-0.05$ & $-0.56$ & $-0.79$ & $-1.19$\\
 & HVC & $0.86$ & $0.86$ & $0.87$ & $0.83$ & $0.80$ & $4.21$ & $4.56$ & $4.75$ & $4.47$ & $4.08$ & $-22.91$ & $-18.63$ & $-20.17$ & $-30.00$ & $-29.46$\\
DRACO & LVC & $1.00$ & $0.98$ & $1.06$ & $1.01$ & $1.02$ & $7.14$ & $7.33$ & $7.64$ & $7.22$ & $7.45$ & $-0.24$ & $-1.26$ & $-0.44$ & $-1.89$ & $-2.74$\\
 & IVC & $1.00$ & $0.99$ & $1.06$ & $1.02$ & $1.03$ & $8.16$ & $7.90$ & $8.91$ & $8.13$ & $7.97$ & $-0.14$ & $-0.47$ & $-0.23$ & $-0.53$ & $-0.47$\\
 & HVC & $0.99$ & $0.97$ & $1.05$ & $1.00$ & $1.01$ & $5.59$ & $5.41$ & $6.01$ & $5.59$ & $5.53$ & $-0.34$ & $-2.68$ & $-3.08$ & $-5.46$ & $-5.56$\\
G86 & LVC & $0.99$ & $0.99$ & $1.02$ & $0.98$ & $1.01$ & $8.46$ & $8.80$ & $8.85$ & $8.92$ & $8.22$ & $0.97$ & $0.35$ & $0.44$ & $0.04$ & $0.36$\\
 & IVC & $1.01$ & $1.01$ & $1.03$ & $1.00$ & $1.02$ & $7.86$ & $8.24$ & $8.39$ & $8.27$ & $7.84$ & $0.18$ & $0.12$ & $-0.24$ & $-0.05$ & $-0.01$\\
 & HVC & $0.93$ & $0.93$ & $0.95$ & $0.92$ & $0.94$ & $5.57$ & $5.65$ & $5.96$ & $5.61$ & $5.53$ & $-14.85$ & $-15.52$ & $-15.52$ & $-16.60$ & $-11.00$\\
MC & LVC & $0.98$ & $0.99$ & $1.00$ & $0.99$ & $0.98$ & $7.61$ & $8.91$ & $9.46$ & $8.88$ & $8.54$ & $-3.36$ & $-4.07$ & $-3.31$ & $-3.51$ & $-1.98$\\
 & IVC & $0.99$ & $1.00$ & $1.00$ & $1.00$ & $0.98$ & $6.18$ & $6.88$ & $7.35$ & $6.96$ & $6.73$ & $0.19$ & $-2.80$ & $-1.74$ & $-2.17$ & $-1.97$\\
 & HVC & $0.99$ & $1.00$ & $1.01$ & $1.00$ & $0.99$ & $6.62$ & $7.42$ & $8.17$ & $7.57$ & $7.60$ & $-4.35$ & $-2.57$ & $-1.38$ & $-3.01$ & $-0.14$\\
N1 & LVC & $0.99$ & $0.99$ & $0.99$ & $1.00$ & $1.02$ & $7.22$ & $7.50$ & $7.52$ & $7.04$ & $6.77$ & $-2.10$ & $-1.98$ & $-2.25$ & $-1.90$ & $-2.02$\\
 & IVC & $0.99$ & $1.00$ & $1.00$ & $1.00$ & $1.03$ & $7.62$ & $7.99$ & $8.07$ & $7.28$ & $7.46$ & $-1.56$ & $-0.95$ & $-1.67$ & $-1.47$ & $-1.54$\\
 & HVC & $0.99$ & $1.00$ & $1.00$ & $1.00$ & $1.03$ & $7.87$ & $8.24$ & $8.25$ & $7.91$ & $7.51$ & $-5.16$ & $-2.03$ & $-8.00$ & $6.82$ & $10.14$\\
NEP & LVC & $0.99$ & $0.98$ & $0.99$ & $0.91$ & $0.75$ & $10.11$ & $10.21$ & $10.23$ & $9.14$ & $7.51$ & $-0.25$ & $-0.20$ & $-0.23$ & $-0.30$ & $-0.15$\\
 & IVC & $0.99$ & $0.99$ & $1.00$ & $0.92$ & $0.76$ & $11.86$ & $12.00$ & $12.52$ & $11.46$ & $9.10$ & $-0.07$ & $-0.25$ & $-0.10$ & $-0.09$ & $-0.03$\\
 & HVC & $0.97$ & $0.97$ & $0.98$ & $0.90$ & $0.74$ & $9.83$ & $10.33$ & $9.76$ & $9.39$ & $7.63$ & $-6.53$ & $-4.58$ & $-3.86$ & $-3.82$ & $-11.00$\\
POL & LVC & $0.98$ & $0.98$ & $0.99$ & $0.98$ & $0.99$ & $6.62$ & $7.09$ & $7.03$ & $7.15$ & $6.78$ & $-0.02$ & $-0.20$ & $-0.23$ & $-0.04$ & $0.20$\\
 & IVC & $0.96$ & $0.97$ & $0.98$ & $0.96$ & $0.98$ & $6.06$ & $6.43$ & $6.52$ & $6.46$ & $6.32$ & $-2.20$ & $-3.05$ & $-1.61$ & $-3.05$ & $-3.38$\\
 & HVC & -- & -- & -- & -- & -- & -- & -- & -- & -- & -- & -- & -- & -- & -- & --\\
POLNOR & LVC & $1.01$ & $1.01$ & $1.02$ & $1.01$ & $1.03$ & $8.97$ & $9.28$ & $9.61$ & $9.11$ & $8.96$ & $-0.17$ & $-0.08$ & $-0.13$ & $-0.21$ & $-0.14$\\
 & IVC & $1.00$ & $1.01$ & $1.01$ & $1.00$ & $1.02$ & $8.14$ & $8.04$ & $7.96$ & $8.34$ & $7.75$ & $-3.14$ & $-1.36$ & $-2.17$ & $-1.66$ & $-1.69$\\
 & HVC & -- & -- & -- & -- & -- & -- & -- & -- & -- & -- & -- & -- & -- & -- & --\\
SP & LVC & $0.98$ & $0.99$ & $1.00$ & $0.99$ & $1.03$ & $8.17$ & $8.63$ & $8.82$ & $7.72$ & $7.65$ & $-2.62$ & $-2.43$ & $-2.35$ & $-2.84$ & $-3.81$\\
 & IVC & $0.98$ & $0.99$ & $1.00$ & $0.99$ & $1.03$ & $7.83$ & $8.55$ & $8.78$ & $7.66$ & $7.74$ & $-3.27$ & $-2.17$ & $-3.25$ & $-2.42$ & $-2.77$\\
 & HVC & $0.98$ & $0.99$ & $1.00$ & $0.99$ & $1.03$ & $7.84$ & $8.46$ & $8.61$ & $7.64$ & $7.27$ & $-16.46$ & $-8.71$ & $-11.32$ & $-6.00$ & $-8.75$\\
SPC & LVC & $1.01$ & $1.01$ & $1.02$ & $1.02$ & $1.02$ & $10.99$ & $11.46$ & $11.61$ & $11.16$ & $11.33$ & $-0.24$ & $-0.05$ & $-0.48$ & $-0.20$ & $-0.33$\\
 & IVC & $0.99$ & $1.00$ & $1.01$ & $1.01$ & $1.01$ & $7.74$ & $7.93$ & $8.38$ & $7.84$ & $8.01$ & $-5.82$ & $-3.23$ & $-1.95$ & $-2.53$ & $-3.36$\\
 & HVC & $1.00$ & $1.01$ & $1.02$ & $1.02$ & $1.02$ & $9.03$ & $9.40$ & $9.49$ & $9.39$ & $8.94$ & $-1.92$ & $-2.27$ & $-3.50$ & $0.48$ & $5.22$\\
SPIDER & LVC & $0.99$ & $1.00$ & $1.02$ & $0.97$ & $1.03$ & $10.59$ & $10.55$ & $11.45$ & $10.61$ & $10.81$ & $0.09$ & $-0.09$ & $0.03$ & $0.06$ & $0.10$\\
 & IVC & $0.98$ & $0.99$ & $1.01$ & $0.96$ & $1.02$ & $7.94$ & $8.49$ & $8.80$ & $8.23$ & $8.37$ & $0.19$ & $-0.40$ & $0.17$ & $-0.41$ & $-0.94$\\
 & HVC & $0.93$ & $0.94$ & $0.96$ & $0.91$ & $0.96$ & $7.57$ & $7.87$ & $7.98$ & $7.69$ & $7.51$ & $-12.02$ & $-10.42$ & $-12.09$ & $-10.06$ & $-8.22$\\
UMA & LVC & $1.01$ & $1.02$ & $1.03$ & $1.01$ & $1.01$ & $8.48$ & $8.51$ & $8.92$ & $8.47$ & $8.35$ & $-0.21$ & $-0.28$ & $-0.12$ & $0.03$ & $-0.02$\\
 & IVC & $1.00$ & $1.02$ & $1.03$ & $1.00$ & $1.00$ & $7.69$ & $8.24$ & $8.29$ & $7.98$ & $7.43$ & $-1.33$ & $-0.86$ & $-1.62$ & $-2.00$ & $-1.63$\\
 & HVC & $1.00$ & $1.02$ & $1.03$ & $1.00$ & $1.01$ & $8.05$ & $8.57$ & $8.49$ & $8.26$ & $7.58$ & $0.23$ & $-0.05$ & $0.21$ & $0.95$ & $0.04$\\
UMAEAST & LVC & $1.00$ & $1.01$ & $1.01$ & $0.99$ & $1.02$ & $6.43$ & $6.73$ & $6.95$ & $6.28$ & $6.32$ & $-0.42$ & $-0.13$ & $-0.21$ & $-0.24$ & $0.11$\\
 & IVC & $1.00$ & $1.01$ & $1.01$ & $0.99$ & $1.02$ & $6.61$ & $7.19$ & $7.23$ & $6.77$ & $6.76$ & $-0.35$ & $-0.89$ & $-0.73$ & $-0.36$ & $-0.71$\\
 & HVC & $1.00$ & $1.01$ & $1.01$ & $0.99$ & $1.02$ & $7.64$ & $7.97$ & $8.46$ & $7.73$ & $7.47$ & $-0.72$ & $-2.69$ & $0.79$ & $-3.52$ & $-0.79$\\
 \bottomrule
\end{tabular}
\caption{\label{table:monte_carlo} Results from the Monte-Carlo
  simulations.  $\delta_\nu = \langle \delta\epsilon^{\prime}_\nu
  \rangle / \delta\epsilon_\nu$ is the ratio of the average statistical
  uncertainty of the emissivities estimated in the fits of the simulated
  data to the statistical uncertainty estimated in the fit of the actual
  data.  A second quantity, with the same denominator, $\phi_\nu =
  \sigma(\epsilon^{\prime}_\nu)/\delta\epsilon_\nu $ is the ratio of the
  standard deviation of the emissivities recovered in the simulations to
  the statistical uncertainty estimated in the fit of the actual data.
  The percentage bias in emissivity $b_\nu= 100 (\langle
  \epsilon^{\prime}_\nu \rangle - \epsilon_\nu) / \epsilon_\nu$.}
\end{sidewaystable*}

\section{Results}

\label{sec:results}

\subsection{\Planck\ and \IRAS\ emissivities}

The present study extends to smaller scales, and to the IVCs and HVCs,
the earlier work done with the FIRAS \cite[]{boulanger1996} or \IRAS\
\cite[]{boulanger1988,reach1998a} data on the dust emission of the
diffuse ISM. It also extends to a much larger sample a similar analysis
done on a diffuse $3^\circ \times 3^\circ$ region at high Galactic
latitude using \IRAS\ and \Spitzer\ data \cite[]{miville-deschenes2005}.
The IR/submm-\hi\ correlation analysis allows us to determine
empirically the spectral dependence of the ratio between the dust
emission and the gas column density.  In addition, the combination of
\Planck, \IRAS, and \nh\ data can be used to trace one elusive component
of the diffuse interstellar medium: the diffuse H$_2$ gas
(Sect.~\ref{sec:transition}).

\begin{figure*}
\begin{center}
\includegraphics[width=\linewidth, draft=false, angle=0]{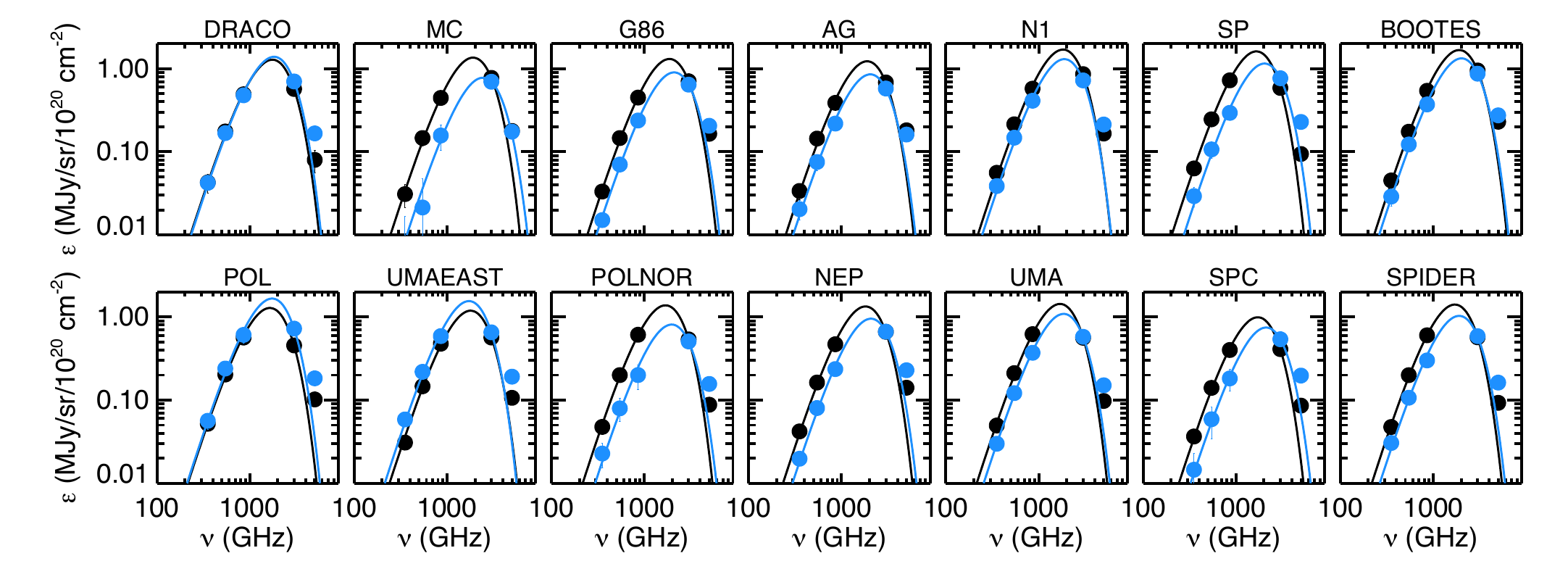}
\caption{\label{fig:all_seds} SEDs from the emissivities of LVC (black) and IVC (blue)
  components for all the fields in our sample. 
  For each \hi\ component in each cloud, 
  the solid line is the modified black body fit
  using 353, 545, 857, and 3000\,GHz data.}
\end{center}
\end{figure*}

As seen in Fig.~\ref{fig:tt_plot_1} there is a clear
correlation between the \IRAS\ or \Planck\ data and \nh\ in all
fields but, as previously seen with \COBE\ and \IRAS\ data, there are
increasing excesses of dust emission with increasing \nh.  The estimated
emissivities for each field/component/frequency are compiled in
Table~\ref{table:emissivities} and shown in Fig.~\ref{fig:all_seds}.  
We note that dust associated with the
LVC and IVC components is detected in each field and at each frequency, 
unlike for HVCs for which we do not report any significant detection.
All HVC emissivities are indeed below 3$\sigma$. 
The results on HVCs are discussed further in section~\ref{sec:dustHVC}.
In the following we analyse what can be drawn from the emissivities for the LVCs and IVCs.

\subsection{Comparison with FIRAS data}

\begin{figure}
\begin{center}
\includegraphics[width=\linewidth, draft=false, angle=0]{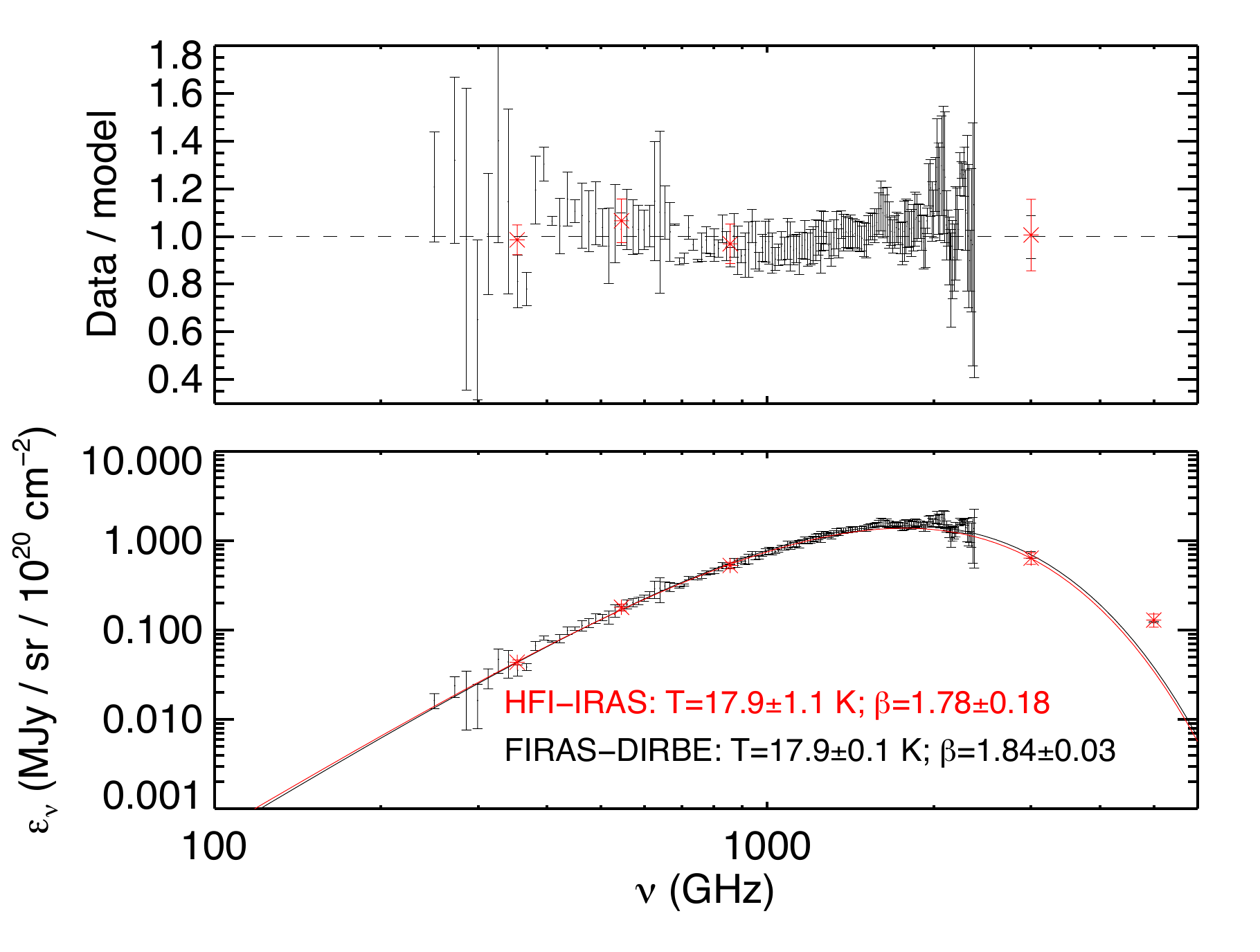}
\caption{\label{fig:firas_spectrum} Bottom panel: Black points
  show the FIRAS spectrum of the diffuse ISM \citep{compiegne2011}.  The
  red points are the average of the \IRAS\ or \Planck\ emissivities for
  the local components of all our fields; the uncertainty is the
  variance of the values divided by $\sqrt{N}$.
The solid curves are modified black body fits to each spectrum -- the
5000\,GHz point was excluded from the fit as it is dominated by
non-equilibrium dust emission.  Top panel: Same as bottom panel
but each data set is divided by its modified black body fit.  }
\end{center}
\end{figure}

The FIRAS data provide a reference for the dust emission spectrum of the
diffuse ISM.  The average FIRAS SED of the high-Galactic latitude sky 
used by \cite{compiegne2011} to set the diffuse ISM dust properties in the DustEM model is shown in
Fig.~\ref{fig:firas_spectrum}.  Also shown in this figure are the
current results (red symbols) for the average of the emissivities for
the LVCs of our sample at 353, 545, 857, 3000, and
5000\,GHz. The \IRAS\ and \Planck\ data points are found to be fully
compatible with the diffuse ISM FIRAS spectrum, showing that our sample
is representative of the diffuse dust emission at high Galactic
latitudes.  We have fit model parameters to both data sets independently
using a modified black body function:
\begin{equation}
\label{eq:bb}
\epsilon_\nu = I_\nu/N_{\rm HI} = \kappa_0 (\nu/\nu_0)^{\,\beta} \mu m_{\rm H}
B_\nu(T),
\end{equation}
where $B_\nu(T)$ is the Planck function, $m_{\rm H}$ the mass of hydrogen, $\mu$ the mean molecular weight
and $\kappa_0$ is the opacity of the dust--gas mixture at some fiducial frequency $\nu_0$.
The higher-frequency (60\,$\mu$m) \IRAS\ datum is not used in the fit
due to contamination by non-equilibrium emission from
stochastically-heated smaller grains \citep{compiegne2011}.
The top panel of Fig.~\ref{fig:firas_spectrum} shows the data divided by
the model to better display the quality of the fit. The values found for
$T$ and $\beta$ with the two data sets are in close accord.  This
analysis shows that the local ISM SED can be fit well with $T=17.9$\,K
and $\beta=1.8$.

\subsection{The spectral energy distribution}

Fig.~\ref{fig:all_seds} shows the dust SED for each field and
\hi\ component separately, with the error bars computed using the
Monte-Carlo simulations.  
{We first note that
all the LVC and IVC SEDs are at a similar level showing that the power emitted per H is comparable 
between fields. 
Second, in many cases, the SED of IVCs peaks at a higher frequency than for the LVCs.
That could be caused by a higher temperature or a larger abundance of smaller grains in the IVCs.
 }

Like for the FIRAS comparison, each SED was fit using a modified black body function (solid lines).  
It has been shown by several studies
\cite[]{dupac2003a,desert2008,shetty2009,veneziani2010} how difficult it
is to estimate separately $T$ and $\beta$ for such an SED fit. These two
parameters are significantly degenerate; their estimate depends greatly
on the accuracy of the determination of the error bars on the SED data
points {and on the correlation of the errors between frequencies.
The estimate of $T$ and $\beta$ also depends} on the spectral range used to make the fit; for
a typical $T=18$\,K dust emission spectrum, the Rayleigh-Jeans range of
the black body curve, where $\beta$ can be well estimated, corresponds
to $\nu < 375$\,GHz ($\lambda > 800$\,$\mu$m). 

Because of the above caveats relating to simultaneous fits of $T$ and $\beta$, 
and because the FIRAS spectrum of the diffuse ISM is compatible with $\beta=1.8$ (in this case
the large number of data points and the broad frequency coverage over
the peak of the SED give more confidence in the value of $\beta$
obtained), we have carried out SED fitting assuming a fixed
$\beta=1.8$. This provides a direct way to compare not only 
with the FIRAS spectrum and
the DustEM model, but also with similar analyses with \Planck\ data on
molecular clouds and in the Galactic plane
\citep{planck2011-7.0,planck2011-7.3,planck2011-7.13} that used the same
convention.

\subsection{Dust properties}

{The modified black-body fit (Eq.~\ref{eq:bb}) provides information on the properties, such as $T$,
of the dust in each \hi\ component.
A useful quantity used below is the emission cross-section of
the interstellar material per H:
\begin{equation}
\sigma_{\rm e} \equiv  \kappa_0 (\nu/\nu_0)^\beta \mu m_{\rm H} =\tau/N_{\rm HI}.
\end{equation}
It is simply the prefactor to the Planck function in Eq.~\ref{eq:bb}.
In what follows we adopt $\nu_0 = 1200$\,GHz ($\lambda_0 = 250$\,$\mu$m)
to compare directly with the value of $\sigma_{\rm e}(1200) = \tau/N_{\rm HI}$ 
at $250\,\mu$m given by \cite{boulanger1996}.

A key quantity is the luminosity per H atom $L$ 
(in W/H) emitted by dust grains (equal to the absorbed power)
computed by integrating the SED over $\nu$: 
\begin{equation}
L = \int 4 \pi \, \kappa_0 (\nu/\nu_0)^\beta \mu m_{\rm H}  B_\nu(T) \, d \nu.
\end{equation}

Complementing the actual SEDs in Fig.~\ref{fig:all_seds},
Fig.~\ref{fig:opacity250} shows the derived values of $\sigma_{\rm e}(1200)$, $T$ and $L$ plotted against 
the velocity of each \hi\ component (LVC and IVC) in our sample. 

\begin{figure}
\begin{center}
\includegraphics[width=\linewidth, draft=false, angle=0]{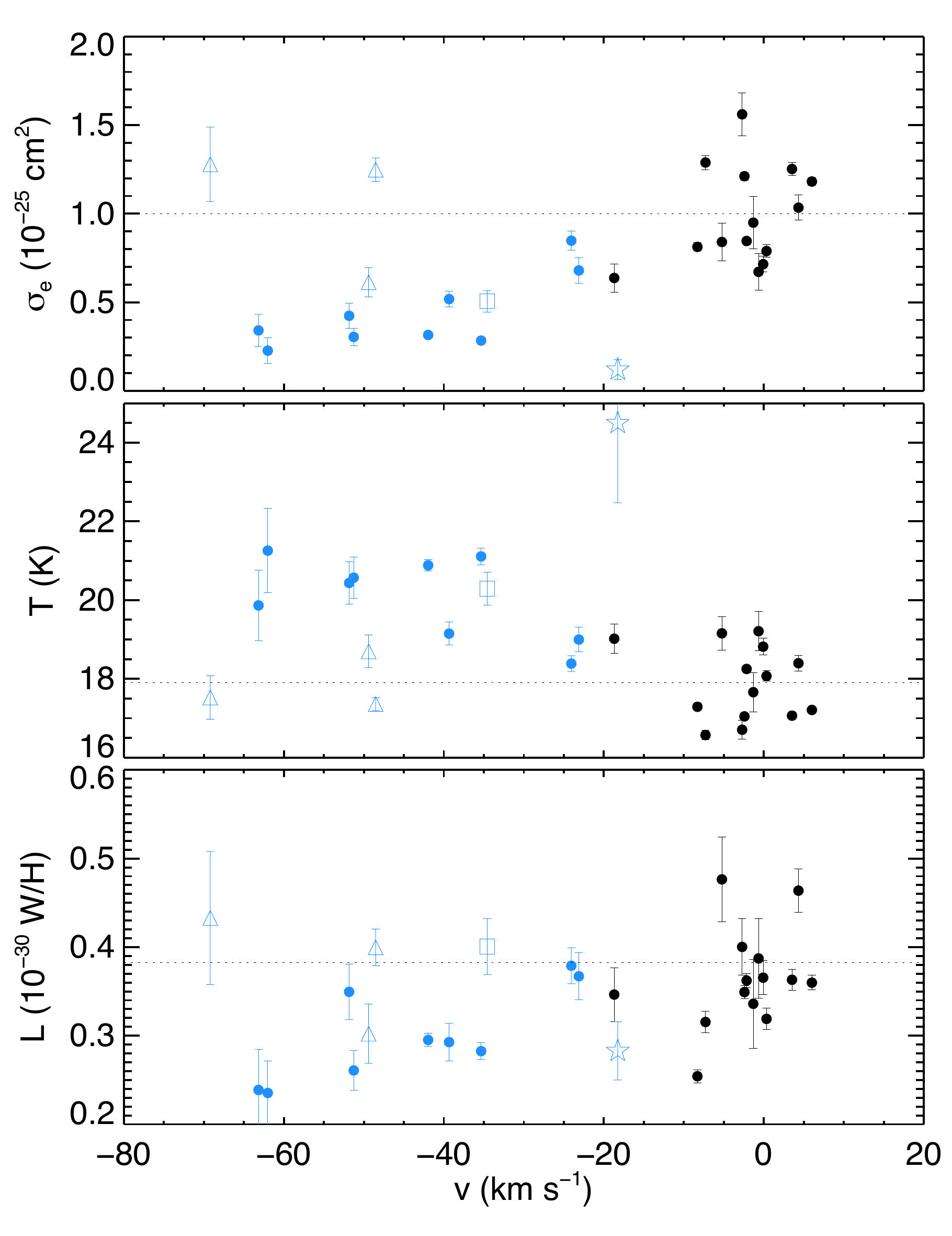}
\caption{\label{fig:opacity250} Values of $\sigma_{\rm e}$ at 1200\,GHz (250\,$\mu$m), $T$ and $L$ versus the
 average velocity of each \hi\ component. These dust parameters were estimated 
 from the SED fit over the range 353 to 3000\,GHz to a modified black body with $\beta=1.8$. 
Black and blue symbols are for LVC and IVC, respectively.  
The different symbols in blue represent IVCs associated with
specific complexes: IV Arch (dot), IV/LLIV Arch (triangle), Complex K (square) and PP Arch (star).
In each panel the dotted line
represent the value for the diffuse ISM obtained with the high-latitude FIRAS spectrum 
(see Fig.~\ref{fig:firas_spectrum}).
  Error bars are given for each data point, some being smaller than the symbol size.}
\end{center}
\end{figure}

The average emission cross-section for the 14 LVC components of our
sample is $1.0 \pm 0.3 \times 10^{-25}$\,cm$^2$, 
in good agreement with the value of $1\times 10^{-25} \, {\rm cm}^2$ obtained by
\cite{boulanger1996}.  The scatter of $\sigma_{\rm e}(1200)$ for the LVCs
(30\%) is not the result of errors.

The emission cross-section for the IVC components is different, often 50\% lower 
compared to the LVCs.
There appear to be differences among the IVCs too, perhaps related to the fact that they
belong to different IVC complexes.
All fields in our sample overlap with the Intermediate Velocity (IV) Arch, 
except MC which is in the southern Galactic sky and belongs to the PP Arch and BOOTES which is part of Complex K
\citep{kuntz1996}.
The North Celestial Loop also overlaps spatially 
with the Low-Latitude Intermediate Velocity (LLIV) Arch \citep{kuntz1996}, an \hi\ feature 
at slightly less negative velocity ($\sim - 50$\,\kms) than the IV~Arch ($\sim -75$\,\kms). 
The fields UMA and UMAEAST
contain clumps identified by \citet{kuntz1996} as being part of the LLIV~Arch 
(specifically LLIV1, LLIV2 and LLIV3).
The field POL also contains emission that can be attributed to the LLIV~Arch.
These different complexes are identified with separate symbols in Fig.~\ref{fig:opacity250}.
The outliers are POL, UMAEAST (high) and MC (low). Excluding these 
$\sigma_e = 0.5 \pm 0.2 \times 10^{-25}$\,cm$^2$ for the rest.

The values of T for the LVCs ($T = 17.9 \pm 0.9$\,K) are in accordance with that obtained 
from the FIRAS spectrum at high Galactic latitude ($T = 17.9 \pm 0.1$\,K; Fig.~\ref{fig:firas_spectrum}) 
and also close to the 17.5\,K found by \cite{boulanger1996} assuming $\beta = 2$.
Like for $\sigma_e$, a systematic difference in $T$ is found 
between LVCs and IVCs, even though it is less statistically significant. On average
the IVC group of clouds\footnote{excluding POL, UMAEAST and MC} has $T=20.0\pm1.0$\,K, a value
greater than in the local ISM at the $2.1\,\sigma$ level. Note the different IVC complexes as well.

Regarding $L$
the striking result here is the small variation observed over all the fields 
and \hi\ components. Combined together, the LVCs and IVCs have 
$<L> = 3.4 \pm 0.6 \times 10^{-31}$\,W/H, representing a variation of only 20\% over all 
clouds. 
The fact that $L$ is rather constant over all fields and \hi\ components can also be appreciated 
in Fig.~\ref{fig:all_seds}, where all SEDs are at about the same level.
The small variation of $L$ is surprising as it indicates that the power absorbed by 
dust is also rather constant, possibly suggesting constancy of the radiation field across all fields
even at the distance of IVCs.

We have made the same analysis with a floating $\beta$ to evaluate the robustness of the results shown 
in Fig.~\ref{fig:opacity250}. 
The greatest impact of a floating $\beta$ is on the uncertainty of $T$, and to a lesser degree on $\sigma_e$.
The values of $L$ are particularly insensitive because the modified black-body still
serves effectively as an interpolation function between the measured data points.
Even though the uncertainties on $T$ and $\sigma_e$ increase with a floating $\beta$, 
the trends seen in Fig.~\ref{fig:opacity250} are still observed, even more pronounced. 
We conclude that using $\beta=1.8$ is a reasonable and conservative approach.

\begin{figure*}
\begin{center}
\includegraphics[width=\linewidth, draft=false, angle=0]{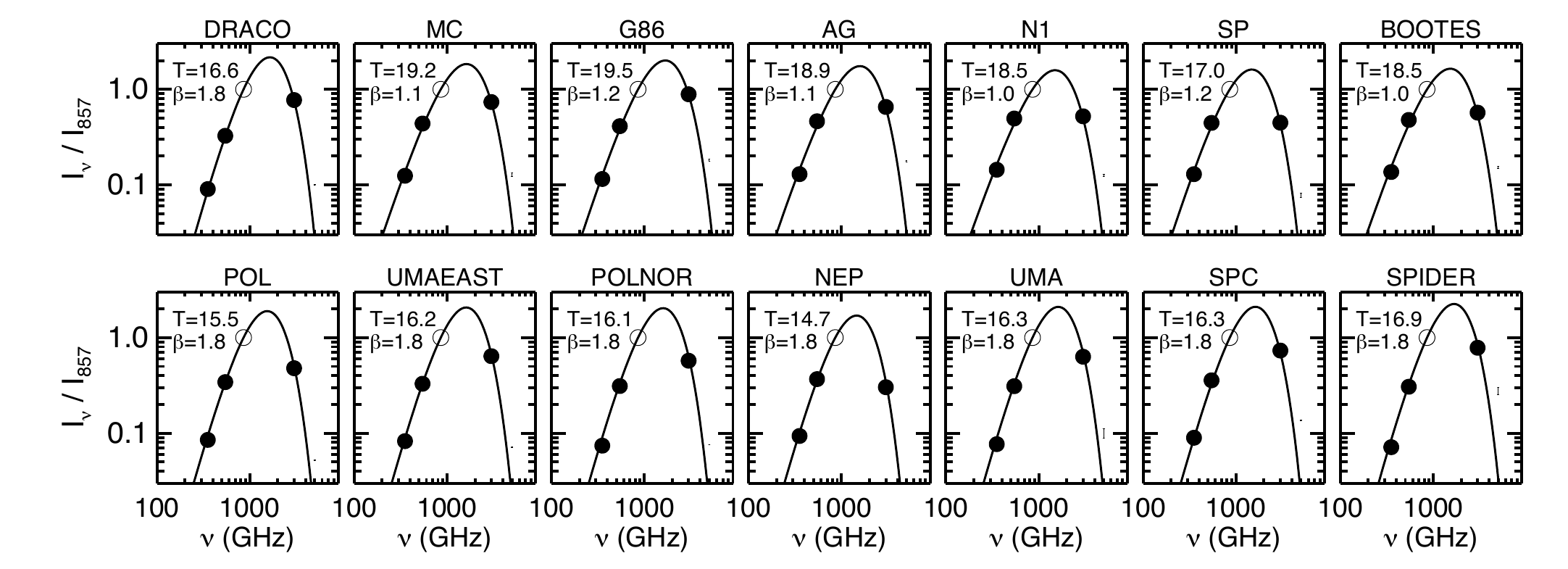}
\caption{\label{fig:all_seds_residu} Correlation coefficient of the
  residual emission with the 857\,GHz residual. Similar to Fig.~\ref{fig:all_seds}, the solid line 
here is the normalised modified black body SED
  fit using 3000, 545 and 353\,GHz data.
  We assumed a fixed $\beta=1.8$, except 
  for the six most diffuse fields (MC, G86, AG, N1, SP and BOOTES) where the residual emission
is dominated by the CIB fluctuations. In these fields $\beta=1.8$, typical of diffuse
Galactic dust emission, does not provide a good fit; these SEDs require a smaller value of $\beta$ (see text).}
\end{center}
\end{figure*}

\subsection{SED of the residual emission}\label{sec:residualsed}

{As seen in Figs.~\ref{fig:N1_dust} to \ref{fig:NEP_dust},
the residual emission once the \hi\ model is subtracted from the \Planck\ and \IRAS\ maps 
exhibits significant spatial coherence which reproduces from frequency to frequency.  
In order to estimate the SED of this residual emission in each field, 
we carried out a linear regression analysis between the residual map 
at each frequency and the residual map at 857\,GHz. 
The resulting slope of the regression (in units of MJy\,sr$^{-1}$ /  MJy\,sr$^{-1}$ and 
equal to 1 at 857\,GHz, by construction) 
can be used to estimate the shape of the spectrum of this residual emission. 
The results are presented in Fig.~\ref{fig:all_seds_residu}. 

As for the SEDs
of the \hi-correlated emission, the SEDs of the residual are well fit by
a modified black body function. 
We note a significant difference in the SED shape between low \hi\ column
density fields and brighter regions. The SEDs of brighter fields,
where the residual emission is likely to come from molecular gas
(Sect.~\ref{sec:transition}), could be fit with $\beta=1.8$ but with a 
slightly lower temperature ($T = 16.1\pm0.6$\,K) than the LVC components.
On the other hand, the SED of fainter fields, where the residual emission is most probably
dominated by CIB anisotropies, could not be fit with $\beta=1.8$. 
It is better described with $T=18.6\pm0.9$ and $\beta=1.1\pm0.1$. 
This should be taken only as a convenient fitting function, nothing physical.
The significantly different value of $\beta$ found here is probably the result of 
the complex composite nature of these fluctuations coming from the combination of 
all galaxies along the line of sight in slices of redshift which change with frequency 
\citep{planck2011-6.6}.}

\section{Discussion}

\label{sec:discussion}

\subsection{The \hi-{\rm H}$_2$ transition}\label{sec:transition}

One possible contribution to the residual emission is dust associated 
with ionized gas. The WIM has a vertical column density of 
about $1 \times 10^{20}$\,cm$^{-2}$ \cite[]{reynolds1989,gaensler2008}, a significant fraction
of the total column density in the most diffuse areas of the sky. 
\cite{planck2011-6.6} showed that, once the emission correlated with \hi\ is removed from
the \Planck\ 353, 545 and 857\,GHz data in faint fields, the residual emission has a power spectrum
compatible with $k^{-1}$, much flatter than any interstellar emissions. 
We also showed that the amplitude of the residual in faint fields 
is constant from field to field  (see Fig.~\ref{fig:pdf_residu}), compatible with an isotropic
extra-galactic emission. These are strong indications
that dust emission associated with the WIM, and not correlated with \hi, is small.

{In the eight fields with bright cirrus, 
the residuals to the IR-\hi\ correlation
are skewed toward positive values, larger than the amplitude of the CIB
fluctuations. 
The most straightforward interpretation is that these
positive residuals trace dust emission within H$_2$ gas.  This
interpretation is reinforced by the fact that in the brightest fields
(e.g., UMA, UMAEAST, and POL) we have checked that the residual emission is very 
well correlated with CO emission 
\citep{dame2001,planck2011-7.0}, confirming the previous study of \cite{reach1998a} 
in the North Celestial Loop region (see their Fig.~11).
The lower dust temperature estimated from the SED of the residual emission in all these
bright fields (see Fig.~\ref{fig:all_seds_residu}) 
is also reminiscent of what is observed in molecular clouds \citep{planck2011-7.13}.
}

{In the following we assume that the submm excess emission provides a way to estimate the
molecular gas column density within the local ISM (LVC) component\footnote{This is
not true for the DRACO field where the excess emission is likely 
to be dominated by Intermediate-Velocity gas which has accompanying CO emission \cite[]{herbstmeier1993}. 
For this reason this field was discarded in the current analysis.}. 
In order to estimate the fraction of H$_2$ in our fields
we carried out an analysis using the $I_{857}$ dust maps and the following
equation explained below:
\begin{equation}
N_{\rm H}^{{\rm LVC}\prime} = \frac{I_{857} \, - \, \epsilon_{857}^{\rm IVC}\, N_{\rm HI}^{\rm IVC} \, -\, \epsilon_{857}^{\rm HVC}\, N_{\rm HI}^{\rm HVC} \, - \, Z_{857}}{\epsilon_{857}^{\rm LVC}}.
\end{equation}
To concentrate on the local (LVC) gas, we removed the IVC and HVC-correlated
emission and the constant term from $I_{857}$, using the above results of the linear
regression in each field.  
Assuming the dust emissivity is the same in the molecular gas as in the atomic gas from which it formed,
we divided this by $\epsilon_{857}^{\rm LVC}$ for each field to
produce estimated maps of the total column density $N_{\rm H}^{\rm LVC\prime}$ 
for the local/low-velocity gas.
We note that this map still includes the fluctuations of the CIB ($C_{857}$) that act like a noise
on the true total column density $N_{\rm H}^{\rm LVC}$:
\begin{equation}
N_{\rm H}^{\rm LVC\prime} = N_{\rm H}^{\rm LVC} + C_{857}/\epsilon_{857}^{\rm LVC}.
\end{equation}

We of course have a map of the LVC atomic column density ($N_{\rm HI}^{\rm LVC}$) 
from the GBT measurements. Therefore, we are able to compute an estimate of 
the molecular column density map $N_{H2}^{\rm LVC} = (N_{\rm H}^{\rm LVC\prime} - N_{\rm HI}^{\rm LVC})/2$ and then 
calculate an estimate of the fraction of mass (or fraction of H nuclei) in molecular form,
$f({\rm H}_2) = 2\, N_{{\rm H}2}^{\rm LVC}/( 2 N_{{\rm H}2}^{\rm LVC} + N_{\rm HI}^{\rm LVC})$,
pixel by pixel. 
The results obtained by combining all pixels in all fields, except DRACO, are plotted 
in Fig.~\ref{fig:nh_over_nhi}. To produce this figure
we binned the data in $N_{\rm H}^{\rm LVC\prime}$ and within each bin examined the PDF 
of $f({\rm H}_2)$, finding its median (dark symbol) 
and half-power points (error bars, which are not necessarily symmetrical).  

\begin{figure}
\begin{center}
\includegraphics[width=\linewidth, draft=false, angle=0]{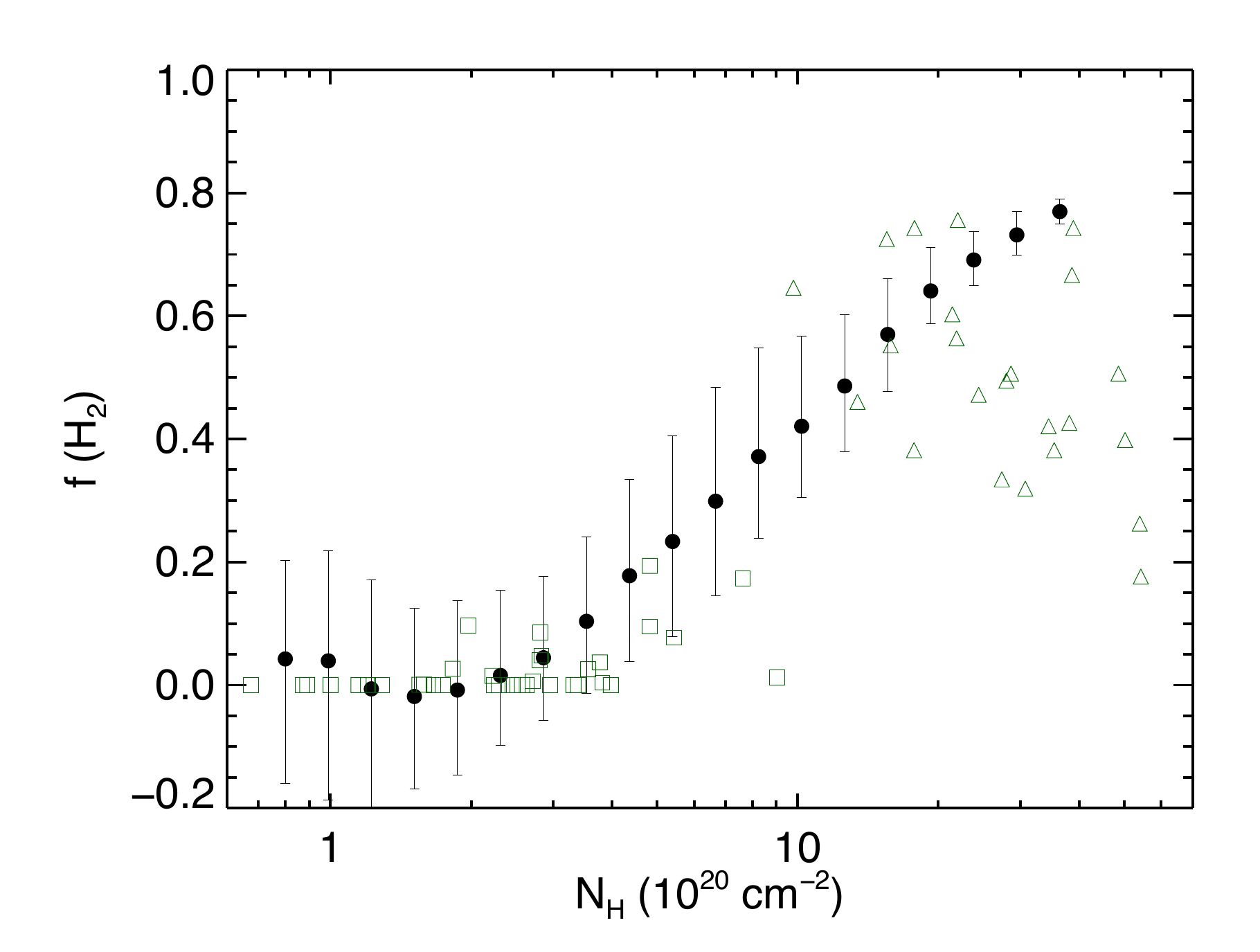}
\caption{\label{fig:nh_over_nhi} 
{Fraction $f({\rm H}_2)$ of hydrogen that is in molecular form in local gas/LVC, 
calculated from emission excess relative to the linear correlation
(see text), versus the total column density estimated using the 857\,GHz dust emission
transformed into gas column density using the emissivities $\epsilon_\nu$.  
Black points show the median value of $f({\rm H}_2)$ in bins of $N_H$ computed
using all lines of sight in our sample. The error bars show the half-width at half maximum
of the PDF computed in each $N_H$ bin. Green symbols show the results
obtained using UV absorption data from high-latitude surveys
(\citealp{gillmon2006, wakker2006}, squares) and toward O stars on lines of sight closer to
the Galactic Plane (\citealp{rachford2002,rachford2009}, triangles).}
}
\end{center}
\end{figure}

On this figure are also plotted estimates of $f({\rm H}_2)$ using {\it FUSE} data
by \citet{gillmon2006,wakker2006} (squares - high-latitude lines of sight) and 
\citet{rachford2009} (triangles - Galactic Plane lines of sight). The combination of these
two datasets covers the column density range probed in our analysis. 

The results plotted in Fig.~\ref{fig:nh_over_nhi} reveal an increase in
$f({\rm H}_2)$ beginning at $N_{\rm H}^{\rm LVC} \sim 3 \times
10^{20}$\,cm$^{-2}$ and reaching 0.8 for $N_{\rm H} \sim 4 \times 10^{21}$\,cm$^{-2}$.
Note that we are not sensitive to the much lower values of $f({\rm H}_2)$ 
found by {\it FUSE} at low column densities.
For pixels with $N_{\rm H}^{\rm LVC} < 3 \times 10^{20}$\,cm$^{-2}$ the dispersion
of our estimate of $f({\rm H}_2)$ is due to the fluctuations of the CIB. For these pixels, 
$N_{\rm H}^{\rm LVC} \approx  N_{\rm HI}^{\rm LVC}$ and therefore 
$N_{{\rm H}2}^{LVC} \approx C_{857}/2\,\epsilon_{857}^{\rm LVC}$. 
Because $C_{857}$ has a mean of zero, it produces both positive and negative
values of $f({\rm H}_2)$.

Note that the UV observations
toward O stars in the Galactic Plane (triangles) 
give some $f({\rm H}_2)$ values at the same level as we
find, but also some much lower values for a given column density.  This
suggests that the UV observations are somewhat affected by clumpiness
and/or are sampling qualitatively different lines of sight than ours at
high latitude.
Indeed \cite{wakker2006} emphasized that it is not straightforward to relate the \hi-H$_2$ 
transition to physical properties of the interstellar gas, like density in the
molecule-producing environment, or even a physical threshold in column
density required for molecule formation, because of the summing over
different environments along the line of sight.  A corollary is that
there can be quite different values of $f({\rm H}_2)$ for the same
$N_{\rm H}$, as is seen in the figure, and so comparison 
of UV data with the complementary infrared/submm analysis is of great
interest.

The overall correspondance between the trends seen independently in the \Planck\ and {\it FUSE} results 
supports our interpretation of the submm excess being caused by dust associated with
molecular hydrogen.  We would call this medium ``dark gas'' if it were not 
detected via CO \citep{planck2011-7.0}. 
Although there are lines of sight where $f({\rm H}_2)$ is quite high,
summed over all lines of sight in our survey, our study shows that the
excess emission at 857\,GHz that is not correlated with \hi\ is only 10\%
of the total emission.  Thus the fraction of the hydrogen gas mass that
is in molecular form in this sample of the high latitude diffuse
interstellar medium in the solar neighbourhood is quite low.
\cite{planck2011-7.0} estimated a value of 35\% for the entire high-latitude sky
above 15$^\circ$, which includes higher column density lines of sight
(this introduces yet another factor, \hi\ self-absorption, to make the medium dark).
Of this 35\%, about half is traced by CO, leaving about 20\% as ``dark
gas'' not traced by CO or \hi.

The structure and nature of the diffuse molecular gas can be studied
using the maps of residual (excess) dust emission.  In SPIDER (see
Fig.~\ref{fig:SPIDER_dust}), an intermediate column density field where
there is very little CO emission detected \cite[]{barriault2010}, we
observe coherent filamentary structures in the residual map that we
interpret as the presence of dust in diffuse H$_2$ gas without CO.
We have checked that the structures cannot be
accounted for by an underestimate of the 21-cm line opacity correction,
being present even with $T_{spin}$ as low as 40\,K. In this field the submm residual
provides a way to map  the first steps of the formation of
molecules in the diffuse ISM.
}

\begin{figure}
\begin{center}
\includegraphics[width=\linewidth, draft=false, angle=0]{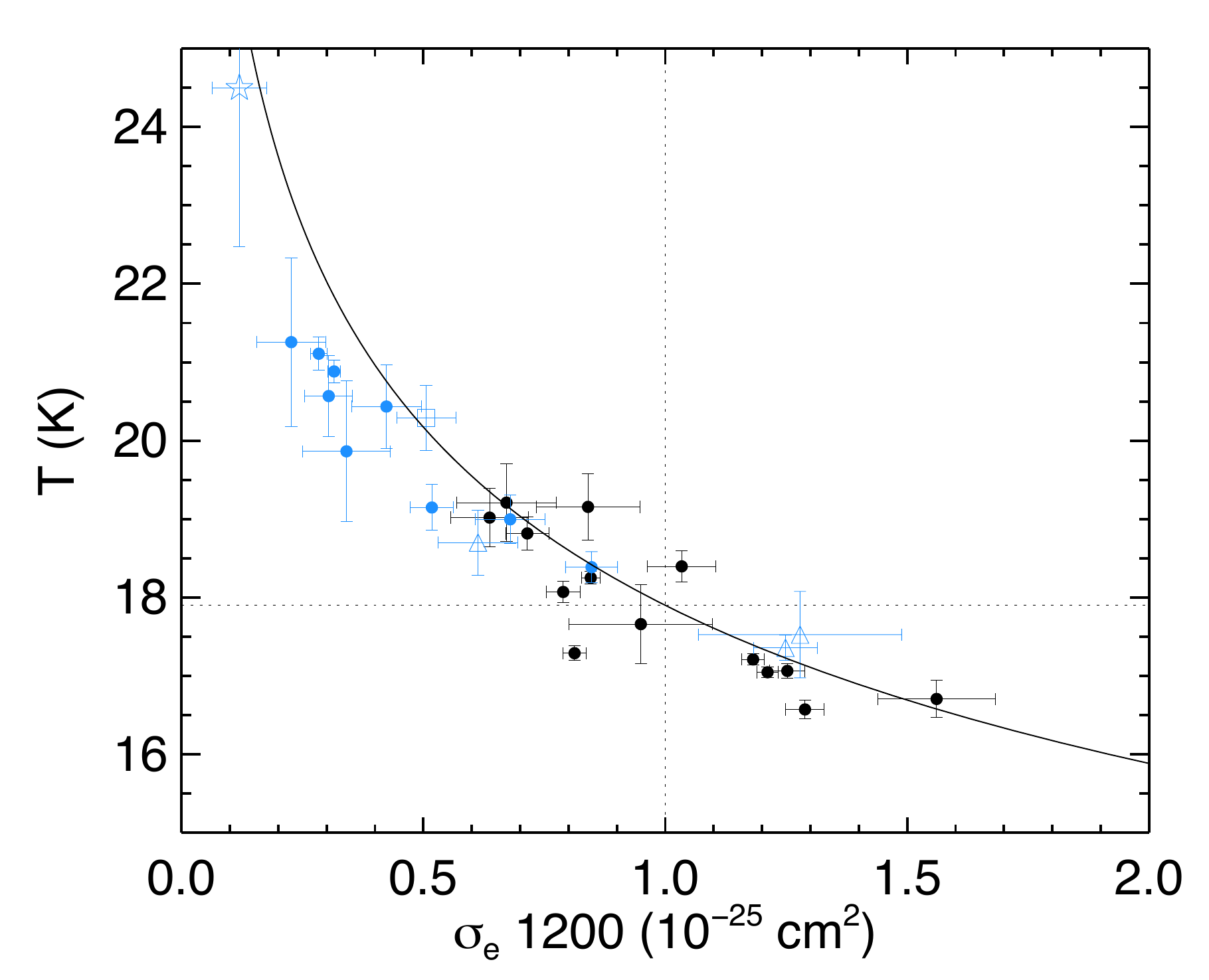}
\caption{\label{fig:sigma_T_taurus} Temperature vs emission cross-section at 1200\,GHz (250\,$\mu$m) estimated from 
modified black-body fit with $\beta=1.8$ to data from 353 to 3000\,GHz.
Black is for LVCs, blue IVCs: IV Arch (dot), IV/LLIV Arch (triangle), Complex K (square) and PP Arch (star).
The solid line respresents a constant
emitted luminosity $L$ corresponding to the diffuse ISM reference values 
($\sigma_{\rm e} = 1 \times 10^{-25}$\,cm$^{-2}$ and $T=17.9$\,K -- dotted lines).}
\end{center}
\end{figure}

\subsection{Evolution of dust}

\subsubsection{Variations of the big grain emission cross-section}

Interstellar dust evolves through grain-grain and gas-grain
interactions.  Fragmentation and coagulation of dust grains are expected
to occur in the ISM, modifying not only the grain size distribution but also the
grain structure.  The data described here provide important evidence for
dust processing in diffuse local clouds and IVCs.  As we will elucidate, the evidence foreshadowed in 
Fig.~\ref{fig:opacity250} can be seen in Figs~\ref{fig:sigma_T_taurus} and \ref{fig:color_color}.

{
Fig.~\ref{fig:sigma_T_taurus} shows the values of $T$ and $\sigma_{\rm e} (1200)$ already shown in Fig.~\ref{fig:opacity250} but here as a scatter plot.
The solid line shows the expected 
$\sigma_{\rm e} (1200)$ as a function of $T$ for a constant emitted luminosity $L$ 
(normalized to $3.8\times 10^{-31}$\,W/H, the average diffuse ISM values for
$T=17.9$\,K and $\sigma_{\rm e}$(1200)=$1.0 \times 10^{-25}$\,cm$^{-2}$). 

The emission cross-section $\sigma_{\rm e}$ reflects the efficiency of thermal dust emission per unit mass.
Dust that emits more efficiently will have a lower equilibrium temperature, the trend seen.
Note that the dust in very different environments has close to the same integrated emission ($L$) 
and therefore is absorbing about the same power (solid line). 
The comparison with the values found in the Taurus molecular cloud by \citet{planck2011-7.13} 
($T\sim 14.5$\,K, $\beta=1.8$, $\sigma_e\sim 2.0 \times 10^{-25}$\,cm$^{-2}$ and
$L=2.3 \times 10^{-31}$\,W/H)
suggests that the $T-\sigma_{\rm e}$ anti-correlation extends to colder dust and denser environments,
although the typical power absorbed is lower because of shielding.

Because $\sigma_e$ depends only on the dust properties
and on the dust-to-gas ratio, the 30\% variation observed in local ISM 
(LVC - see Fig.~\ref{fig:opacity250}), 
where we expect little variations of the metallicity, is interpreted as a
genuine variation of the dust properties from one cloud to another.
An alternative explanation would be the presence of different amounts of
H$_2$ gas spatially correlated with the \hi.  In this case the dust
opacity $\sigma_{\rm e}(1200)$ would be overestimated due to an underestimate
of the actual gas column density.  Although this scenario cannot be
excluded formally, the data shown here does not support
an increase of $\sigma_{\rm e}(1200)$ with \nh\ through the \hi-H$_2$ transition, 
as one might expect in this case. For this reason we favor an interpretation of the variations
of $\sigma_{\rm e}(1200)$ related to modifications of the grain properties.

\begin{figure}
\begin{center}
\includegraphics[width=\linewidth, draft=false, angle=0]{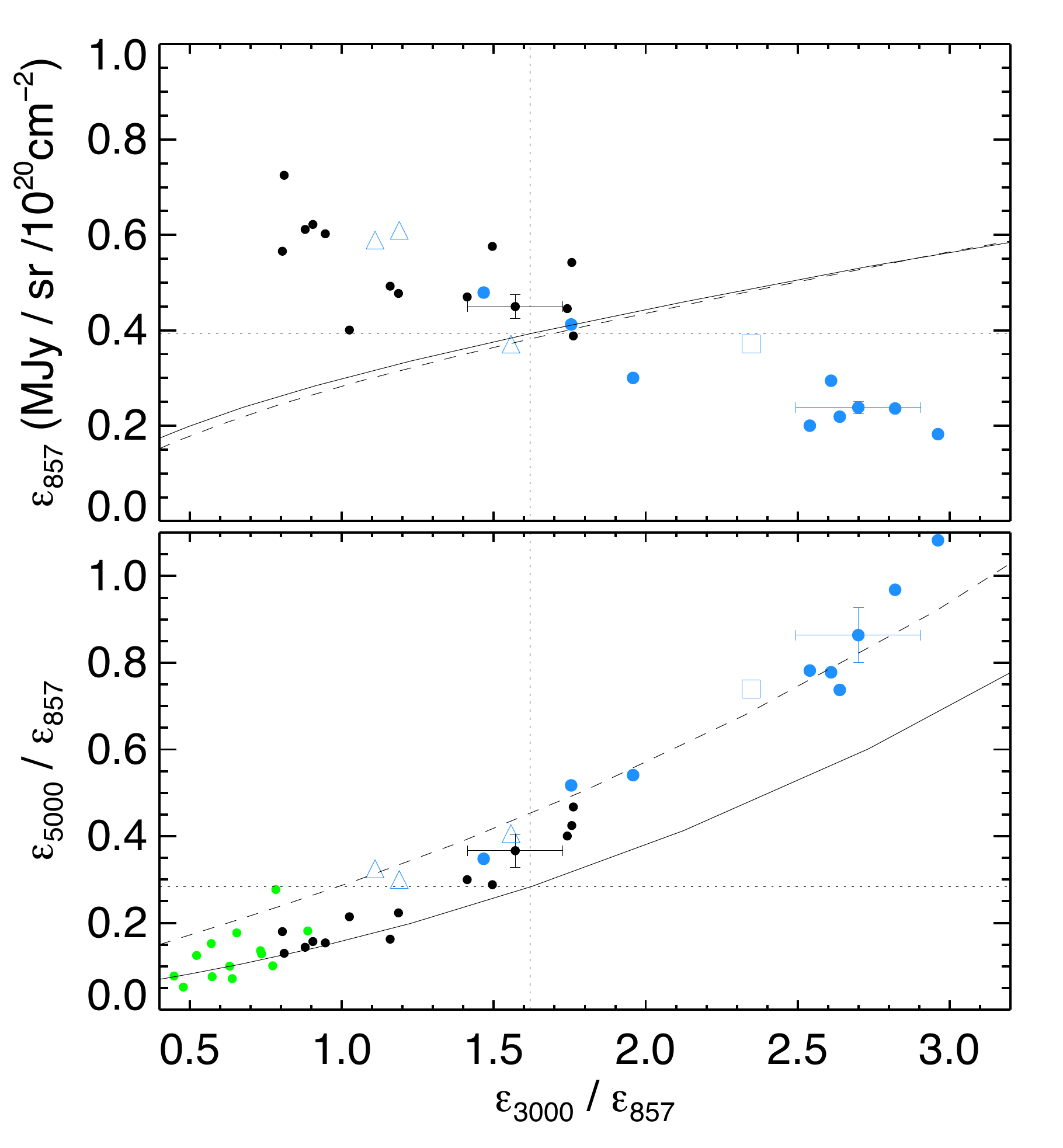}
\caption{\label{fig:color_color} Top: dust emission per N$_{\rm HI}$ at 857\,GHz versus 
  the 3000/857 GHz (100/350\,$\mu$m) ratio. 
  Local (black), IVC (blue - dot is IV Arch, triangle IV/LLIV Arch, square Complex K), residual (green).
  Solid line is the DustEM
  model for the diffuse ISM \citep{compiegne2011}, with radiation field variations from
  $G=0.1$ to $G=5$. Dashed line is the same model but with a {\em
    relative} abundance of VSGs four times higher than the standard
  diffuse ISM value.  Dotted lines gives the local ISM fiducial values
  ($G=1.0$). Typical uncertainties are shown for each \hi\ component.
  Bottom: 5000/857 GHz (60/350\,$\mu$m) ratio as a
  function of the 3000/857 GHz (100/350\,$\mu$m) ratio. }
\end{center}
\end{figure}

Fig.~\ref{fig:color_color} complements Fig.~\ref{fig:sigma_T_taurus} by comparing directly measured 
emissivities ($\epsilon_\nu$)
with the DustEM model of the average diffuse high latitude emission. 
This comparison, independent of any modified black-body fit, 
also shows strong evidence for dust evolution. The top plot of Fig.~\ref{fig:color_color} shows the
857\,GHz emissivity versus the 3000\,GHz to 857\,GHz ratio. This is compared to a simple DustEM model 
prediction for constant dust properties and a variation of the radiation
field strength from $G=0.1$ to $G=5$ ($G=1$ being the fiducial value). 
The prediction of DustEM is an increase of both $\epsilon_{857}$
and $\epsilon_{3000}/\epsilon_{857}$ with increasing radiation field strength, and therefore $T$. 
The former increases with $G$ simply because of the increase of $B_\nu(T)$. 
The ratio  $\epsilon_{3000}/\epsilon_{857}$ increases with $T$ as the peak of 
the black-body shifts towards higher frequencies.
The data points do not follow this trend showing clearly that the variations 
in the SEDs found here in LVCs and IVCs cannot be explained by local variations 
of the radiation field strength. 
The data are consistent with a decrease of the 
dust emission cross-section ($\sim \epsilon_{857}$) with temperature 
($\sim \epsilon_{3000}/\epsilon_{857}$), the same trend seen in Fig.~\ref{fig:sigma_T_taurus}.
An evolutionary model in which dust structure changes due to 
aggregation (and the reverse process, fragmentation) is qualitatively consistent with these results: 
grains with a fluffy structure will absorb about the same amount 
of optical and ultraviolet radiation per unit mass as more compact grains, 
but compared to these more homogeneous and spherical grains they are more emissive at submm wavelengths 
because of their more complex structure and therefore cool more efficiently \citep{stepnik2003}. 
}

\begin{table*}
\centering
\begin{tabular}{lccccc}\toprule
 & $\epsilon_{353}$ & $\epsilon_{545}$ & $\epsilon_{857}$ & $\epsilon_{3000}$ & $\epsilon_{5000}$ \\ \midrule
AG & $<0.010$ & $<0.029$ & $<0.055$ & $<0.071$ & $<0.024$ \\
BOOTES & $<0.079$ & $<0.258$ & $<0.519$ & $<0.289$ & $<0.106$ \\
DRACO & $<0.021$ & $<0.075$ & $<0.168$ & $<0.194$ & $<0.054$ \\
G86 & $<0.035$ & $<0.096$ & $<0.232$ & $<0.342$ & $<0.087$ \\
MC & $<0.003$ & $<0.009$ & $<0.017$ & $<0.027$ & $<0.037$ \\
N1 & $<0.016$ & $<0.047$ & $<0.094$ & $<0.054$ & $<0.018$ \\
NEP & $<0.014$ & $<0.040$ & $<0.084$ & $<0.085$ & $<0.076$ \\
SP & $<0.015$ & $<0.041$ & $<0.068$ & $<0.040$ & $<0.015$ \\
SPC & $<0.012$ & $<0.040$ & $<0.090$ & $<0.060$ & $<0.030$ \\
SPIDER & $<0.017$ & $<0.061$ & $<0.144$ & $<0.113$ & $<0.054$ \\
UMA & $<0.007$ & $<0.021$ & $<0.056$ & $<0.053$ & $<0.024$ \\
UMAEAST & $<0.017$ & $<0.054$ & $<0.135$ & $<0.105$ & $<0.028$ \\
\midrule
LVC & $0.045 \pm 0.006$ & $0.18 \pm 0.03$ & $0.52\pm0.09$ & $0.55\pm0.10$ & $0.10\pm0.02$ \\
IVC & $0.023\pm0.015$ & $0.09\pm0.04$ & $0.27\pm0.09$ & $0.66\pm0.06$ & $0.20\pm0.03$\\
HVC & $<0.013$ & $<0.040$ & $<0.085$ & $<0.077$ & $<0.034$\\
\bottomrule
\end{tabular}
\caption{\label{table:hvc_upper_limits} Upper limit for the HVC emissivities in each field. 
For the emissivities of the LVC and IVC components, weighted average and standard deviation of the 14 fields.
Last row: average HVC upper limit, excluding BOOTES, G86 and SPIDER for which the \hi\ column density of the HVC is less than
$10^{19}$\,cm$^{-2}$ inside the mask. All values are in MJy\,sr$^{-1}$/$10^{20}$\,cm$^{-2}$}
\end{table*}

\subsubsection{Dust shattering in Intermediate Velocity clouds?}

{The bottom plot of Fig.~\ref{fig:color_color} shows the 5000\,GHz to 857\,GHz
emissivity ratio as a function of the 3000\,GHz to 857\,GHz ratio for all
\hi\ components and the residual emission SEDs.  The locus from standard ISM
DustEM models shows an effect of increasing colours with increasing $G$.
The bright cirrus clouds of the LVC components (black symbols at the left
end of the plot) and the molecular residuals (green) have
colours in good agreement with the standard DustEM model, consistent with some variation of $G \lesssim 1$.

Both ratios are higher for the IVCs and the trend is offset relative to
the standard locus. The qualitative interpretation is that the emission
in these higher-frequency bands is more contaminated by non-equilibrium
emission from an increased relative abundance of very small grains
(VSGs) compared to the larger grains (BGs) in thermal equilibrium.  To
make this more quantitative, the dashed curve passing closer to the IVC
data shows the colours for a DustEM diffuse ISM model with a {\em
  relative} abundance of VSGs four times higher than the standard value.

Because of their non-equilibrium emission, VSGs dominate the
emission for $\nu >5000$\,GHz and, according to the standard DustEM model and 
for the standard radiation field ($G=1$), 
contribute to 30\% of the emission in the \IRAS\
60\,$\mu$m band even with an abundance of
only 1.6\% of the total dust mass.  By contrast, the BGs make up
90.7\% of the total dust mass, 14.2\% carbon rich and 76.5\% silicates.
Therefore an increase of a factor four of the mass
in VSGs is easy to accommodate by shattering of a small fraction of the BG dust mass.  
The fact that the emission cross-section of IVCs is a factor of two lower than for the local ISM 
can be due to a decrease of the mass in BGs (i.e., the dust-to-gas mass ratio) 
and/or to a decrease in the emission cross-section of a single grain, 
the latter affecting the equilibrium dust temperature. 
For the average temperature found in IVC$_b$ ($T=20.4$~K), and considering 
a variation of the dust-to-gas mass ratio from 1 to 1/2 the local ISM value, 
the radiation field strength needed to explain these results would be $G=1-2$,
not an unreasonable range for IVCs. It is thus difficult to conclude at this point
on the exact nature of the variation of $\sigma_e$ in IVCs.
On the other hand these results are clearly compatible with a 
modification of the BGs (total mass and/or grain properties) and a significant 
increase of the abundance of VSGs.
The details would be linked to the specific dynamical
and shock history of this interstellar matter that is part of the
Galactic fountain \citep{shapiro1976,houck1990}.}

\subsection{Dust in high-velocity clouds}

\label{sec:dustHVC}

Dust emission in HVCs is expected to be weak for several reasons.
First, their great distance from the Galactic disk 
\citep[typically several kpc --][]{wakker2001,van_woerden2004a}
implies a faint radiation field, hence a low dust equilibrium 
temperature and a low luminosity 
\citep{wakker1986,trifalenkov1993,miville-deschenes2005}.
But the main reason comes from gas phase
abundance measurements obtained from absorption lines 
of oxygen, silicon and iron in the far-UV.
Because oxygen is mostly in the gas phase and 
because its ionization potential is similar to hydrogen, 
the ratio \Oi/\hi\ provides the most robust measure of metallicity
in diffuse clouds. 
Unlike oxygen, iron and silicon are depleted 
into dust grains in the local ISM \citep{savage1996};
low ratios of Si/O and Fe/O thus trace the presence of dust.
In most HVCs the \Oi/\hi\ ratio shows that 
the metallicity is significantly sub-solar
\citep[of the order 0.1-0.5 solar, see][]{lu1998a,wakker1999,richter2001a,tripp2003,sembach2004,van_woerden2004a,fox2006,collins2007}, 
indicative of their extragalactic origin.
In addition the Si/O and Fe/O ratios in Complex~C are infered to be 
close to solar \citep{richter2001a,collins2007} indicating little silicon and 
iron locked in dust grains.
Together the overall low abundance of metals and the fact that Si and Fe seem to be
mostly in the gas phase imply a low dust to gas ratio in HVCs.

Nevertheless, because of the unknown enrichment history of HVCs that 
might set different intrinsic abundances than in the solar 
sytem\footnote{O and Si are mostly produced in Type II supernovae, Fe is produced primarly in Type Ia supernovae.}, 
the difficulty to estimate the effect of ionization on \Siii\ and \Feii\ to 
infer the abundances of Si and Fe\footnote{the true abundance of Si and Fe are 
less than the ones infered from \Siii/\hi\ and \Feii/\hi. 
To take this bias into account, an ionization correction is applied \citep{collins2003} 
that depends on gas density and radiation field strength, 
both not well constrained.},
and the unkown nature of dust in HVCs\footnote{the amount of carbonaceous dust is not constrained by 
gas phase abundances.}, 
the link between the metallicity, the depletions and the amount of dust remains indicative.
One illustrative related example, albeit for an IVC, is the finding by \cite{richter2001a} of only 
mild depletion of Si and Fe for the IVC IV-Arch which nevertheless has
dust emission at a comparable level to the local ISM. 
Thus the search for dust emission from HVCs is
an interesting complementary way to put contraints on the physical 
conditions and nature of matter in the Galactic halo.

After unsuccesful attemps to detect dust emission from HVCs using IRAS data 
\citep{wakker1986,fong1987,bates1988,desert1988},
\cite{miville-deschenes2005} claimed a detection in Complex~C
using a combination of {\it Spitzer}, {\it IRAS} and GBT data.
Compared to these previous searches, the analysis presented here extend greatly the number of fields, 
the surface covered and the wavelength coverage.
The fields used here overlap with four HVC complexes: 
Complex C (DRACO, G86, N1, NEP, SP, SPIDER, BOOTES\footnote{BOOTES is on the side of Complex C. 
It has faint emission from -80 to -100\,\kms\ that could also be associated with the IVC Complex K.}), 
Complex M (AG), Complex A (SPC, UMA, UMAEAST) and the Magellanic Stream (MC). 
HVC-correlated emission is not detected significantly in any of these fields.

At each frequency the emissivities found are compatible with zero within 3$\sigma$,
with about as many negative as positive values. 
At first glance, examination of the emissivities in Table~\ref{table:emissivities}
shows systematic behaviour across frequencies that could be taken as 
encouraging evidence for an HVC detection in some fields, like AG.
However, this does not necessarily bolster the statistical significance of the detection in a
single pass band.  This is because the dust emission is intrinsically
highly correlated frequency to frequency, as is the CIB, and so given
the generally small noise, the fits to the correlations with (the same)
\nh\ are to first order just scaled versions of each other.
This analysis establishes the main difficulties hampering the detection of dust emission from HVCs,
the contaminating foreground emission of LVCs and
IVCs and the background CIB fluctuations which greatly
increase the uncertainties in $\epsilon_\nu^{HVC}$
(Sect.~\ref{sec:monte_carlo} and Table~\ref{table:monte_carlo}).

In Table~\ref{table:hvc_upper_limits} we focus on the upper limits to the HVC emissivities.
They were computed using a Bayesian approach in the case of Gaussian errors with a bounded physical region \citep{feldman1998};
we assumed that the true emissivities can only be zero or positive. 
Table~\ref{table:hvc_upper_limits} gives the value of $\epsilon_\nu$ for which
there is 99.87\% probability (equivalent to a $3\sigma$ upper limit) 
that the true emissivity is $\leq \epsilon_\nu$. 
These upper limits are compared in Table~\ref{table:hvc_upper_limits}
to the weighted average of the measured emissivites for the LVC and IVC components;
any dust in HVCs has an emissivity, on average,
no more than 15\% of the value found in the local ISM at 857 and 3000\,GHz.

We considered the null hypothesis that the HVC emissivity is zero (or a small uniform value) in all fields and find
that, at any given frequency, the weighted standard deviation of the measured HVC emissivities 
($\sigma(\epsilon)$) is somewhat larger than expected 
given the uncertainties on each measurement. 
This is true even when considering only AG, N1, MC and SP, the fields with significant HVC \hi\ column density
and almost no masking.
For example, the probability of obtaining the measured $\sigma(\epsilon)$ is only 0.2\% at 857\,GHz for these four fields.
One explanation could be that we have underestimated the uncertainties on each measurement; to increase
the probability of the measured $\sigma(\epsilon)$ to 10\% at 857\,GHz, the uncertainties would need to be multiplied by 1.6.
But even such a moderate increase of the uncertainties seems unlikely in the most diffuse fields considered here because
the noise is dominated by the strongly constrained and well modeled CIB fluctuations \citep{planck2011-6.6}. 
A second explanation is that the $\sigma(\epsilon)$ reflects a variation of the
true $\epsilon$ from field to field. 
For these four fields we verified that, if one allows for a variation of $\epsilon$ from field to field
following a uniform distribution in the range between zero and the upper limit of each field, 
the probability of obtaining the measured $\sigma(\epsilon)$ reaches 15\% at 857\,GHz without having to increase the uncertainties
on the individual measurements. If this is the right explanation, the measured value of $\sigma(\epsilon)$ 
implies that $\epsilon$ is small but non-zero in some fields.

Finally, there have been many interesting discussions of the possible
relationships between IVCs and HVCs \citep[e.g.,][]{albert2004,wakker2001}.
Here the relative emissivity is of some relevance; as our data show,
\hi -correlated emission is readily detected in IVCs, but not in HVCs.  A
specific example is provided by our AG field which encompasses the cloud
MI in HVC Complex M \citep{wakker2001}.  The low HVC emissivity in
contrast to the fairly normal IVC emissivity (Table~\ref{table:emissivities}) 
does not support any direct relationship between MI and the intermediate 
velocity gas in the IV Arch in this direction \citep{wakker2001}.

\section{Conclusions}

\label{sec:conclusion}

We have presented results of a first comparison of \Planck\ and
\IRAS\ with new \hi\ 21-cm line GBT data for 14 high Galactic latitudes
fields, covering about 825\,deg$^2$ on the sky.  Using the
velocity information of the 21-cm data we made column density maps of
the LVC, IVC, and HVC in each field. By correlating these
\hi\ maps with the submm/infrared dust emission maps we estimated the
distinct dust emissivities for these three high-latitude components and
made corresponding SEDs from 353 to 5000\,GHz.

On average, the dust SED of the local ISM (LVC) deduced from the \IRAS\ and \Planck\ data
is compatible with that from the FIRAS data over the high-latitude sky
and is well fit with a modified black body with parameters $\beta=1.8$
and $T=17.9$\,K. On the other hand, even though the energy emitted by dust is rather
constant in our sample, we report significant variations of the dust SED shape, compatible
with variations of the dust temperature anti-correlated with the emission cross-section. 
We interpret these variations as a signature of active evolution of the dust grain structure 
through coagulation and fragmentation in the diffuse ISM.

For faint cirrus fields with average \hi\ column density lower than
$2\times10^{20}$\,cm$^{-2}$, the residual FIR-submm emission, after
removal of the \hi-correlated contributions, is normally distributed
with a standard deviation compatible with the expected level of CIB
fluctuations. The SED of the residual is unlike any interstellar
component.  For such diffuse fields we also show that the interstellar
dust emission is dominated by the contribution from atomic gas.

For fields with larger \hi\ column density there are significant
FIR-submm emission excesses that, for the brightest fields, follow the
structure of the CO emission.  For intermediate column density regions,
the residual emission shows coherent spatial structures not seen in CO,
revealing the presence of H$_2$ gas. The SED of this residual shows that
it is slightly colder than the dust in the local \hi.  There is a lot of
variation in $f({\rm H}_2)$, the fractional mass in the form of H$_2$,
but over the whole survey of these diffuse fields the fraction is only
10\%.

We report strong detection of dust emission from all IVCs in our sample.
The dust emission cross-section is typically two times lower than for the local ISM and
the relative abundance of small grains having non-equilibrium emission
is about four times higher than normal.  This evolution of the dust
properties is indicative of dust shattering in halo gas
with the dynamics of a Galactic fountain.  We also find that, compared to the local ISM, 
several of these clouds have a higher dust temperature 
which could also be the result of globally smaller grains.
The total energy emitted by dust in IVCs is comparable to what is observed in the local ISM
suggesting similar radiation field strength.

Finally, we have attempted to detect HVC-correlated dust emission. We show why this is very challenging, because of 
the uncertainties induced by foreground contamination and the CIB anisotropies. 
The average of the 99.9\% confidence upper limits on the emissivity, 
$0.15$ times the average emissivity found
in the local ISM at 857 and 3000\,GHz, is compatible with the low level of dust emission expected from HVCs based on their low 
metallicity and relatively low depletion of elements like Si and Fe.

\acknowledgements
A description of the Planck Collaboration and a list of its members can be found 
at \url{http://www.rssd.esa.int/index.php?project=PLANCK&page=Planck_Collaboration}.
We thank the anonymous referee for many helpful comments.

\bibliographystyle{aa}
\bibliography{Planck_bib,16485}
\raggedright

\begin{appendix}

\section{Noise of the \hi\ components}

\label{sec:noise_hi}

The analysis presented here relies on a correlation between
far-infrared/submm data and the \hi\ components deduced from 21-cm
emission. In order to estimate properly the uncertainties on the
correlation coefficients the noise on the \hi\ components needs to be
evaluated.  We have made this estimate using two different methods which
address different contributions to the noise \citep[see][]{boothroyd2011}.

First we applied the traditional method in which the end channels of the
21-cm spectrum, where there is no emission. The velocity channels are
independent, i.e., the spectrometer resolution is better than the
0.8\,\kms\ channel width.  A large number of end channels can be used to
compute a map of the standard deviation of the noise at each $(x,y)$
position:
\begin{equation}
\label{eq:noisehi}
\delta_{Tb}(x,y) = \sqrt{\frac{1}{N_\delta-1}\sum_{v=v1}^{v2}
T_b^2(x,y,v) - \frac{1}{N_\delta} \left( \sum_{v=v1}^{v2} T_b(x,y,v)
\right)^2},
\end{equation}
where $N_\delta = v2-v1+1$ is of the order of 200 channels.  For the GBT
data, this is a very flat map, whose typical value is the same as the
standard deviation of a single end channel.  This is about 0.17\,K for a
single visit to a field.  Some fields were mapped two or three times.

The uncertainty map $\delta N_{\rm HI}(x,y)$ for the column density of a
given \hi\ component summed over $N$ channels is
\begin{equation}
\delta N_{\rm HI}(x,y) \mbox{ [cm$^{-2}$] } = A \sqrt{N} \delta_{Tb}(x,y)
\Delta v,
\end{equation}
where $\Delta v$ is the channel width in km\,s$^{-1}$.  For $N \sim 60$,
this amounts to $\delta N_{\rm HI} \sim 0.02 \times 10^{20}$\,cm$^{-2}$ for a
single visit.  This assumes, as is usual, that the noise properties
estimated using the end channels are representative of the noise in
channels summed to build the column density maps.  Actually, the channel
noise scales as roughly $(1 + T_b/20{\rm K})$ and this can be taken into
account.

We developed a second method which exploits the fact that the
\hi\ brightness-temperature cubes were built by averaging independent
data taken in different polarisations (called $XX$ and $YY$). Each
polarisation observation can be reduced separately, which requires
separate baseline estimates for the spectra in the $XX$ and $YY$ cubes.
Baseline fitting is another source of error.  Taking the difference
between the $XX$ and $YY$ cubes removes the common unpolarized \hi
emission, leaving in each velocity channel only the thermal noise and
systematic effects from the baseline subtraction. The difference cube is
\begin{equation}
\Delta(x,y,v) = (T_{XX}(x,y,v) - T_{YY}(x,y,v))/2,
\end{equation}
where we divide by two to give the same statistics as in the average of
the cubes.

We make a map of the column density differences (divided by two) over
the same channel range as for \nh:
\begin{equation}
\label{eq:noisehi_2}
\Delta N_{\rm HI}(x,y) = A\, \Delta v \, \sum_{v=v1}^{v2} \Delta(x,y,v).
\end{equation}
The estimate of $\delta N_{\rm HI}$ is then the standard deviation of this
map.

This method gives an uncertainty typically 1.3 times the simple estimate
using end channels (it can reach up to 2 times for the brightest
components) because it includes baseline uncertainties and the increase
of thermal noise with signal.  The column density uncertainties
evaluated with this second method for each component in each field are
given in Table~\ref{table:mainHI}). There are additional errors from the
stray radiation correction that can affect the local and IVC components.
The total uncertainty including this additional contribution can be
estimated in a similar way to the second method if there are two or more
visits to compare \citep{boothroyd2011}.
Because in cases with multiple visits these
are found to be not much larger, and because the uncertainties in
\nh\ are not the major source of uncertainty in the correlation analysis
(Sect.~\ref{sec:monte_carlo}), adopting the values in
Table~\ref{table:mainHI} is satisfactory for our purposes.

\section{Estimating the noise of the \Planck\ and \IRAS\ maps}

\label{sec:noise_hfi_iris}

To estimate the noise level of the \IRAS\ and \Planck\ maps we use the
difference between independent observations of the same sky. For \IRAS,
the original ISSA plates \citep{wheelock1993} were delivered with three
independent set of observations (called HCONs), each obtained at
different periods during the life of the satellite.  Each HCON has its
own coverage map. The ISSA final product is the coverage-weighted
co-addition of the three HCONs.  The IRIS product
\cite[]{miville-deschenes2005a} also contains the three HCONs.

A similar approach was used to estimate the noise of each \Planck\ map.  
For each pointing period \Planck\ scans the sky about 50
times. The Data Processing Centre delivered two maps made with data from
the first half and second half of each pointing period, respectively. In
this case the number of samples used to estimate the signal at a given
sky position is the same in the two maps (i.e., the coverage map is
identical for the two maps). Again we used the difference between 
these two maps of the same region of the sky to estimate the noise.

\begin{table}
\begin{center}
\begin{tabular}{lccccc}\toprule
Field & $\sigma_{353}$ & $\sigma_{545}$ & $\sigma_{857}$ & $\sigma_{3000}$ & $\sigma_{5000}$\\ \midrule
AG & 0.0055 & 0.0095 & 0.0094 & 0.0295 & 0.0156\\
BOOTES & 0.0078 & 0.0122 & 0.0124 & 0.0243 & 0.0156\\
DRACO & 0.0044 & 0.0075 & 0.0073 & 0.0234 & 0.0115\\
G86 & 0.0056 & 0.0085 & 0.0086 & 0.0243 & 0.0147\\
MC & 0.0088 & 0.0140 & 0.0139 & 0.0306 & 0.0172\\
N1 & 0.0050 & 0.0084 & 0.0083 & 0.0218 & 0.0120\\
NEP & 0.0040 & 0.0065 & 0.0063 & 0.0265 & 0.0138\\
POL & 0.0061 & 0.0095 & 0.0098 & 0.0333 & 0.0141\\
POLNOR & 0.0054 & 0.0087 & 0.0088 & 0.0252 & 0.0134\\
SP & 0.0055 & 0.0086 & 0.0088 & 0.0268 & 0.0153\\
SPC & 0.0065 & 0.0104 & 0.0104 & 0.0280 & 0.0142\\
SPIDER & 0.0058 & 0.0093 & 0.0096 & 0.0277 & 0.0143\\
UMA & 0.0065 & 0.0101 & 0.0108 & 0.0340 & 0.0156\\
UMAEAST & 0.0072 & 0.0111 & 0.0113 & 0.0365 & 0.0167\\ \bottomrule
\end{tabular}
\end{center}
\caption{\label{table:noise_hfi_iris} \Planck\ and \IRAS\ noise levels
  (in MJy\,sr$^{-1}$) for each field and at each frequency. The value
  given here is the noise level in the map that has been convolved to
  the 9.4\arcmin\ GBT resolution.}
\end{table}

Here we want to estimate the average noise level $\delta I_\nu$ of an
\IRAS\ or \Planck\ map $I_\nu$ used for the analysis given a difference map
\begin{equation}
\Delta_\nu(x,y) = (I_{\nu,1}(x,y) - I_{\nu,2}(x,y))/2
\end{equation}
obtained from independent observations, each $I_{\nu,i}$ map having its
own coverage map $N_{\nu,i}$.  In the general case
\citep{miville-deschenes2005a}, $\delta I_\nu$ is obtained by taking the
standard deviation of the map
\begin{equation}
\Delta_\nu^\prime(x,y) = \Delta_\nu(x,y) \sqrt{ \frac{4\,N_{\nu,1}(x,y)\,
N_{\nu,2}(x,y)}{ N_\nu(x,y) \, (N_{\nu,1}(x,y) + N_{\nu,2}(x,y) } }.
\end{equation}
In the case of \Planck, $N_{\nu,1}=N_{\nu,2}=N_\nu/2$ and so $\Delta_\nu^\prime =
\Delta_\nu$ and the standard deviation of the average map is the same
for the difference map, as expected.  Because the ISSA plates are the
combination of three observations with different coverages such a
simplification is not possible and the above more general equation has
to be used.

In our analysis we used the \IRAS\ and \Planck\ maps convolved to the
9.4\arcmin\ resolution of the GBT.  Therefore, the appropriate $\delta I_\nu$
is obtained by first convolving the $\Delta_\nu^\prime(x,y)$ map and
then taking the standard deviation.  These are the noise levels
tabulated for each field in Table~\ref{table:noise_hfi_iris}.

\end{appendix}

\end{document}